\documentclass{aa}  
\usepackage[varg]{txfonts}
\usepackage{natbib}
\bibpunct{(}{)}{;}{a}{}{,} 
\usepackage{graphicx}
\usepackage{caption}
\usepackage{subcaption}
\usepackage{float}
\usepackage[dvipsnames]{xcolor}
\usepackage{longtable}
\usepackage{pdfpages}
\usepackage{txfonts}

\newcommand{\kms} {$\mathrm{km\,s}^{-1}$}

\usepackage[hidelinks]{hyperref}
\hypersetup{
  colorlinks   = true, 
  urlcolor     = blue, 
  linkcolor    = blue, 
  citecolor   = cyan 
}

\newcommand{\revone}{}
\newcommand{\revtwo}{} 
\begin{document}

   \title{Discovery of $\sim$2200 new supernova remnants in 19 nearby star-forming galaxies with MUSE spectroscopy}

   \subtitle{}

   \author{Jing Li
        \inst{1}\fnmsep\thanks{\email{Jing.Li@uni-heidelberg.de}},
          K. Kreckel\inst{1}, S. Sarbadhicary\inst{2}
          \and
          Oleg V. Egorov\inst{1}, B. Groves\inst{3}, K. S. Long\inst{4} 
          \and
          Enrico Congiu\inst{5} 
          \and
          Francesco Belfiore\inst{6} 
          \and
          Simon C.~O. Glover\inst{7} 
          \and
          Ashley~.T~Barnes\inst{8}
          \and
          Frank Bigiel\inst{9}
          \and
          Guillermo A. Blanc\inst{10,11}
          \and
          Kathryn Grasha\inst{12}
          \and
          Ralf S.\ Klessen\inst{7,13}
          \and
          Adam Leroy\inst{2}
          \and
          Laura A. Lopez\inst{2}
          \and
          J.Eduardo Méndez-Delgado \inst{1}
          \and
          Justus Neumann\inst{14}
          \and
          Eva Schinnerer\inst{14}
          \and
          Thomas G. Williams\inst{15}
          \and
          PHANGS collaborators\inst{99}}

   \institute{Astronomisches Rechen-Institut, Zentrum für Astronomie der Universität Heidelberg, Mönchhofstraße 12-14, 69120 Heidelberg, Germany
   \and
   Department of Astronomy, The Ohio State University, 140 West 18th Avenue, Columbus, Ohio 43210, USA
   \and
   International Centre for Radio Astronomy Research, University of Western Australia, 7 Fairway, Crawley, 6009 WA, Australia
   \and
  Space Telescope Science Institute, 3700 San Martin Drive, Baltimore, MD 21218, USA
   \and
    European Southern Observatory (ESO), Alonso de Córdova 3107, Casilla 19, Santiago 19001, Chile
   \and
   INAF -- Arcetri Astrophysical Observatory, Largo E. Fermi 5, I-50125, Florence, Italy
   \and
   Zentrum f\"{u}r Astronomie, Universit\"{a}t Heidelberg, Institut f\"{u}r Theoretische Astrophysik,  Albert-Ueberle-Str. 2, 69120 Heidelberg, Germany
   \and 
   European Southern Observatory (ESO), Karl-Schwarzschild-Stra{\ss}e 2, 85748 Garching, Germany
   \and
   Argelander-Institut für Astronomie, Universität Bonn, Auf dem Hügel 71, 53121 Bonn, Germany
   \and
   Observatories of the Carnegie Institution for Science, 813 Santa Barbara Street, Pasadena, CA 91101, USA
   \and
   Departamento de Astronom\'{i}a, Universidad de Chile, Camino del Observatorio 1515, Las Condes, Santiago, Chile
   \and
   Research School of Astronomy and Astrophysics, Australian National University, Canberra, ACT 2611, Australia
   \and
   Universit\"{a}t Heidelberg, Interdisziplin\"{a}res Zentrum f\"{u}r Wissenschaftliches Rechnen, Im Neuenheimer Feld 205, 69120 Heidelberg, Germany
   \and
   Max-Planck-Institut f\"{u}r Astronomie, K\"{o}nigstuhl 17, D-69117 Heidelberg, Germany
   \and
   Sub-department of Astrophysics, Department of Physics, University of Oxford, Keble Road, Oxford OX1 3RH, UK}

   \date{Received September 15, 1996; accepted March 16, 1997}

 
  \abstract
   {
   In this paper, we present the largest extragalactic survey of supernova remnant (SNR) candidates in nearby star-forming galaxies using exquisite spectroscopic maps from MUSE. Supernova remnants (SNRs) exhibit distinctive emission-line ratios and kinematic signatures, which are apparent in optical spectroscopy. Using optical integral field spectra from the PHANGS-MUSE project, we identify SNRs in 19 nearby galaxies at $\sim$100~pc scales. We use five different optical diagnostics: (1) line ratio maps of [\ion{S}{ii}]/H$\alpha$; (2) line ratio maps of [\ion{O}{i}]/H$\alpha$;  (3) velocity dispersion map of the gas; (4) and (5) two line ratio diagnostic diagrams from BPT diagrams to identify and distinguish SNRs from other nebulae. \revtwo{Given that our SNRs are seen in projection against \ion{H}{ii} regions and diffuse ionized gas, in our line ratio maps we use a novel technique to search for objects with [\ion{S}{ii}]/H$\alpha$ or [\ion{O}{i}]/H$\alpha$ in excess of what is expected at fixed H$\alpha$ surface brightness within photoionized gas.} In total, we identify 2,233 objects using at least one of our diagnostics, and define a subsample of 1,166 high-confidence SNRs that have been detected with at least two diagnostics. The line ratios of these SNRs agree well with the MAPPINGS shock models, \revtwo{and we validate our technique using the well-studied nearby galaxy M83, where all SNRs we found are also identified in literature catalogs and we recover 51\% of the known SNRs.} The remaining 1,067 objects in our sample are detected with only one diagnostic and we classify them as SNR candidates. We find that $\sim$ 35\% of all our objects \revtwo{overlap with the boundaries of \ion{H}{ii} regions from literature catalogs, highlighting the importance of using indicators beyond line intensity morphology to select SNRs}. 
   \revtwo{We find that the [\ion{O}{i}]/H$\alpha$ line ratio is responsible for selecting the most objects  (1,368; 61\%), however, only half are classified as SNRs, demonstrating how the use of multiple diagnostics is key to both increasing our sample size and improving our confidence in our SNR classifications.  }  
   }


   \maketitle
%

\section{Introduction}
\label{sec:intro}

The interstellar media (ISM) of galaxies are dotted with supernova remnants (SNRs) -- emission nebulae produced by supernova (SN) shocks propagating through the ambient ISM. Over 300 SNRs have been discovered in the Milky Way \citep{Green2019}, and more than a thousand SNRs and SNR candidates have been found in external galaxies from multi-wavelength surveys \citep[see][and references therein]{vuvcetic2015optical, long2017galactic}. SNRs are vital to our understanding of collisionless shocks \citep[e.g.][]{Raymond1979, Chevalier1980,  Mckee1980, Marcowith2016}, cosmic ray acceleration \citep{Blandford1987, Blasi2013, Bell2013}, progenitor models and explosion physics of SNe \citep[e.g][]{Chevalier2005, Schaefer2012, Krause2008, Patnaude2017}, the formation and survival of dust grains in SN ejecta \citep[e.g.][]{Dwek1992, Williams2017, Sarangi2018}, the properties of central compact objects \citep[e.g.][]{Dubner1998, Holland2017, Katsuda2018}, and the effects of shocks on interstellar clouds \citep[e.g.][]{White1991, Reach2006, Slane2015, Koo2020}. 

Extragalactic SNR surveys provide a unique perspective into the evolution of SNR shocks, their stellar progenitors, and impact on the surrounding ISM. They are complementary to Galactic SNRs, which undoubtedly offer the most detailed multiwavelength views of shocks \citep[e.g.][]{milisavljevic2024jwst}, but are also affected by distance uncertainties, variable line-of-sight extinction, and source confusion at low latitudes where most objects are located \citep[e.g.][]{Green2005,reach2006spitzer, Green2008}. Examples of well-studied extragalactic SNR populations include those in the Magellanic Clouds \citep[e.g.][]{mathewson1973supernova, Long1981, maggi2016population, Maggi2019}, M31 \citep{Magnier1995, Lee2014a, Galvin2014}, M33 \citep{Gordon1999, Long2010, Lee2014b, long2018mmt, White2019}, M83 \citep[e.g.][]{dopita2010supernova, blair2014expanded, Winkler2017, long2022supernova}, NGC 6946 \citep[e.g.][]{Matonick1997, Lacey1997, Long2019, Long2020}, M51 \citep{winkler2021optical}, and the Sculptor Group galaxies \citep[e.g.][]{blair1997identification, Pannuti2000, Pannuti2002,kopsacheili2024supernova}.

With extragalactic surveys, one can observe the distribution of SNRs of different ages across entire galaxies for face-on galaxies, at the known, fixed distances of the galaxies. This leads to reliable estimates of SNR sizes and luminosities at different wavelengths, from which one can infer a variety of statistical information about the SNR population, such as the distribution of evolutionary stages, visibility times, ambient densities, and magnetic field properties \citep{Gordon1999, Chomiuk2009, Thompson2009, badenes2010size, Long2010, Asvarov2014, Sarbadhicary2017, Elwood2019}. The spatial correlation of these SNRs with ISM and stellar population surveys of nearby galaxies have led to unique constraints on the progenitor age distribution of SNe \citep[e.g.][]{Badenes2009, Maoz2010, Jennings2014, Williams2014, DiazRod2018, Auchettl2019, williams2019masses, Koplitz2021}, highly relevant to the many open questions about progenitor models of Type Ia and core-collapse SNe \citep[see][for detailed discussion]{Maoz2014, Smartt2015}. 

The discovery and identification of these extragalactic SNRs have been primarily at optical wavelengths from observations of forbidden emission lines. The most widely used diagnostic to distinguish SNRs from photoionized nebulae has been the [\ion{S}{ii}]/H$\alpha$>0.4 criteria \citep[e.g.][]{mathewson1973supernova,dodorico1980catalogue}, since low-ionization species like S$^+$ are more abundant in the extended recombination zones of radiative shocks compared to \ion{H}{ii} regions \citep{Raymond1979,dopita2005modeling,allen2008mappings}. This contrasts with the Milky Way, where SNRs are primarily identified with radio continuum observations that can pierce through the midplane dust, and the power-law index of their synchrotron spectra can be measured \citep{green2014catalogue,anderson2017galactic,dokara2021global,cotton2024meerkat}. The Magellanic Clouds also have a significant number of SNRs identified at X-ray (for young SNe) and radio wavelengths \citep[e.g.][]{maggi2016population, Maggi2019, Kavanagh2022, Bozzetto2023, Zangrandi2024}. Beyond the Local Group, however, radio and X-ray observations have yet to reach sufficient spatial resolution and sensitivity to independently discover large numbers of SNRs like optical surveys \citep{Russell2020}.

Most optical SNR surveys have relied on narrow-band filter imaging which, despite its success in discovering many SNRs, has not been without its limitations. A major issue is the continuum subtraction, which is done by scaling the flux from a broadband image or an off-band narrow-band image. This was often sub-optimal in surveys due to variable seeing conditions and color properties of stars in the fields \citep[e.g.][]{dopita2010supernova, blair2014expanded}, which have particularly affected the recovery of faint, low-surface brightness SNRs \citep{long2022supernova}. Typically the filters blended adjoining lines, such as the [\ion{N}{ii}]$\lambda$6549,6583 and H$\alpha$, which compromised the [\ion{S}{ii}]/H$\alpha$ ratio \citep{blair2014expanded}. As a result, narrow-band imaging and band ratios alone were often not sufficient to confirm objects as SNRs, and additional medium/high-resolution follow-up spectroscopy of individual objects was necessary \revtwo{to provide additional line ratios and kinematic indicators} \citep{Winkler2017, long2018mmt}. 

The proliferation of wide-field integral field unit (IFU) spectroscopy with instruments such as MUSE \citep{bacon2010muse} and SITELLE \citep{drissen2019sitelle} has significantly improved our ability to survey extragalactic nebulae with comparable efficiency as narrow-band imaging, while at the same time circumventing many of its shortcomings. The ability to obtain a spectrum per pixel at high sensitivity and medium spectral resolution ($R > 10^3$) has led to more accurate modeling and subtraction of the underlying stellar continuum, and separation of adjoining emission lines. IFS has also opened up access to additional line diagnostics for identifying SNRs, such as [\ion{O}{i}] and [\ion{O}{ii}] \citep{Fesen1985}, linear combinations of S, N, and O forbidden lines \citep{kopsacheili2020diagnostic}, and the widths of forbidden emission lines, which are expected to be broader than in \ion{H}{ii} regions due to the presence of high-velocity shocked material \citep{points2019kinematics}. The promise of IFU-based SNR searches have already been demonstrated in individual galaxies such as NGC 300 \citep{Roth2018}, NGC 3344 \citep{moumen20193d}, NGC 4030 \citep{Cid2021}, M83 \citep{long2022supernova}, and NGC 4214 \citep{Vincens2023}, \revtwo{with more recent work building towards cataloging thousands of shock-ionized nebulae across larger galaxy samples \citep{congiu2023phangs}, hereafter C23.}

In this paper, we push the boundaries of extragalactic SNR studies even further, to a sample of 19 galaxies between $5-20$ Mpc, as part of the PHANGS-MUSE survey \citep{emsellem2022phangs}. We have already characterized and constructed catalogs of likely \ion{H}{ii} regions and diffuse ionized gas (DIG) in these galaxies \citep[see e.g.][]{emsellem2022phangs, belfiore2022tale, santoro2022phangs, groves2023phangs, congiu2023phangs, egorov2023phangs}, which is critical for identifying SNRs. Although the SNRs are unresolved or barely resolved (average spatial resolution of 67 pc) at these distances, this is the largest single survey of extragalactic SNRs ever conducted, with a diverse sample of host galaxies that span a range of stellar mass and star formation rate.  These galaxies have substantial multi-wavelength data at $\sim$1\arcsec resolution or better from facilities including, but not limited to, HST, JWST, ALMA, VLA, AstroSAT, and Chandra. This multi-wavelength synthesis provides the most complete observational census of the multi-phase ISM and the multi-generation stellar population that are associated with SN explosions.

The focus of the current paper is to introduce the catalog, the selection procedure adopted, and the basic properties of the PHANGS-MUSE SNRs. Subsequent papers will investigate the correlation of these SNRs with stellar population and ISM maps of the galaxies. The paper is organized as follows -- In Section \ref{sec:data}, we introduce the PHANGS-MUSE dataset for our sample. Then in the following Section \ref{sec:methods}, we explore methods to identify SNRs in these galaxies and describe how we construct our catalogs. The resulting catalogs and SNRs properties are presented in Section \ref{sec:results}. Further discussion and comparison with literature works are included in Section \ref{sec:discussion}. Finally, we summarise the main conclusions in Section \ref{sec:conculsion}.
~
\begin{table*}
\caption{Properties of PHANGS--MUSE galaxies used in this work. }
\label{tab:property}
\centering
\begin{tabular}{lrccccccccc}
\hline\hline\noalign{\vskip 0.05in}
Name & parent & SNRs&Distance & plate scale & ${\rm log}(M_\star)$ & $\rm log(SFR)$  & PA & $i$ & Survey Area\\
& number &number& [Mpc] &pc/$''$ & [$M_\odot$]	& [$M_\odot \mathrm{yr}^{-1}$] & [deg]  & [deg] & [kpc$^2$]\\
\hline\noalign{\vskip 0.05in}
IC~5332 & 36&24&9.0 & 43.7& 9.68 & -0.39 & 74.4 & 26.9  & 34 \\
NGC~628 & 120&79&9.8 & 47.7& 10.34 & 0.24 &20.7 & 8.9  & 89 \\
NGC~1087 & 123&29&15.9 & 76.8& 9.94 & 0.11 &359.1 & 42.9  & 126\\
NGC~1300 &23&15& 19.0 & 92.1& 10.62 & 0.07 & 278.0 & 31.8  & 356\\
NGC~1365 & 63&38&19.6 & 94.9& 11.00 & 1.24 &201.1 & 55.4  & 409\\
NGC~1385 & 114&63&17.2 & 83.5& 9.98 & 0.32 &181.3 & 44.0  & 101\\
NGC~1433 &46&24& 18.6 & 90.3& 10.87 & 0.05 &199.7 & 28.6  & 426\\
NGC~1512 &25&20& 18.8 & 91.3& 10.72 & 0.11 &261.9 & 42.5  & 266\\
NGC~1566 &180&101& 17.7 & 85.8& 10.79 & 0.66 &214.7 & 29.5  & 208\\
NGC~1672 &61&28& 19.4 & 94.1& 10.73 & 0.88 &134.3 & 42.6  & 250\\
NGC~2835 &187&119& 12.2 & 59.2& 10.00 & 0.10 &1.0 & 41.3  & 87\\
NGC~3351 &46&24& 10.0 & 48.3& 10.37 & 0.12 &193.2 & 45.1  & 73\\
NGC~3627 &152&86& 11.3 & 54.9& 10.84 & 0.59 &173.1 & 57.3  & 85\\
NGC~4254 &332&154& 13.1 & 63.5& 10.42 & 0.49 &68.1 & 34.4  & 169\\
NGC~4303 &303&130& 17.0 & 82.4& 10.51 & 0.73 &312.4 & 23.5  & 214\\
NGC~4321 &132&80& 15.2 & 73.7& 10.75 & 0.55 &156.2 & 38.5  & 191\\
NGC~4535 &75&54& 15.8 & 76.5& 10.54 & 0.34 &179.7 & 44.7  & 124\\
NGC~5068 &194&92& 5.2 & 25.2 & 9.41 & -0.56 &342.4 & 35.7 & 23\\
NGC~7496 &21&6& 18.7 & 90.8& 10.00 & 0.35 &193.7 & 35.9  & 92\\
\hline\noalign{\vskip 0.05in}
\end{tabular}\\
\textbf{Note:} For column \textit{parent} and \textit{SNRs}, see Section \ref{DAP} for their detailed classification. Distances of PHANGS galaxies come from \cite{anand2021distances}, with a complete set of references provided in the acknowledgments. Their stellar masses and SFRs are from  \cite{leroy2021phangs}. Position angle (PA) and inclination ($i$) come from the CO (2-1) dynamics of PHANGS-ALMA  \citep{lang2020phangs}. Survey areas are using the values from \cite{belfiore2022tale}.
\end{table*}
\section{Data}
\label{sec:data}
We select SNRs using data from the PHANGS-MUSE survey of 19 nearby ($\lesssim$20 Mpc) galaxies, conducted using the Multi Unit Spectroscopic Explorer (MUSE) instrument mounted on the Very Large Telescope (VLT) in Chile \citep{bacon2010muse}. MUSE provides $\sim$ arcsecond resolution integral field spectroscopy in the wavelength range of 4800 to 9300 \AA\ at a spectral resolution of $R\sim3000$. 
Full details about the sample, observations, reductions, and data products are found in \citet{emsellem2022phangs}. These 19 galaxies (see Table \ref{tab:property}) populate the star-forming main sequence of galaxies, with stellar masses in the range of 10$^{9.4}-10^{11}$M$_\odot$. 
The pixel scale is 0.2"/pixel and the typical seeing in R-band is 0.8" \citep{emsellem2022phangs}, corresponding to an average spatial resolution of 67 pc at the distance of our targets \citep{anand2021distances,scheuermann2022planetary}. 
Our $\sim$100~pc resolution is larger than the typical size of SNRs \citep{long2010chandra,tullmann2011chandra,asvarov2014size}, but is well suited to spatially distinguish SNRs from nearby unrelated \ion{H}{ii} regions and surrounding DIG. 
\revtwo{To limit the effects of spatial blending between SNRs and surrounding line emitting regions we used the native resolution data, which provide the best possible spatial resolution at the expense of a lightly varying PSF between fields. As many of the regions are found to be blended with luminous \ion{H}{ii} regions (see Section \ref{sec:overlapHII}), our source identification is limited more by seeing than by sensitivity.  Using smoothed data (e.g. the `copt' data products), we obtain a sample of objects that is 60\% smaller. }

The main PHANGS-MUSE data products we use are the total line intensity and kinematic (velocity and velocity dispersion) maps for the set of strong lines discussed below. A detailed description of the datasets, including the Data Analysis Pipeline (DAP) used to extract these parameters, is given in \cite{emsellem2022phangs}, and here we provide a summary. 
Spectral fitting is performed by using the penalized pixel-fitting code  \texttt{pPXF} 
 \citep{cappellari2004parametric,cappellari2017improving}. 
The stellar continuum in the spectra was identified and removed using 
stellar population templates from the E-MILES simple stellar population models \citep{vazdekis2016uv}. 
Emissions lines are fitted with Gaussian templates to recover line fluxes as well as line kinematics (velocity and velocity dispersion). \revtwo{We note that at the MUSE spectral resolution, we generally find that the line emission associated with SNRs is well described by single Gaussian fits}. In \texttt{pPXF}, we performed fitting for 23 lines along with their corresponding gas kinematics. The kinematics of some emission lines were grouped together and fitted simultaneously, such as Balmer lines, and both low and high ionization lines. In this work we use [\ion{S}{ii}]$\lambda$6716, [\ion{S}{ii}]$\lambda$6730, H$\alpha$, [\ion{O}{i}]$\lambda$6300, [\ion{O}{iii}]$\lambda$5006 fluxes and [\ion{S}{ii}]$\lambda$6716 velocity dispersion. Note that H$\alpha$ and [\ion{S}{ii}], lines commonly used for SNR identification, are fit independently.  \revtwo{Variations in seeing due to the differences in wavelength between these line maps are negligible ($<$0.01\arcsec; \citealt{emsellem2022phangs}).} For fluxes, a signal-to-noise (S/N) cut of S/N $\geq$ 5 is applied while for the velocity dispersion, a higher cut S/N $\geq$ 20 is used \citep{egorov2023phangs}, hereafter E23. These S/N cut are applied before identifying SNRs in all maps. ~


We also make use of the  \citet{groves2023phangs} and \citet{santoro2022phangs} catalog of \ion{H}{ii} regions to compare with our SNR properties. 
\citet{groves2023phangs} identified 23,244 nebulae as \ion{H}{ii} regions, imposing on a S/N cut of five for all emission lines and using BPT diagnostics ([\ion{O}{iii}]/H$\beta$ vs. [\ion{O}{i}]/H$\alpha$ and [\ion{O}{iii}]/H$\beta$ vs. [\ion{S}{ii}]/H$\alpha)$ for classification \citep{baldwin1981classification, kewley2001theoretical, kauffmann2003host}. 

\subsection{Star and environment masks}
Before selecting objects from the images, we also mask and exclude regions that suffer from contamination by foreground stars (provided as star masks by \citealt{emsellem2022phangs}), as well as regions where we suspect the ionized gas may have shocks from processes other than SNRs. For example, regions close to the galactic centers and bars can exhibit line ratios and physical conditions consistent with widespread shocks caused by large-scale supersonic inflows or outflows of gas \citep{Athanassoula1992, LopezCoba2022}. In addition, some of the galaxies in our sample (e.g.\ NGC~1365) contain AGN, which also strongly affects the line ratios in the central regions. As a result, we use the environments defined in  \citet{querejeta2021stellar} to mask \revtwo{the centers of all galaxies to avoid shocked emissions from potential AGN, and the bar region for galaxies showing increased gas turbulence over large areas. For bar regions that were without increased shocked regions, we did not mask them off}. 
See Table \ref{tab:mask} in Appendix \ref{sec:env_masks} for a complete list of which environmental mask is applied for each galaxy.

\section{Methods}
\label{sec:methods}
\subsection{Parent sample identification}\label{identify}
Distinguishing SNR from other nebulae is a long-standing problem  in nearby galaxies \citep{Sabbadin1977, rhea2023machine}. At $\sim$100~pc resolution, most ionized nebulae appear to be luminous and compact sources. However, with multiple emission-line diagnostics, it is possible to find and classify SNRs. 

The [\ion{S}{ii}]/H$\alpha$ line ratio is historically the most commonly used diagnostic \citep{blair1981supernova,blair2004optical} to separate SNRs from \ion{H}{ii} regions. This diagnostic works because the shocked region is richer in warm [\ion{S}{+}] with respect to \ion{H}{ii} regions, which translate into a much higher  [\ion{S}{ii}]/H$\alpha$ line ratio \citep{allen2008mappings}. 
A similar situation applies to [\ion{O}{i}]. Indeed, the [\ion{O}{i}]/H$\alpha$ line ratio has been predicted to be a better diagnostic for distinguishing SNRs and \ion{H}{ii} regions than the more commonly used [\ion{S}{ii}]/H$\alpha$ line ratio \citep{kopsacheili2020diagnostic}. However, this emission line is normally weaker than [\ion{S}{ii}], and in the very local universe, it can be affected by blending with skyline airglow emission. 
Thanks to the  redshift of our sample, the intrinsic [\ion{O}{i}] emission is shifted  from sky emission, and the depth of our MUSE data allows us to obtain sufficient S/N in our [\ion{O}{i}] fluxes to consider it in our selection of SNRs. 

Another prominent feature of SNRs is their broad emission-line profiles arising from the presence of fast radiative shocks of $\sim$ 200 km~s$^{-1}$ \citep{long2017galactic}, which significantly exceeds the typical  $\sim$20 km~s$^{-1}$ velocity dispersion of \ion{H}{ii} regions \citep{winkler2015interstellar}. This kinematic information provides a promising alternate identification method \citep{points2019kinematics,long2022supernova,winkler2023supernova,egorov2023phangs}. 
Finally, by combining multiple line ratios (such as in BPT diagrams) it is possible to determine the dominant ionization mechanism, whether it is photoionization-dominated or shock-dominated \citep{congiu2023phangs, makarenko2023supernova}. 

We explore the use of five different criteria in this paper. These are the  [\ion{S}{ii}]/H$\alpha$ line ratio, the [\ion{O}{i}]/H$\alpha$ line ratio, the [\ion{S}{ii}]$\lambda$6716 velocity dispersion ($\sigma$) and two BPT diagnostics: [\ion{O}{i}]-[\ion{O}{iii}] and [\ion{S}{ii}]-[\ion{O}{iii}]. We identify objects using each of these five different criteria separately and combine these catalogs to construct our \textit{parent sample} of objects, which we later classify as either SNRs or SNR candidates. All these criteria are explained below in more detail. 

The Python package \texttt{astrodendro}\footnote{\url{https://dendrograms.readthedocs.io/en/stable/index.html}} was used to select objects \revtwo{from line ratio maps and line kinematic maps.} \texttt{astrodendro} is a tool used for calculating dendrograms \citep{Rosolowsky2008}, which visually represent the hierarchical clustering of data points, particularly in astronomical data analysis. 
\revtwo{We use astrodendro to identify regions around local maxima (above a threshold  \texttt{min\_value}), selected based on their contrast with the surrounding pixels (\texttt{min\_delta}) and required to have a minimum associated area that indicates the line ratio to be coherent over about the area of a PSF.}  
\revtwo{We require regions used in further analysis to contain at least 10 pixels, which is the approximate area of the PSF. This ensures that all considered regions are detected over the whole area of the PSF and do not reflect an artifact or `hot pixel.' We further require all regions to have a diameter of less than 200~pc, as otherwise they would be unphysically large compared to literature catalogs of SNRs \citep{asvarov2014size}.}

In this analysis, we do not aim to produce a `complete' catalog of SNRs in these galaxies but rather try to provide a sample which is as pure as possible. Many of the choices made in the following sections are designed to minimize the selection of spurious objects, such that all of the objects in our final catalog are credible SNRs and SNR candidates, as judged in a sub-sample of objects by visual inspection. 
The well studied nearby galaxy M83 is used as a benchmark to validate our method in Section \ref{sec:m83}, and we compare with other approaches to identify `shocked' nebulae in Section \ref{sec:shocked_regions}. 

\subsubsection{Selection via [\ion{S}{ii}]/H$\alpha$ and [\ion{O}{i}]/H$\alpha$} \label{residual}
\begin{figure*}
    \centering
    \includegraphics[width=\textwidth]{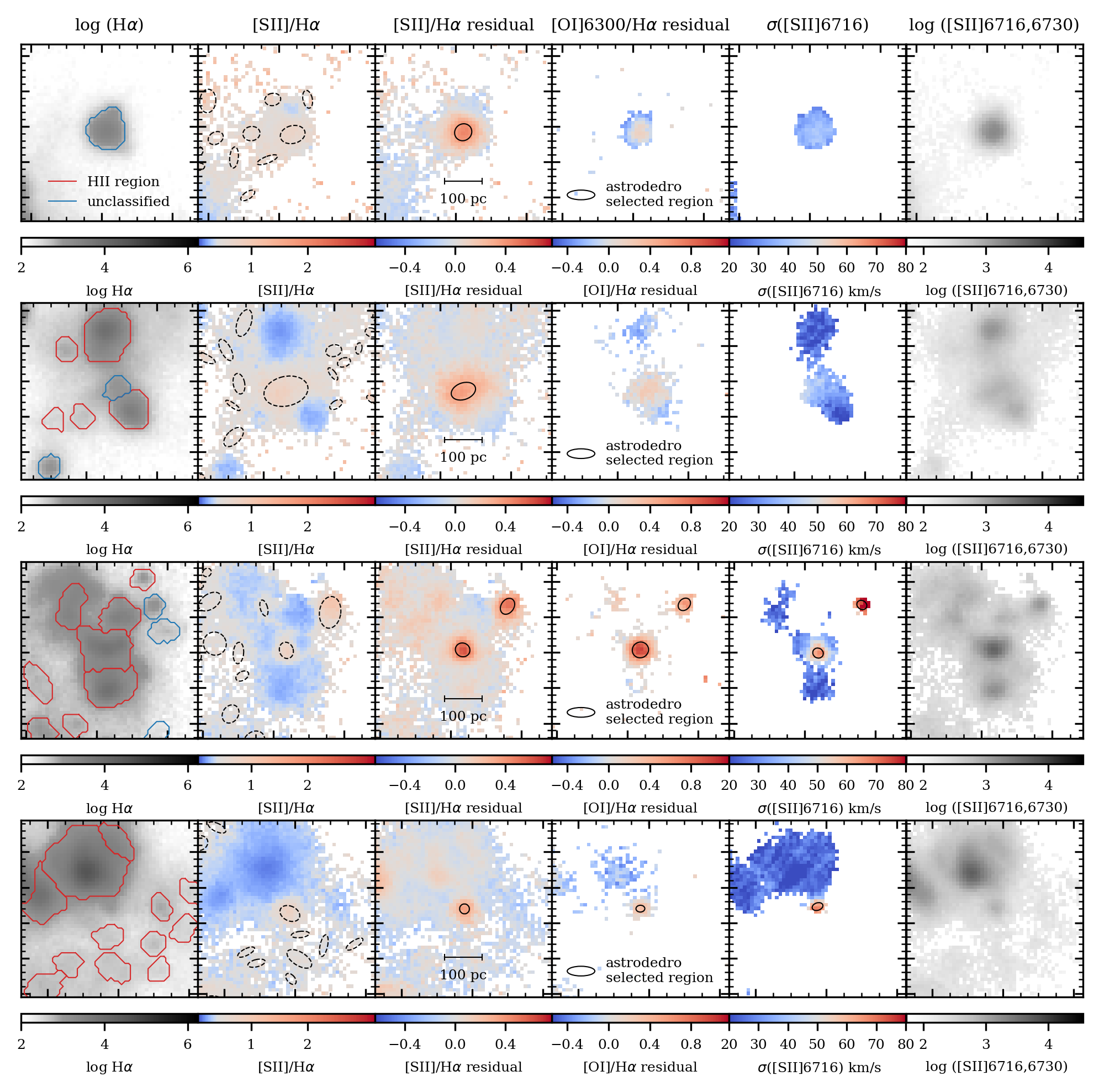}
    \caption{Zoom in on four regions, with images showing (from left to right) log H$\alpha$ flux (10$^{-20} cm^{-2} erg s^{-1}$), [\ion{S}{ii}]/H$\alpha$, [\ion{S}{ii}]/H$\alpha$ residual, [\ion{O}{i}]/H$\alpha$ residual, [\ion{S}{ii}] velocity dispersion maps and log [\ion{S}{ii}] flux. Objects identified by our method as SNRs and SNR candidates are marked with black solid ellipses. Objects identified in the \cite{groves2023phangs} nebular catalog as \ion{H}{ii} regions are enclosed by red solid lines, while regions unclassified in that catalog are outlined with blue solid lines. Region contours in the first column are from \citet{groves2023phangs}. The second column shows region contours identified by \texttt{astrodendro} without any photoionization correction. The remaining columns show regions identified by \texttt{astrodendro} after using our final selection methods.  It is apparent that many of the regions in the second column are spurious. }
    \label{fig:SNRs_zoom}
\end{figure*}
[\ion{S}{ii}] emission arises not only from shocks in SNRs but also through photoionization, such as in \ion{H}{ii} regions or the DIG. We initially tried to identify SNRs directly from the line ratio maps, 
and ran \texttt{astrodendro}, identifying peaks where the \texttt{min\_value} was set to a threshold of 0.4 in [\ion{S}{ii}]/H$\alpha$ and a threshold of 0.017 in [\ion{O}{i}]/H$\alpha$, based on shock models \citep{kopsacheili2020diagnostic}. We found 36,826 objects in the [\ion{S}{ii}]/H$\alpha$ maps and 15,891 objects in the [\ion{O}{i}]/H$\alpha$ maps. This implies nearly $1000-2000$ SNRs per galaxy, which appears unphysically high \revtwo{(although see Section \ref{sec:snrates})} considering that even galaxies like M83, NGC 6946, and M51, which have similar star formation rates as the PHANGS sample and are about a factor of 2-4 closer, only have on the order of a few hundred SNRs that have been spectroscopically identified \citep{winkler2021optical,long2022supernova}. 

Visual inspection reveals that the majority of these objects found in the PHANGS-MUSE sample are at the boundaries of \ion{H}{ii} regions 
and at low H$\alpha$ surface brightness, \revtwo{implying they are either artifacts due to noise or associated with DIG.} 
Figure \ref{fig:SNRs_zoom} shows cutout images of four areas in NGC~628 and the regions identified within them. 
\revone{These images compare the region boundaries identified in the \cite{groves2023phangs} nebular catalog, where objects were identified via their H$\alpha$ morphology, with the objects identified by \texttt{astrodendro} using the [\ion{S}{ii}]/H$\alpha$ and [\ion{O}{i}]/H$\alpha$ ratio maps. The first column illustrates objects identified by \cite{groves2023phangs} and classified as \ion{H}{ii} regions (red lines) or unclassified regions \footnote{These regions do not have a definitive classification in \cite{groves2023phangs}. This is principally because they are composite or shock-dominated objects in the BPT diagrams (e.g.\ AGN outflows, SNRs, planetary nebulae), although in some cases they simply have low S/N in one or more of the key diagnostic lines.} (blue lines), overlaid on the H$\alpha$ image. The second to fifth columns show peaks identified by \texttt{astrodendro} overlaid on the corresponding line-ratio or velocity dispersion map. The last column indicates the [\ion{S}{ii}] flux in log scale. }
It is apparent that the 0.4 threshold in [\ion{S}{ii}]/H$\alpha$ selects not only SNRs but also spurious regions, 
\revtwo{which in shallower imaging would not have been detected \citep[see also][]{winkler2023supernova}.}  
\begin{figure*}
    \centering
    \includegraphics[width=\textwidth]{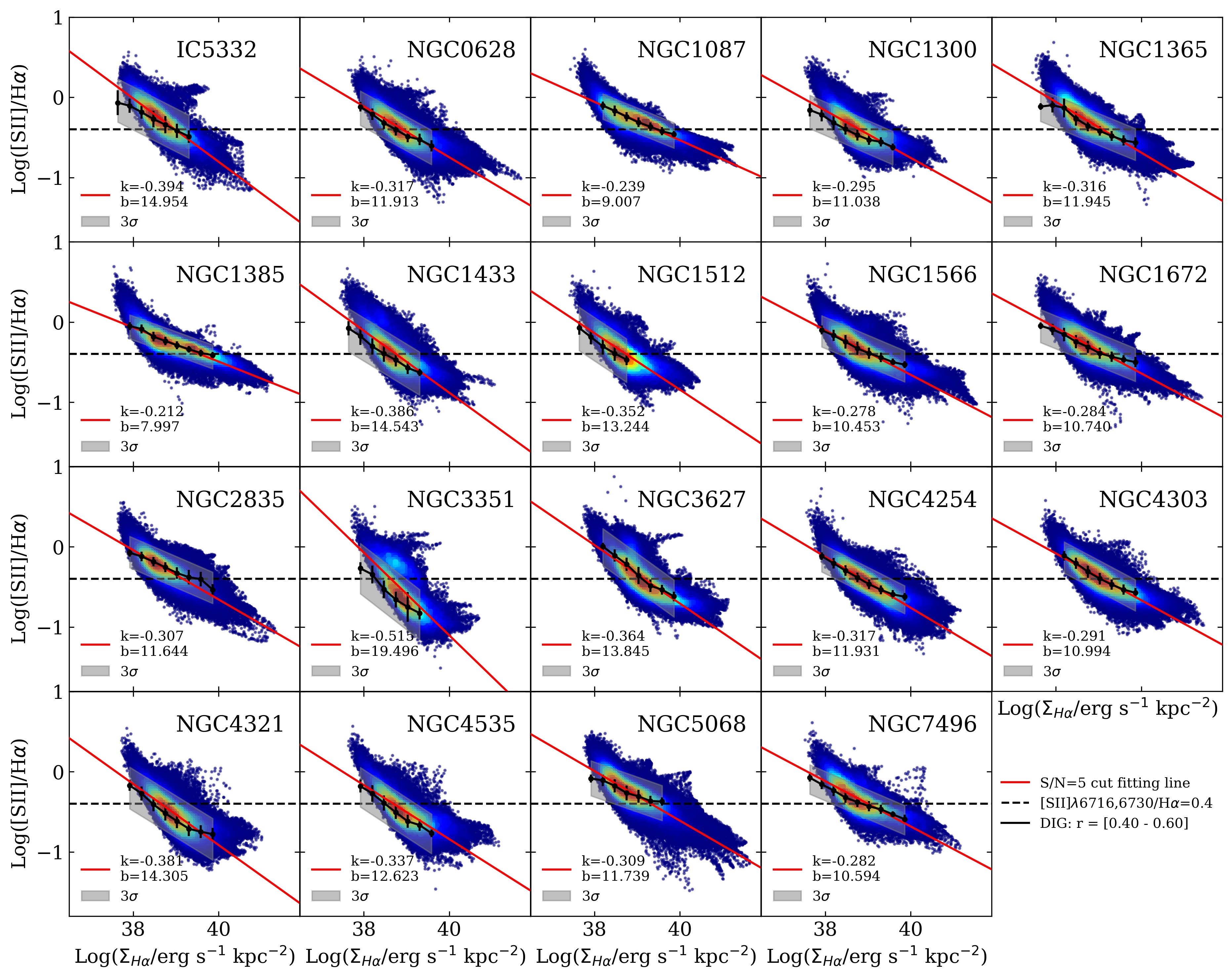}
    \caption{Log([\ion{S}{ii}]/H$\alpha$) as a function of logarithm H$\alpha$ surface brightness for all pixels in each  galaxy. The red solid line is the fitted correlation for each galaxy, with slope and intercept values given in the legend of each panel. \revone{The black solid line is representative of the DIG relation, as measured between 0.40 R$_{max}$ (maximum radial coverage) and 0.60 R$_{max}$ for each galaxy from \citet{belfiore2022tale}. The gray area indicates the 3$\sigma$ range of this relation.}  SNRs typically lie above the fitted lines and show looping structure towards the upper right. The horizontal black dashed line indicates the typical value of [\ion{S}{ii}]/H$\alpha$ = 0.4 used to select SNRs. Many low H$\alpha$ surface brightness pixels associated with the DIG have [\ion{S}{ii}]/H$\alpha \geq$ 0.4. }
    \label{fig:SII_DIG}
\end{figure*}
\begin{figure*}
    \centering
    \includegraphics[width=\textwidth]{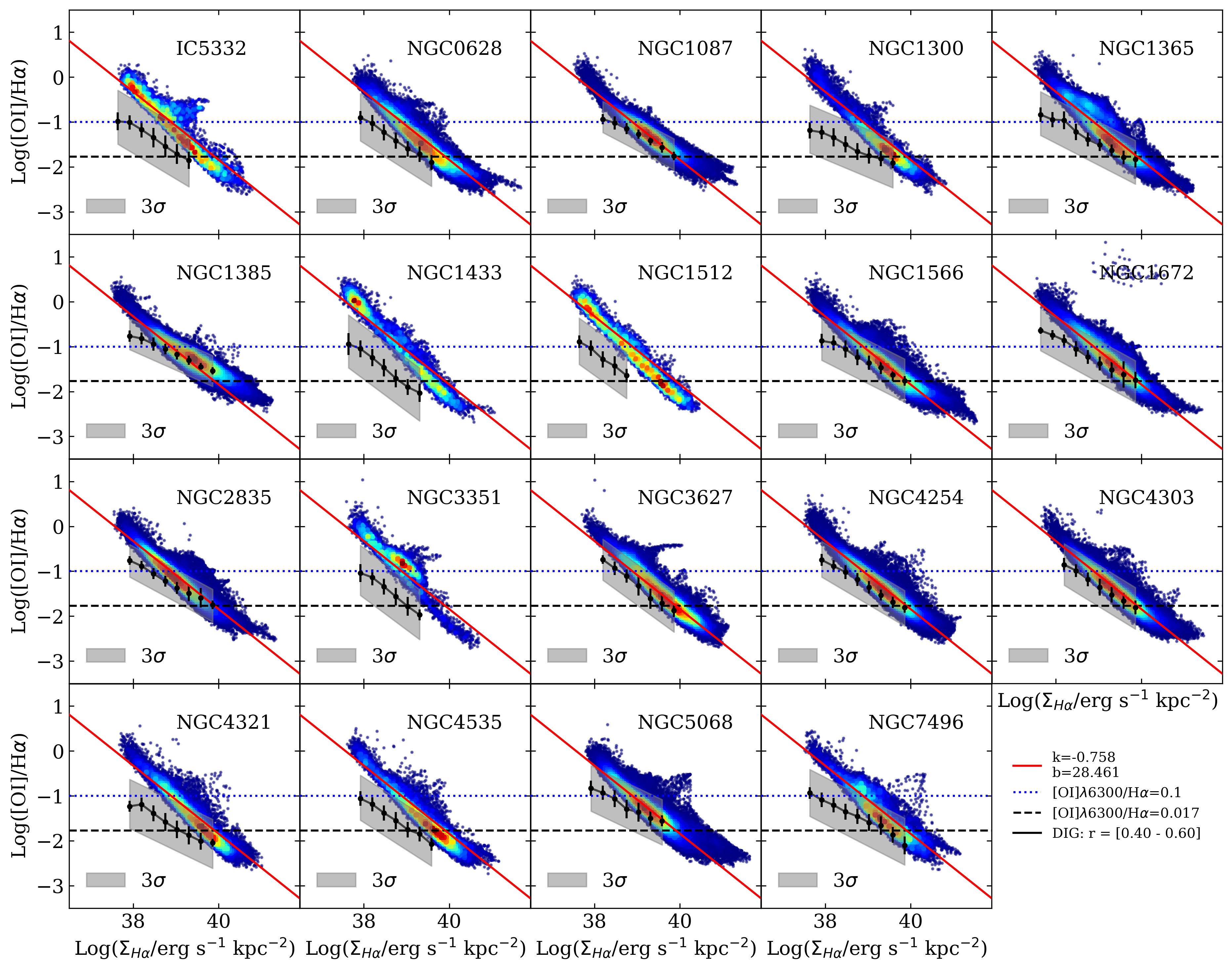}
    \caption{log([\ion{O}{i}]/H$\alpha$) as a function of log H$\alpha$ surface brightness for all pixels in all 19  galaxies. The red solid line is the fitted global correlation for all galaxies, with slope and intercept provided in the lower right corner of this plot. \revone{The black solid line is representative of the DIG relation, as measured between 0.40 R$_{max}$ (maximum radial coverage) and 0.60 R$_{max}$ for each galaxy from \citet{belfiore2022tale}. The gray area indicates the 3$\sigma$ range of this relation.} The black dashed line indicates the theoretical value of [\ion{O}{i}]/H$\alpha$ = 0.017 to select SNRs \citep{kopsacheili2020diagnostic}. The horizontal blue dotted line indicates the empirical value of [\ion{O}{i}]/H$\alpha$ = 0.1 used select SNRs. For NGC1672, the individual pixels lying above the majority come from an SNe that just happened several days before our observation. }
    \label{fig:OI_DIG}
\end{figure*}
After accounting for the contributions of photoionized emission to our maps (see below), our region selection becomes much cleaner, and can often be more directly associated with the `unclassified' nebular objects. 

\revtwo{At low surface brightness, most star-forming galaxies show widespread line emission arising from DIG, which makes up $\sim$50\% of the integrated H$\alpha$ emission of galaxies \citep{Zurita2000}.} 
DIG typically shows elevated levels of [\ion{S}{ii}]/H$\alpha$, [\ion{N}{II}]/H$\alpha$ and [\ion{O}{I}]/H$\alpha$ \citep{haffner2009warm}, and can thus be confused for low surface-brightness SNRs \citep[e.g.][]{blair1997identification,  long2022supernova}. As shown in \cite{belfiore2022tale}, both the [\ion{S}{ii}]/H$\alpha$ and [\ion{O}{I}]/H$\alpha$ line ratios in the DIG can be modeled by leaking radiation from \ion{H}{ii} regions, suggesting the DIG is principally powered by leakage of radiation from \ion{H}{ii} regions with a typical mean free path for the ionizing radiation of $\sim$2~kpc. 

At high surface brightness, SNRs might be blended in projection with giant \ion{H}{ii} regions (see the third row in Figure \ref{fig:SNRs_zoom}). 
\revtwo{SNe often overlap with \ion{H}{ii} regions \citep{maykerchen2024}, and many SNRs are identified in and around young star-forming regions and along spiral arms \citep{long2022supernova}. \ion{H}{ii} regions are typically characterized by high H$\alpha$ surface brightness but  lower [\ion{S}{ii}]/H$\alpha$  and [\ion{O}{I}]/H$\alpha$ than is seen in SNRs \citep{kopsacheili2020diagnostic}, such that  
if a SNR is seen in projection against an \ion{H}{ii} region, there is the possibility that the SNR will be outshone by the \ion{H}{ii} region, and thus be challenging to distinguish. }
To exclude the contribution from DIG and \ion{H}{ii} regions to the [\ion{S}{ii}]/H$\alpha$ and [\ion{O}{i}]/H$\alpha$  line ratio maps, we build on the results of \cite{belfiore2022tale}, who carried out a detailed characterization of DIG in the PHANGS-MUSE galaxies.  \citet{belfiore2022tale} spatially binned the line emission to study the faint DIG, and found strong anti-correlations between the H$\alpha$ surface brightness and common diagnostic line ratios, \revtwo{which they show can be well fitted by photoionization models as a function of ionization parameter (in this context, a proxy for H$\alpha$ intensity). } 

We examine the correlations between line ratio and H$\alpha$ surface brightness on a pixel-by-pixel level in each of the 19 galaxies (Figures \ref{fig:SII_DIG} and \ref{fig:OI_DIG}), \revtwo{in order to search for regions exhibiting anomalously high line ratios at fixed H$\alpha$ surface brightness.} We expect that most pixels are dominated by photoionization, as SNRs typically only contribute $\sim$ 5\% of the H$\alpha$ flux of an entire galaxy \citep{vuvcetic2015optical}. \revtwo{Overall, we find very good agreement with the binned relations found in \citet{belfiore2022tale} for the [\ion{S}{ii}]/H$\alpha$ line ratios, as our data is sufficiently deep that the [\ion{S}{ii}] line is detected in a large fraction of the pixels. For the fainter [\ion{O}{i}] line, it is apparent that we are limited by sigma-clipping at low H$\alpha$ surface brightness, where [\ion{O}{i}] is less frequently detected, and the majority of our pixels reflect $\sim3\sigma$ outliers in the binned distribution.  
At high surface brightness, however, the relations agree. 
 As is apparent in Figures \ref{fig:SII_DIG} and \ref{fig:OI_DIG}, we expect that in many ways our pixel-based fits are more conservative, as they generally result in steeper slopes that do not correctly model the DIG line ratios at low surface brightness.}
\revtwo{We emphasize here that as a result of our sensitivity limits, we expect that our sample is both incomplete and biased towards selecting SNR with high line ratios. } 

Given the observed trends, we derive linear scaling relations to characterize the expected line ratios for log [\ion{O}{i}]/H$\alpha$ (taking all galaxies jointly) and log [\ion{S}{ii}]/H$\alpha$ (taking each galaxy separately) as a function of log H$\alpha$ surface brightness. We used a linear regression model to fit these pixels to get a slope and an intercept, provided in Figures \ref{fig:SII_DIG} and \ref{fig:OI_DIG}.  
 \revtwo{As we find that the [\ion{O}{i}]/H$\alpha$ vs. H$\alpha$ surface brightness shows uniform trends for all 19 galaxies, we fit all galaxies together, although we caution that these fits are not necessarily representative of the full DIG due to the S/N limitations on detecting [\ion{O}{i}].  Pixels belonging to SNRs are apparent in some cases, as shown in the horizontal stripes above the fitted relation. We are generally searching for spatially coherent objects that are outliers with high line ratios, and we tailor our object selection to identify regions based on the pixel distribution and our own visual inspection.} 

Using these  relations, we construct line-ratio \emph{residual} maps as
\revtwo{
\begin{equation}
    \left(\frac{[\ion{S}{ii}]}{H\alpha}\right)_{\rm residual} = \log \left(\frac{[\ion{S}{ii}]}{H\alpha}\right)_{\rm obs} - \log \left(\frac{[\ion{S}{ii}]}{H\alpha}\right)_{\rm fit},
\end{equation}
where the subscripts $obs$ denotes the observed line flux and $fit$ is calculated as a function of the H$\alpha$ surface brightness from the linear fit equations. ([\ion{O}{i}]/Ha)$_{\rm residual}$ is calculated in a similar way.} 


In our residual maps, we measure the standard deviation ($\sigma$) 
and use \texttt{astrodendro} to identify peaks assuming a \texttt{min\_value} of 2$\sigma$ in [\ion{S}{ii}]/H$\alpha$$_{\rm residual}$ and 1$\sigma$ in [\ion{O}{i}]/H$\alpha$$_{\rm residual}$. \texttt{min\_delta} is set to 0.1 dex to avoid selecting the second maximum peak that is close to the highest peak. We chose these thresholds to guarantee we did not miss objects (as judged by visual inspection). 
We identify 880 objects selected from the [\ion{S}{ii}]/H$\alpha$ line ratio residual, and 1368 objects selected from the [\ion{O}{I}]/H$\alpha$ line ratio residual. 

\revtwo{To judge the impact of any sigma-clipping bias at low H$\alpha$ surface brightness, which principally affects the [\ion{O}{i}]/H$\alpha$ line ratios, we repeated this analysis using the fit and $\sigma$ determined from the DIG binned data,  and identifying 3$\sigma$ pixel-scale outliers. This resulted in a very similar number of objects selected by their [\ion{S}{ii}]/H$\alpha$ residuals, and only slightly fewer objects selected by their [\ion{O}{i}]/H$\alpha$ residuals, indicating that our fine-tuning of the \texttt{astrodendro} parameter \texttt{min\_value} results in a roughly equivalent selection of objects.}

\subsubsection{Selection via velocity dispersion maps}\label{sigma}
As mentioned earlier, the spectral resolution of our MUSE data also enables the selection of SNRs from broadened emission lines in their spectra. 
In particular, we use the velocity dispersion of the [\ion{S}{ii}] line, corrected for instrumental broadening, to select our SNRs.  Note the kinematics of  [\ion{S}{ii}]$\lambda$6716 has been tied in the fit with other low-ionization lines to further improve the reliability of the measurement, however, all of these (e.g.  [\ion{S}{ii}]$\lambda$6730, [\ion{N}{ii}]$\lambda\lambda$6548,84, [\ion{O}{i}]$\lambda\lambda$6300,64; see \citealt{emsellem2022phangs}) should be strongly associated with shocks. While the H$\alpha$ velocity dispersion can, in principle, be utilized, we find that the [\ion{S}{ii}] line kinematics can be more directly and uniquely associated with SNR shocks. On average, the line kinematics typically agree between both tracers within 5~km~s$^{-1}$.  

An S/N cut of 20 for [\ion{S}{ii}]$\lambda$6716 is applied to ensure errors in the measured velocity dispersion remain below 10\% when deconvolved from the $\sim$50 \kms\ instrumental velocity dispersion \citep{bacon2017muse, egorov2023phangs}. \revtwo{With smoothing and binning, higher S/N is achieved across the map, but SNRs become blended with \ion{H}{ii} regions and DIG, and it is thus more challenging to identify by their kinematic signatures.} 
At low S/N, DIG might be hard to distinguish kinematically from SNRs, as it also produces broadened line emission due to its turbulent nature and increased scale height \citep{haffner2009warm, levy2019}. After masking galaxy centers, we do not see evidence that we are biased towards selecting DIG emission with this method \revtwo{as our S/N cut on [\ion{S}{ii}] has effectively masked most of the DIG}, however, with sufficiently deep data it remains challenging to distinguish these components \citep{winkler2023supernova}.  

We take a conservative approach and set our \texttt{astrodendro} threshold to select peaks with velocity dispersion larger than 50 \kms\ (full width at half maximum (FWHM) = 118 \kms) for our parent sample. We visually tested different thresholds and then chose this value that optimized the number of detected SNR candidates without including too many spurious detections. A \texttt{min\_delta} of 10 \kms\ is used to make sure local peaks are identified properly. A total of 352 objects are identified using this technique.

\subsubsection{Selection via BPT distances}
\begin{figure*}
    \centering
    \includegraphics[width=0.6\textwidth]{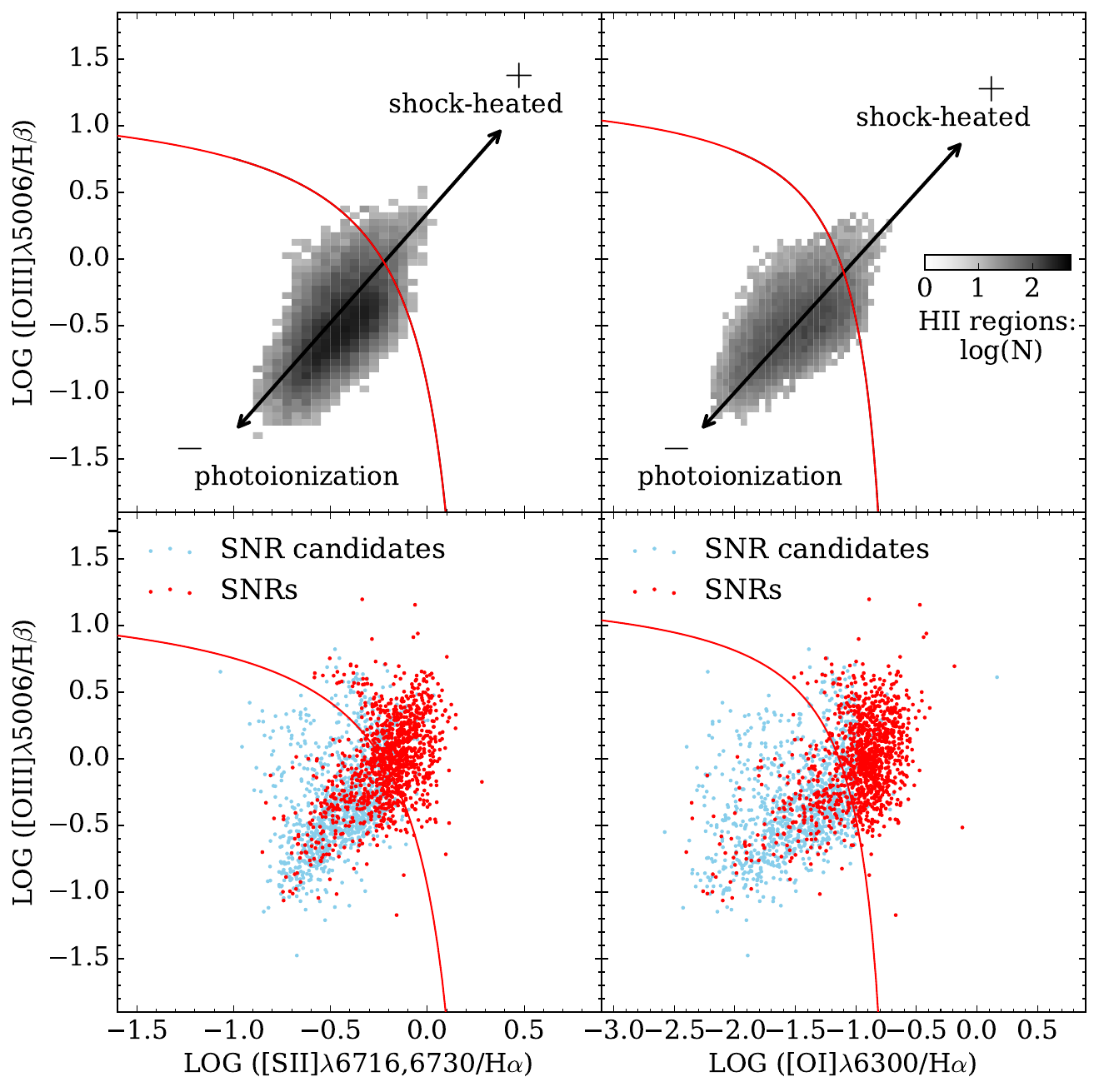}
    \caption{BPT diagrams with demarcation line from \citet{kewley2006host} with an indication of BPT distances as defined as the distance to the red solid extreme starburst lines. "+" indicates positive distance where shock-heating dominates while "--" indicates negative distance where photoionization dominates. The top-left and top-right plots show the distribution density of nebular regions in these 19 galaxies \citep{groves2023phangs}. In the bottom-left and bottom-right plots, SNRs and SNR candidates are indicated by red and blue dots, respectively. Nebular regions are prone to occupy the photoionized region while SNRs concentrate in the shocked areas.}
    \label{fig:BPT_dist}
\end{figure*}
In addition to the aforementioned criteria of [\ion{S}{ii}], [\ion{O}{i}] and velocity dispersion that are traditionally used for selecting SNRs, we also explore the properties, and the potential for selecting SNRs using Baldwin, Phillips \& Terlevich (BPT) diagrams, widely used to assess the line emission mechanism in galaxies \citep{baldwin1981classification, veilleux1987spectral}. Here we did not consider the [\ion{O}{iii}]/H$\beta$ vs. [\ion{N}{ii}]/H$\alpha$ diagram as [\ion{N}{ii}]/H$\alpha$ varies more with metallicity (see Section \ref{sec:results}). 
On the $\sim$100 pc scale of an individual nebula (top row of Figure \ref{fig:BPT_dist}), regions that are dominated by shock heating lie above the extreme starburst line \citep{kewley2001theoretical}, while regions dominated by photoionization from OB stars lie below this line \citep{congiu2023phangs}. These diagrams make use of the same [\ion{S}{ii}]/H$\alpha$ and [\ion{O}{i}]/H$\alpha$ line ratios as we describe in Section \ref{residual}, however, from the shape of the extreme starburst line it is apparent that a single fixed threshold may not be appropriate in all physical conditions. 

For this reason, we select pixels as shock-heated vs. photoionized based on their distances from the extreme starburst line in the [\ion{O}{iii}]/H$\beta$ vs. [\ion{O}{i}]/H$\alpha$ and [\ion{O}{iii}]/H$\beta$ vs. [\ion{S}{ii}]/H$\alpha$  BPT diagrams \citep{kewley2006host}, as shown in Figure \ref{fig:BPT_dist}. Here `distance' refers to the Euclidean distance (square root of the sum of vertical and horizontal separation squared) to the closest point on the extreme starburst line in both BPT diagrams. 
\revtwo{While the DIG also populates a similar parameter space in such diagrams \citep{Zhang2017}, the [\ion{O}{iii}] emission is particularly faint and is effectively excluded with our S/N$>$5 cut.} 

We construct maps of these two \emph{BPT distances}, and run \texttt{astrodendro} on each, with a \texttt{min\_delta} of 0.05 dex to avoid selecting the local maxima. We use a minimum distance of $\pm$~0.2 dex to classify pixels as shocks (positive distance) vs. photoionized (negative distance), where the chosen threshold of 0.2 is roughly the 3$\sigma$ uncertainty in the ratios measured for all pixels with S/N$\geq$5.   Figure \ref{fig:BPT_dist} (bottom row) shows the integrated line ratios for all objects in our final catalog.  SNRs do preferentially populate the `shock-heated' regime, and those SNRs found to be populating the `photionized' regime have typically been selected via one of our other criteria, or are blended with \ion{H}{ii} regions(see Section \ref{sec:overlapHII}). A total of 635 objects are identified using the BPT distance in the [\ion{O}{iii}]/H$\beta$ vs. [\ion{O}{i}]/H$\alpha$ diagram, and a total of 372 objects are identified using the BPT distance in the [\ion{O}{iii}]/H$\beta$ vs. [\ion{S}{ii}]/H$\alpha$ diagram. 

\subsubsection{Final parent sample construction and integrated line flux measurements}\label{DAP}
We employ five criteria to select objects, and using the central coordinates determined through \texttt{astrodendro}, we perform a crossmatch within a 1" radius across all objects. This process yields a  $parent ~sample$ consisting of 2233 objects selected by at least one of our criteria. In cases where the \texttt{astrodendro} centers differ across the five maps, we adopt the center of the first criterion that identified the object as an SNR or SNR candidate (following the order in which we present the criteria in this paper). \revtwo{A summary of how these different criteria each contribute to the total sample is presented in Section \ref{sec:classification} and revisited in Section \ref{sec:lessons}. }

\begin{figure*}
    \centering
    \includegraphics[width=0.9\textwidth]{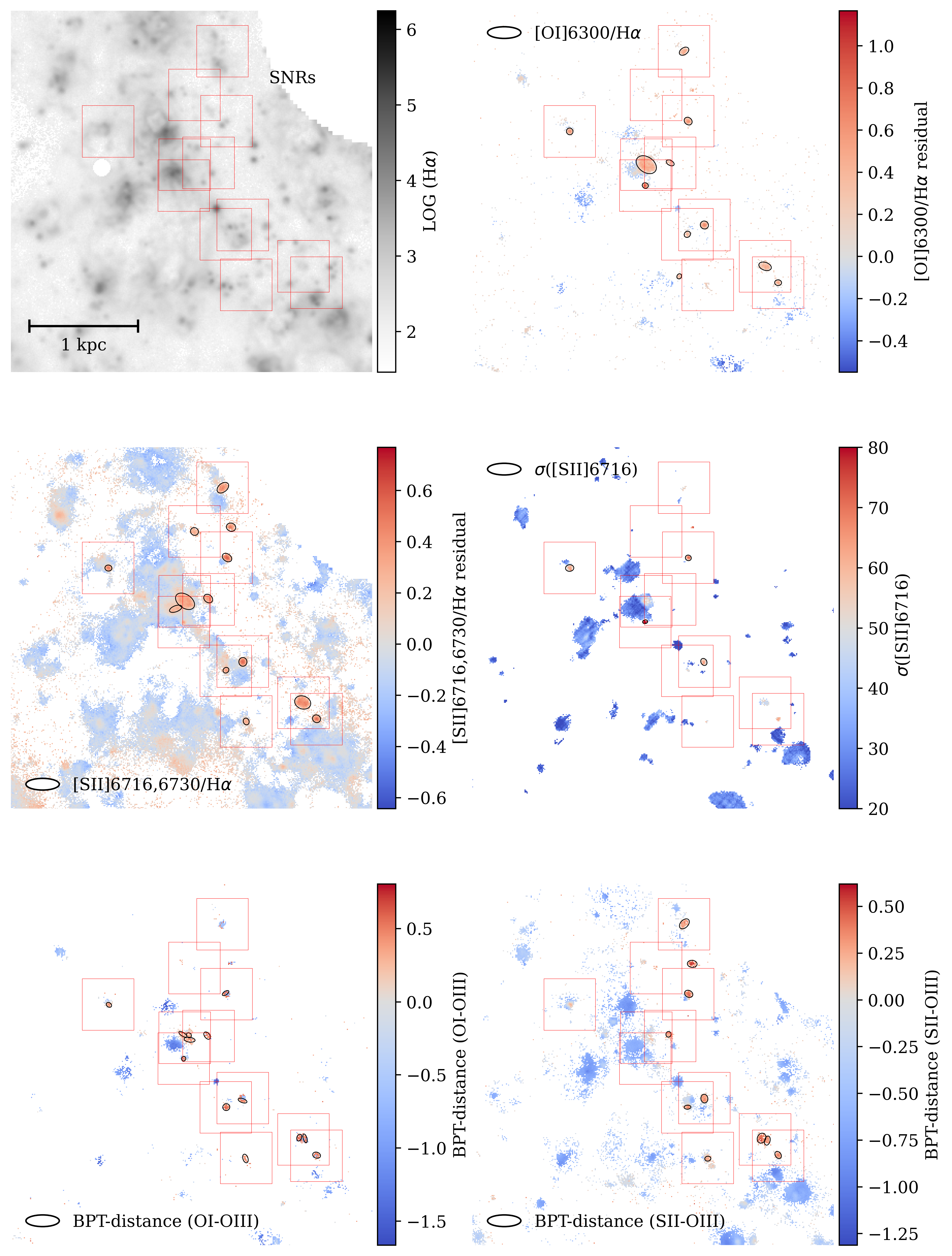}
    \caption{A 3~kpc $\times$ 3~kpc zoom-in map of the SNR population recovered in one of the galaxies in our sample, NGC 628, using our five selection criteria (described in Section \ref{sec:methods}). Panels from top left to bottom right show: the log H$\alpha$ emission; [\ion{O}{i}]/H$\alpha$ residual map; [\ion{S}{ii}]/H$\alpha$ residual map; [\ion{S}{ii}]$\lambda$6716 velocity dispersion; BPT distance (OI-OIII) map, and the BPT distance (SII-OIII) map. Objects selected by each individual criterion are marked with black ellipses in the corresponding subplot
    \revtwo{ and the locations of objects classified as SNR are marked with red boxes.} 
    Pixels with low S/N ($\leq$5), and those corresponding to stellar sources, and the central region of the galaxy, have been masked (see text for details).}
    \label{fig:SNRs-NGC0628}
\end{figure*}

\begin{figure*}
    \centering
    \includegraphics[width=\textwidth]{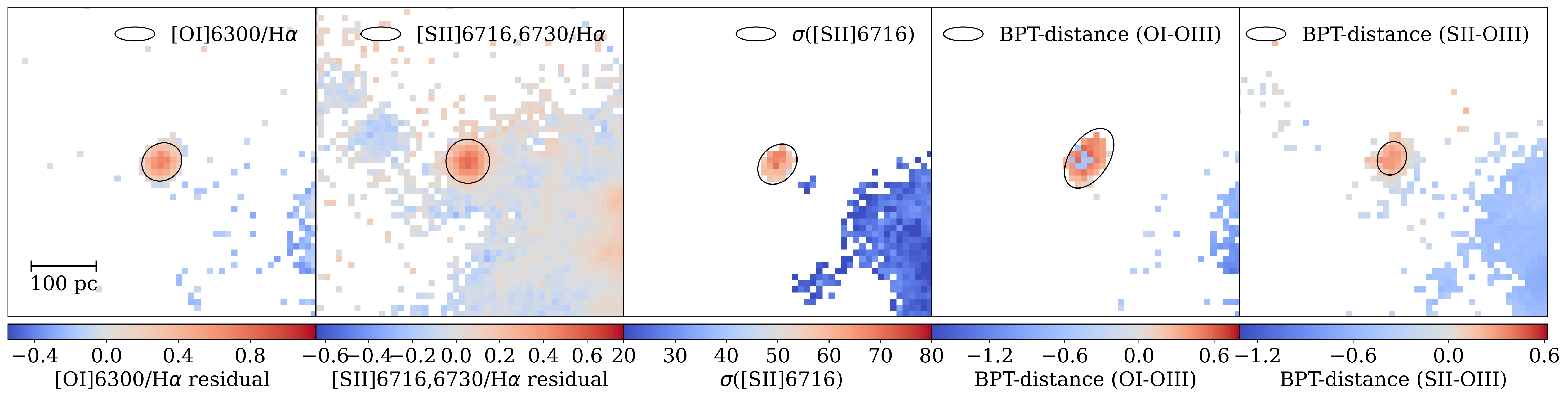}
    \caption{Example of what one of one SNRs (in NGC 628) looks like in the five criteria. This SNR is selected in the [\ion{O}{i}]/H$\alpha$ residual map, [\ion{S}{ii}]/H$\alpha$ residual map, [\ion{S}{ii}]$\lambda$6716 velocity dispersion, BPT distance (OI-OIII) and BPT distance (SII-OIII) maps. The selection is marked with a black ellipse.}
    \label{fig:SNRs-NGC0628-part}
\end{figure*}

\begin{figure*}
    \centering
    \includegraphics[width=\textwidth]{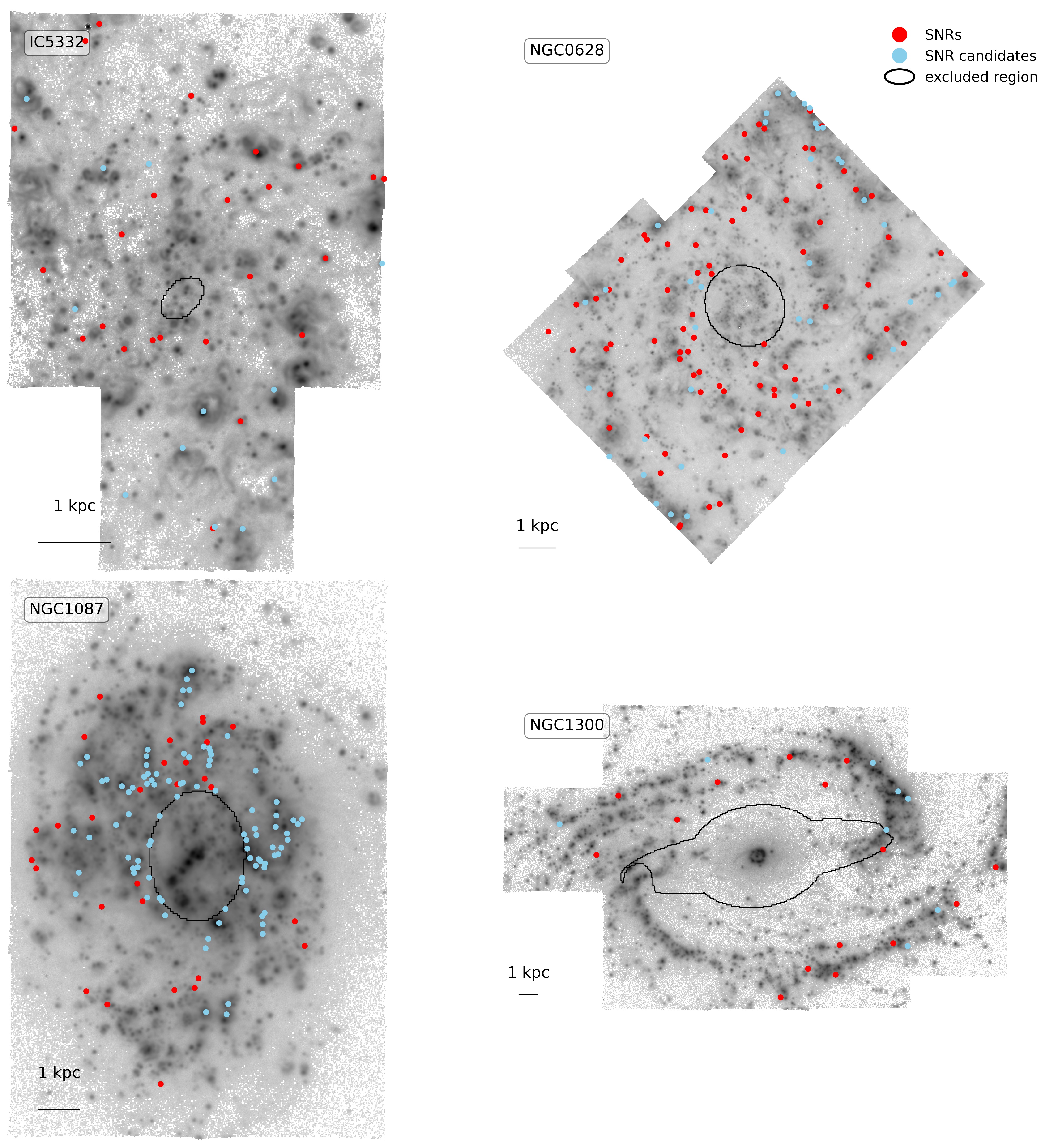}
    \caption{Parent sample of 2233 objects identified across the 19 PHANGS-MUSE galaxies, compared to the H$\alpha$ emission. The galaxy name is in the upper left corner. SNRs are indicated by red dots while SNR candidates are in blue. The background shows the H$\alpha$ map. Regions excluded from our search are outlined in solid black. Most galaxies have the center masked. See Appendix \ref{sec:env_masks} for more details. We observe SNRs to be distributed across the full field of view, and good qualitative correspondence with H$\alpha$ bright sites of star formation. See Appendix \ref{rest} for the rest of the galaxies.}
    \label{fig:19gal}
\end{figure*}

Figure \ref{fig:SNRs-NGC0628} shows a map of the parent sample identified within NGC 628 using each criterion, and Figure \ref{fig:SNRs-NGC0628-part} shows a zoomed-in SNR of NGC 628, demonstrating the distinct range of values of each criterion at the SNR location compared to the surrounding region.  For all 19 PHANGS-MUSE galaxies, the locations of 2233 objects in the parent sample are shown in Figure \ref{fig:19gal} and Appendix \ref{rest}. 
We find on average 118 objects per galaxy, which are located throughout the disks of each galaxy.

For this parent sample we measure line fluxes, velocities, and velocity dispersions from the integrated spectra of these objects (see Table \ref{tab:lines} in Appendix \ref{lines}) using the PHANGS-MUSE Data Analysis Pipeline or DAP \citep[described in][]{emsellem2022phangs}. We expect most SNRs will not be not resolved in our galaxies, so we use a fixed radius of 50 pc for all objects, comparable to or slightly larger than the PSF in all galaxies, and measure emission line fluxes and kinematics integrated over this area. See \citet{emsellem2022phangs} and \citet{groves2023phangs} for more details on how emission line fluxes and gas dynamics are measured using the DAP.

It is important to note a caveat in the interpretation of these emission line fluxes. The integrated line flux ratios [\ion{S}{ii}]/H$\alpha$ and [\ion{O}{i}]/H$\alpha$ that we measure from the integrated spectra here are different from the [\ion{S}{ii}]/H$\alpha$ and [\ion{O}{i}]/H$\alpha$ residual maps that we use to identify objects in Section \ref{residual}. While we use \texttt{astrodendro}  to identify SNRs from line flux ratios residual maps,  when running the DAP we are unable to apply such a correction. Corrections for background photoionized emission have not been extensively undertaken in previous research, so we believe the line measurements in our catalog are directly comparable to literature results.  
We emphasize that these line fluxes may be biased by the DIG background, or any \ion{H}{ii} region they overlap with. However, disentangling these background effects is challenging as they do not exhibit a smooth morphology. As can be seen in Figure \ref{fig:SNRs_zoom}, when the identified object overlaps with a \ion{H}{ii} region, the [\ion{S}{ii}]/H$\alpha$ ratio is less distinct from the environment, however, after our photoionization correction, the [\ion{S}{ii}]/H$\alpha$ ratio becomes more prominent. Hereafter, line fluxes for our SNRs and SNR candidates all refer to integrated flux without background subtraction.

\subsection{Parent sample classification}
\label{sec:classification}
\begin{table*}
    \centering
    \caption{Number of identified SNRs and SNR candidates in 19 galaxies. The columns from left to right are -- the category of SNR sample, sample size, and the number of SNRs satisfying each of our five criteria: [\ion{O}{i}]/H$\alpha$, [\ion{S}{ii}]/H$\alpha$, [\ion{S}{ii}] velocity dispersion, BPT distance as defined in [\ion{O}{iii}]/H$\beta$ vs. [\ion{O}{i}]/H$\alpha$ diagram, and in the [\ion{O}{iii}]/H$\beta$ vs. [\ion{S}{ii}]/H$\alpha$ diagram.}
    \label{tab:number}
    \begin{tabular}{ccccccc}
    \hline\hline\noalign{\vskip 0.05in}
         Sample& number & [\ion{O}{i}]/H$\alpha$ residual &  [\ion{S}{ii}]/H$\alpha$ residual& $\sigma$([\ion{S}{ii}]$\lambda$6716) & BPT distance ([\ion{O}{i}]-[\ion{O}{iii}]) & [\ion{S}{ii}]-[\ion{O}{iii}] \\
         \hline\noalign{\vskip 0.05in}
         parent sample&2233  &1368  &880  &352  &635& 372  \\
         \hline\noalign{\vskip 0.05in}
         SNRs& 1166 & 658 & 727 &255  & 569 & 325\\
         -- isolated& 964 & 464 & 549 & 189 & 543 &322 \\
         -- blended&202  & 194 &178  & 66 & 26 & 3\\
         \hline\noalign{\vskip 0.05in}
         SNR candidates& 1067 & 710 & 153 & 97 &66  & 47\\
         -- isolated& 498 & 226 & 86 & 80 & 65 &46 \\
         -- blended& 569 & 484 & 67 &17  & 1 & 1\\
         \hline\noalign{\vskip 0.05in}
    \end{tabular}
\end{table*}


\begin{figure*}
    \centering
    \includegraphics[width=\textwidth]{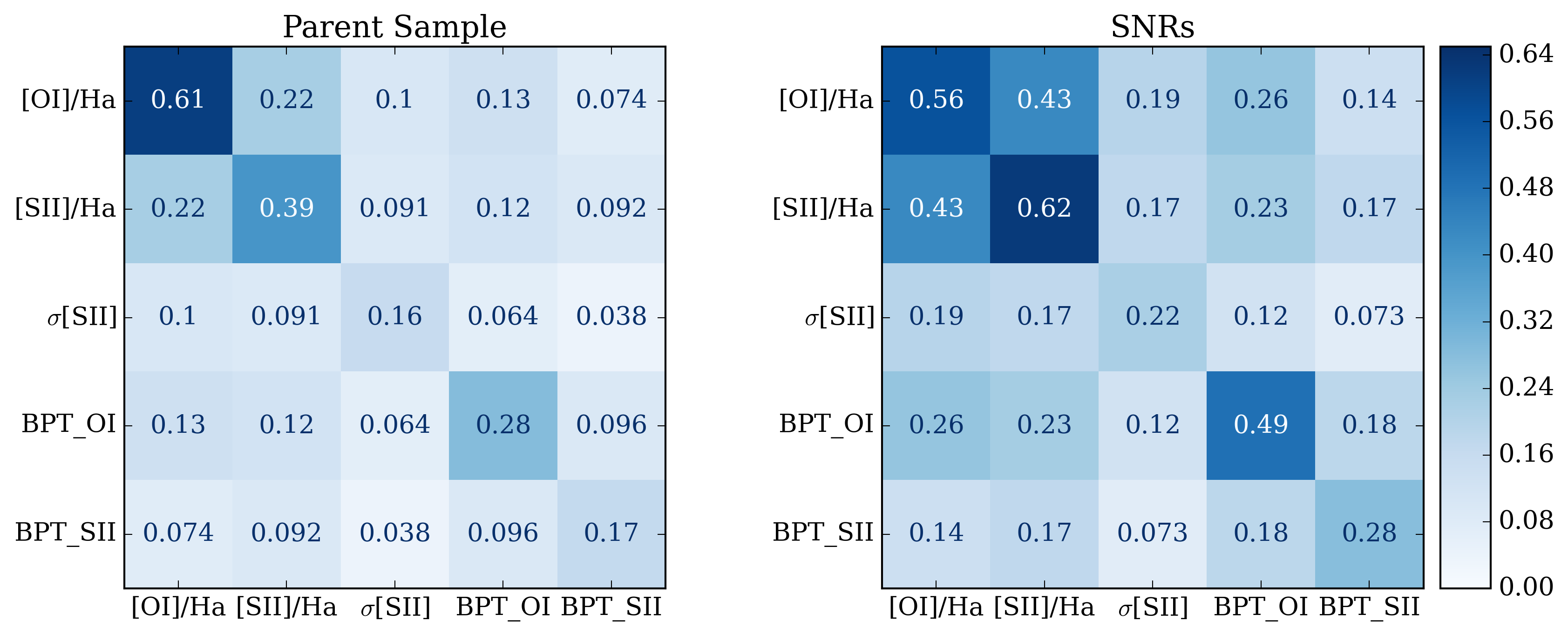}
    \caption{\revtwo{Fraction of objects in the parent sample (left) and SNR sample (right) that are identified by specific pairwise combination of our five criteria ([\ion{O}{i}]/H$\alpha$ residual, [\ion{S}{ii}]/H$\alpha$ residual, [\ion{S}{ii}]$\lambda$6716 velocity dispersion, BPT distance (OI-OIII) and BPT distance (SII-OIII).} }
    \label{fig:confusion}
\end{figure*}

We have used five criteria in Section \ref{identify} to identify objects for our parent sample. \revtwo{While the [\ion{S}{ii}]/H$\alpha$ line ratio has most commonly been used in the literature to identify SNRs \citep[e.g.][]{long2022supernova}, all of the diagnostics we employ are valid indicators of SNR shocks. To increase our confidence in our identification of these objects as true SNRs, we consider how to combine our different criteria and the integrated line fluxes to identify a robust sub-sample of objects that we can classify as SNRs}.

\revtwo{Table \ref{tab:number} presents a breakdown of how many objects are identified by each criterion. We identify the largest number of objects using the [\ion{O}{i}]/H$\alpha$ line ratio maps (1368), and the smallest number of objects using the [\ion{S}{ii}]$\lambda$6716 velocity dispersion (352). Among our five criteria, 1476 objects were only selected by only one criterion, 757 objects were selected by at least two criteria, 365 objects were selected by at least three criteria, 184 objects were selected by at least four criteria and 68 objects were selected by all five criteria. Given the range of characteristics apparent in our parent sample, we investigate using different pairs of criteria, as this provides a secondary confirmation of our SNR classification.}

\revtwo{Figure \ref{fig:confusion} (left) presents a visualization of what fraction of objects are matched pairwise between the five criteria. Clearly, the [\ion{O}{i}]/H$\alpha$ residual and [\ion{S}{ii}]/H$\alpha$ residuals produce the most overlap ($\sim$20\% of the parent sample), but are still not perfectly in agreement, and also show non-negligible $\sim$10\% overlap with the remaining indicators. We also note that a large number of objects (414) do not match between any two criteria, but exhibit high line ratios in their integrated line fluxes ([\ion{O}{i}]/H$\alpha$>0.1 and [\ion{S}{ii}]/H$\alpha$>0.4 ), such that in previous work, they would have resulted in clear SNR classifications. }

Based on this analysis of the 2233 objects from our parent sample,  we define two sub-samples. The ``SNR candidates'' are those objects only selected by one criterion, and either [\ion{S}{ii}]/H$\alpha$<0.4 or [\ion{O}{i}]/H$\alpha$<0.1. The rest of the objects we classify as ``SNRs'', which means that at least two of the five methods identify the object, or both the integrated [\ion{S}{ii}]/H$\alpha$ and [\ion{O}{i}]/H$\alpha$ are high enough that the line emission is best explained by shock excitation. 
\revtwo{Figure \ref{fig:confusion} (right) presents a visualization of how objects are matched pairwise for just the SNR sample. [\ion{O}{i}]/H$\alpha$ residual and [\ion{S}{ii}]/H$\alpha$ residuals select about half of the objects, but other pairs of diagnostics are also contributing significantly (10-20\%) to the SNR sample. }

\begin{figure*}
    \centering
    \includegraphics[width=0.8\textwidth]{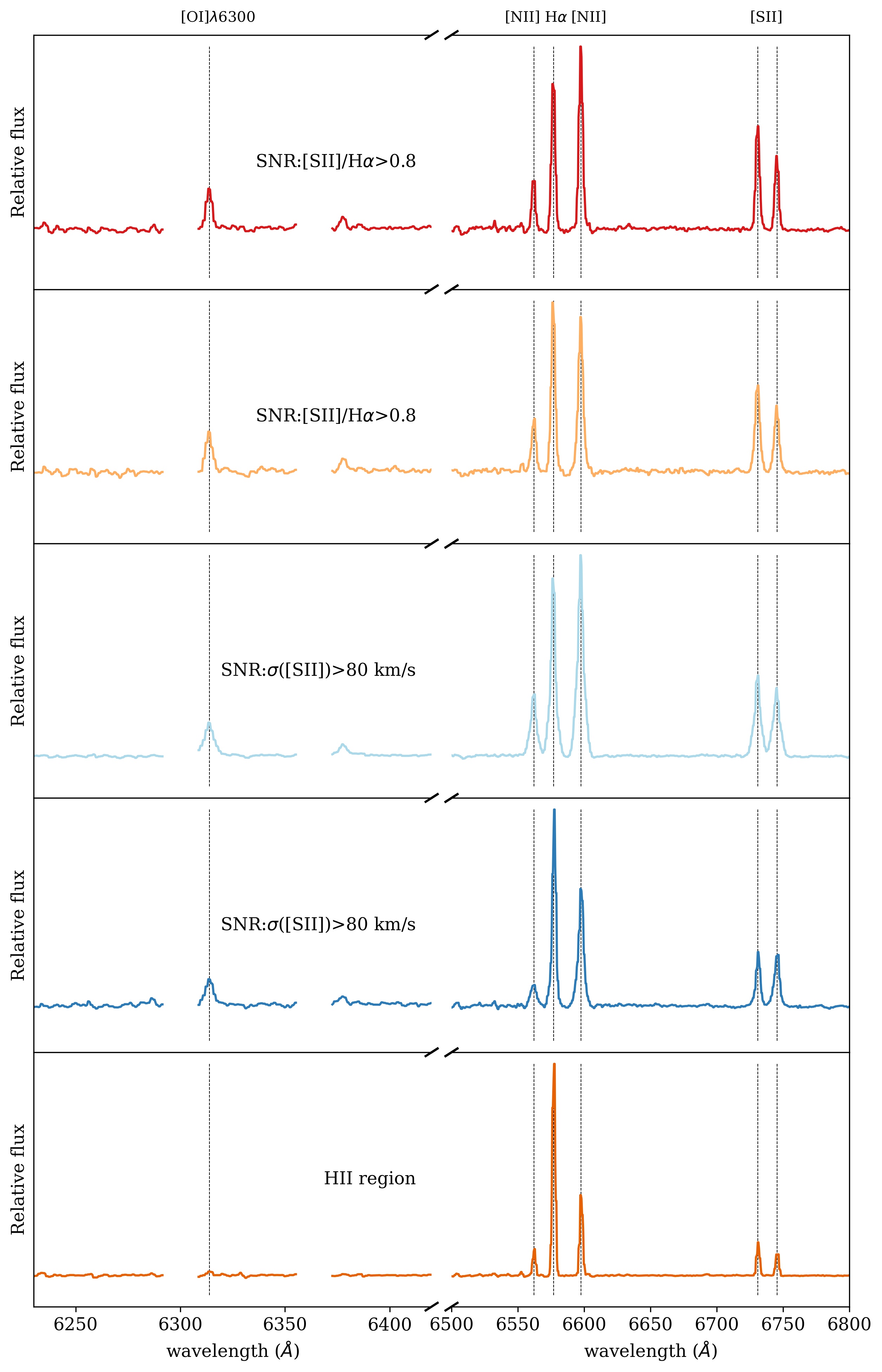}
    \caption{Spectra for four SNRs and an \ion{H}{ii} region in NGC 628. Top two: SNRs with high [\ion{S}{ii}]/H$\alpha$ value > 0.8; middle two: SNRs with high velocity dispersion > 80 \kms; bottom one: \ion{H}{ii} region. Characteristic lines, like [\ion{O}{i}], [\ion{S}{ii}], [\ion{N}{ii}] and H$\alpha$ are labeled in the plot. All SNRs have broader emission lines than the \ion{H}{ii} region, \revtwo{though this is most apparent in the two SNRs (NGC0628\_17 and NGC0628\_48) as they were selected to have particularly broad lines.} Sky emission at [\ion{O}{i}]$\lambda6300\AA,\lambda6363\AA$ has been masked.}
    \label{fig:spectra}
\end{figure*}

We present spectra for four representative SNRs and one typical \ion{H}{ii} region in Figure \ref{fig:spectra} \revtwo{in order to show the typical sensitivity and spectral features that we observe. In all panels, the strong skylines at 6300\AA\,and 6363\AA\, are masked. } The top two panels show the spectra for two SNRs with high [\ion{S}{ii}]/H$\alpha$ $>$ 0.8, the middle two panels for two SNRs with high velocity dispersion $>$ 80 \kms, and the bottom panel for a representative \ion{H}{ii} region. The broadened emission lines are apparent by eye for the two high velocity dispersion SNRs compared to the \ion{H}{ii} region, as is the increased [\ion{S}{ii}]/H$\alpha$ ratio in the top two SNR. 

\begin{figure}[h!]
    \begin{subfigure}[b]{0.5\textwidth}
    \centering
    \includegraphics[width=\textwidth]{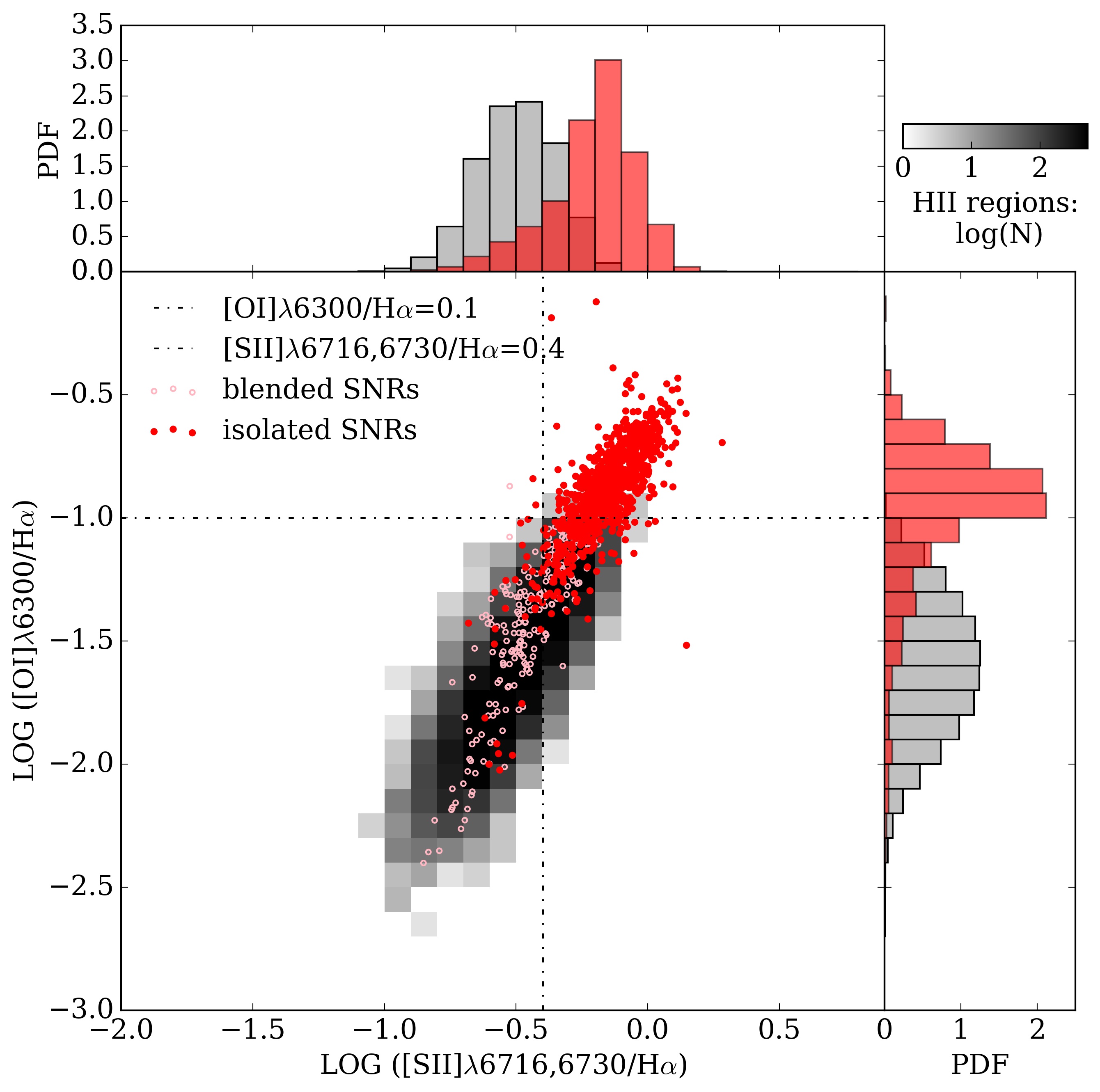}
    \caption{}
    \label{fig:OI_SII_clean}
    \end{subfigure}
    \hfill
    \begin{subfigure}[b]{0.5\textwidth}
    \includegraphics[width=\textwidth]{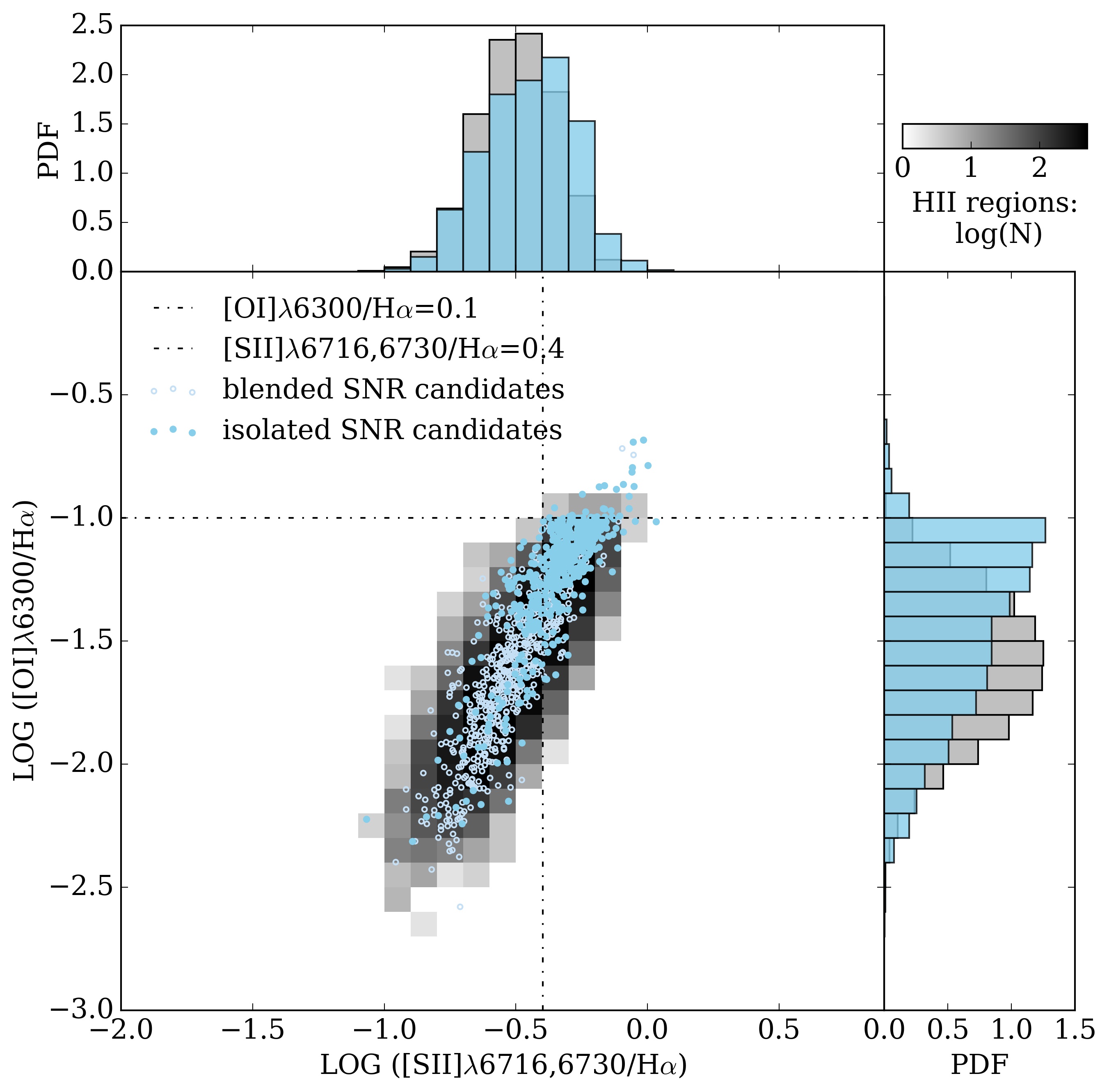}   
    \caption{}
    \label{fig:OI_SII_can}
    \end{subfigure}
    \caption{Distribution of identified objects and \ion{H}{ii} regions density (in grayscale) in the [\ion{O}{i}]/H$\alpha$ vs. [\ion{S}{ii}]/H$\alpha$ plane. \emph{(a)}: SNRs (red dots); \emph{(b)}: SNR candidates (blue dots). For objects that overlap with a \ion{H}{ii} region, the symbol is an empty circle with light red or blue. Identified objects are prone to occupy different spaces than \ion{H}{ii} regions in the [\ion{O}{i}]/H$\alpha$ vs. [\ion{S}{ii}]/H$\alpha$ plane. However, this tendency is more obvious for a SNR than a SNR candidate. }
\end{figure}

For the SNRs and SNR candidates, when we place them in the [\ion{O}{i}]/H$\alpha$ v.s [\ion{S}{ii}]/H$\alpha$ parameter space, we can clearly see that (by design) the majority of SNRs lie above [\ion{O}{i}]/H$\alpha$>0.1 and [\ion{S}{ii}]/H$\alpha$>0.4 (see Figure \ref{fig:OI_SII_clean}) while the SNR candidates distribution overlaps with the \ion{H}{ii} regions (see Figure \ref{fig:OI_SII_can}) making it difficult to distinguish SNR candidates from \ion{H}{ii} regions. 
The distinct boundary at 0.1 for [\ion{O}{i}]/H$\alpha$ in \ion{H}{ii} regions arises from the definition of these objects in \citet{groves2023phangs} using BPT diagram cuts. 
\begin{table*}
	\centering
	\caption{Parent sample catalog generated in this work, see text for details.}
	\begin{tabular}{c  c  c} 
	\hline\hline\noalign{\vskip 0.05in}
		Column Name& Unit & Description\\
	\hline\noalign{\vskip 0.05in}
\texttt{Index}&--&Global SNR ID\\
\texttt{gal$\_$name}&--& Galaxy name\\
\texttt{sample$\_$name}&SNRs/SNR candidates& Classification of parent sample\\
\texttt{gal$\_$dist}&Mpc& galaxy distance\\
\texttt{snr$\_$dist}&arcsecond&  deprojected distance to the galaxy center \\
\texttt{r$\_$over$\_$reff}&--&ratio of  distance to the effective radius of the galaxy\\
\texttt{environment}&--& environment mask\\
\texttt{Ha6562$\_$lumi}&erg s$^{-1}$&  H$\alpha$ luminosity, has been corrected for Milky Way foreground extinction\\
\texttt{ID}&--&local SNR ID\\
\texttt{match}&--&numbers of matched criteria\\
\texttt{OI}&boolean&whether identified in [\ion{O}{i}]/H$\alpha$ residual map\\
\texttt{SII}&boolean&whether identified in [\ion{S}{ii}]/H$\alpha$ residual map\\
\texttt{SII$\_$sigma}&boolean&whether identified in [\ion{S}{ii}]$\lambda$6716\AA\ velocity dispersion map\\
\texttt{BPT$\_$OI}&boolean&whether identified in [\ion{O}{i}] BPT distance map\\
\texttt{BPT$\_$SII}&boolean&whether identified in [\ion{S}{ii}] BPT distance map\\
\texttt{SII$\_$Ha}&ratio&the value of [\ion{S}{ii}]/H$\alpha$\\
\texttt{iso$\_$flag}&boolean&blended is False, isolated is True\\
\texttt{region$\_$ID}&number&if blended, ID of HII region object is blended with\\
\texttt{RA}&degree&right ascension coordinate\\
\texttt{DEC}&degree&declination coordinate\\
\texttt{Emission$^*$$\_$FLUX}&$10^{-20}$cm$^{-2}$erg\,s$^{-1}$&emission line flux, has been corrected for Milky Way foreground extinction\\
\texttt{Emission$\_$FLUX$\_$ERR}&$10^{-20}$cm$^{-2}$erg\,s$^{-1}$&emission line flux error\\
\texttt{Emission$\_$VEL}&\kms&emission line velocity\\
\texttt{Emission$\_$VEL$\_$ERR}&\kms&emission line velocity error\\
\texttt{Emission$\_$SIGMA}&\kms&emission line velocity dispersion\\
\texttt{Emission$\_$SIGMA$\_$ERR}&\kms&emission line velocity dispersion error\\
		\hline
	\end{tabular}	\\
\vskip 0.05in
\textbf{Note:} (*) Measured emission lines are in Table \ref{tab:lines} in Appendix \ref{lines}.
 \label{tab:catalog}
\end{table*}

Using the region footprints identified by \cite{groves2023phangs}, we measure how many of the \revtwo{objects in our parent sample fall in projection on top of nebulae. In total, $\sim$35\% of our objects overlap with nebulae classified as \ion{H}{ii} regions.  $\sim$30\% overlap with unclassified nebulae (typically labeled `unclassified' as they are found to be inconsistent with photoionization based on diagnostic line ratios), indicating that these objects are likely also identifiable by their H$\alpha$ morphology. The remaining $\sim$35\% of the parent sample is not included in the nebular catalog at all, and half of these are classified as SNRs, indicating how many objects would be missed if only line flux intensity is used for object selection.}  Within our catalog, we label those objects that overlap with \ion{H}{ii} regions as `blended', while those that do not overlap (or overlap with unclassified regions in the Nebular catalog) we identify as `isolated'. 

We provide integrated properties for all objects in our parent sample as a machine readable catalog with the list of columns provided in Table \ref{tab:catalog}. The catalog includes the host galaxy name and distance, object classification, object distance to the galaxy center in units of effective radius, environment mask that SNR lies on, coordinates, which criterion selected the object, emission lines fluxes, gas kinematics, isolated flag, [\ion{S}{ii}]/H$\alpha$ value, and H$\alpha$ luminosity.  

\subsection{Validation of methods -- comparing with SNRs in M83}
\label{sec:m83}
Our SNR identification method differs somewhat from past SNR searches in nearby galaxies, which were done using visual identification of H$\alpha$ bright sources that exhibit high [\ion{S}{ii}]/H$\alpha$ line ratios. In addition to our automated source detection, we also differ by using line-ratio \emph{residual} maps to account for diffuse emission, and make use of multiple optical line diagnostics for selection, including the kinematics (from the velocity dispersion maps) of our sources.

To test how well our identification method can recover SNRs, we apply our technique to the well-studied SNR population in M83.
This galaxy has an extensive MUSE mosaic \citep{dellabruna2022} and is well matched to our sample in terms of stellar mass, star formation rate, and morphology, but is closer in distance (4.61 Mpc; \citealt{saha2006cepheid}) and thus reaches significantly higher physical resolution ($\sim$20~pc).  M83 is one of the best-studied galaxies in terms of its SNR population, with a total of 366 SNRs identified using a combination of X-ray, IR and optical data \citep{dopita2010supernova, blair2012magellan, blair2014expanded, long2014deep, williams2019masses}, including recent work that used MUSE data to expand the SNR catalog \citep{long2022supernova} (L22 hereafter).  

\begin{figure*}
    \centering
    \includegraphics[width=\textwidth]{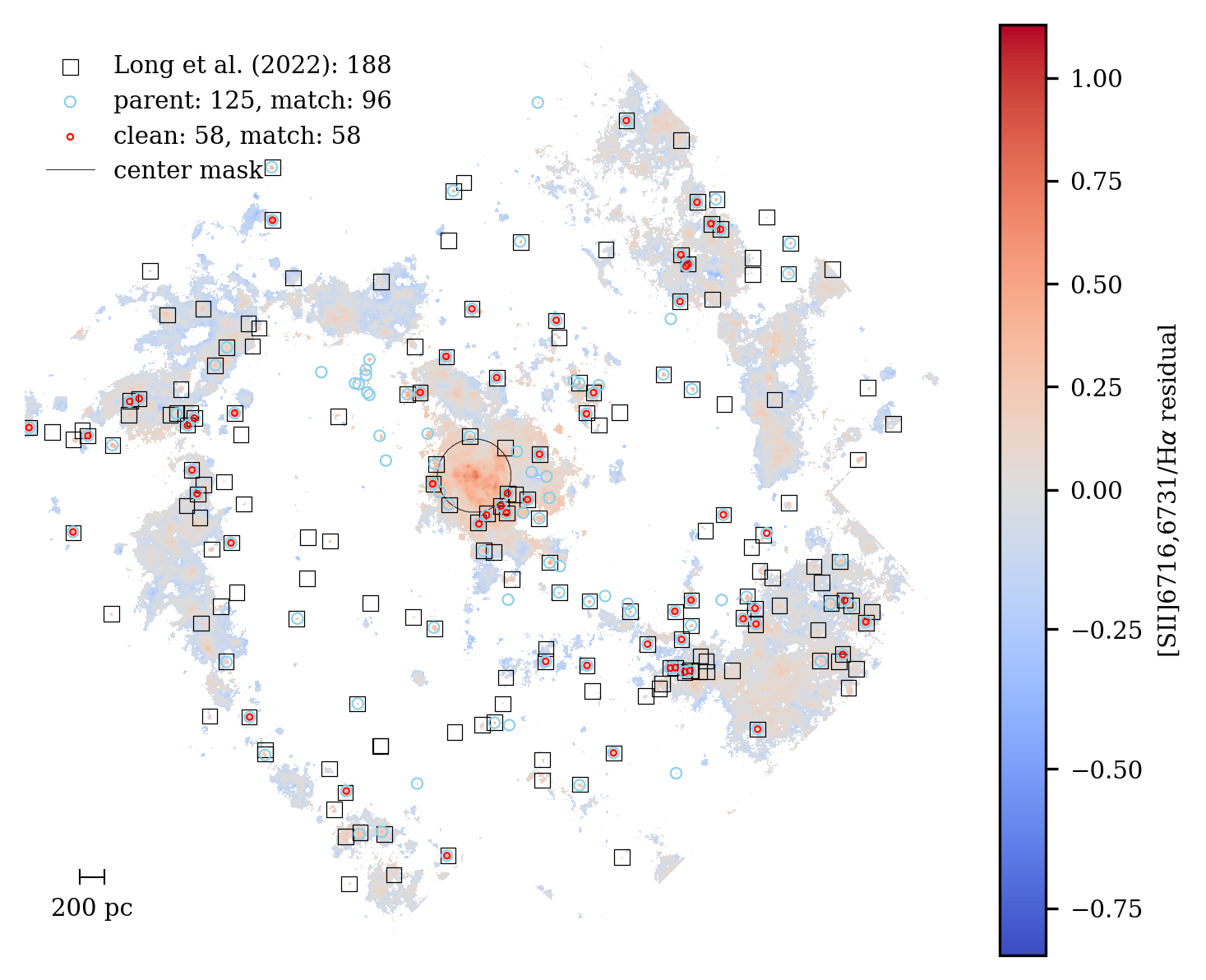}
    \caption{SNRs in M83. We mark the locations of SNRs that are identified by \citet{long2022supernova} (black boxes) and by applying methods used in this work (red circles for SNRs and blue circles for SNR candidates). All of the objects we identify as SNRs are also identified as SNRs in \citet{long2022supernova}. Overall, 77\% of our parent sample were previously identified as SNRs in the literature. }
    \label{fig:M83_SII}
\end{figure*}

The comparison between the objects we detect and the SNRs in the L22 catalog is visualized in Figure \ref{fig:M83_SII}. After masking the galaxy center, we identify a parent sample of 125 objects that satisfy at least one of our five criteria (see Appendix \ref{app:m83} for a complete table). Within the MUSE footprint (excluding the center), L22 identifies 188 SNRs and we recovered 96 of these (51\%). If we adjust our threshold for selecting sources in the [\ion{S}{ii}]/H$\alpha$ residual map then we can recover more of the known SNRs, but introduce a lot ($>$200) of additional SNR candidates that are not in the L22 catalog and may just be spurious detections or dynamically induced shock networks in the ISM. In all, 96 out of 125 (76.8\%) of our parent sample correspond to known SNRs.  

Statistically, the unrecovered 92 objects have lower median H$\alpha$ surface brightness and [\ion{S}{ii}]/H$\alpha$ than the recovered 96 ones, which makes their identification difficult using our method. In addition, the SNRs listed in L22 include 16 SNRs selected using methods other than optical lines (e.g. X-ray and [\ion{Fe}{ii}]1.644$\mu$m). We do not expect to detect these objects in the optical, \revtwo{and do not recover any of them}.

If we construct a SNR sample in M83, consisting of objects that satisfy at least two criteria using our methods, then 58 SNRs are identified. All of them are also identified as SNRs in L22.  This demonstrates that our method can recover SNRs with high confidence by leveraging multiple-line diagnostics. 

The 67 objects in the parent sample that are identified using only one method make up our SNR candidate sample.  57\% of these are contained in L22. This suggests that we can generally consider many of these SNR candidates being true SNRs. Of the objects not in the L22 catalog, based on visual inspection some look like reasonable SNR candidates, while others do not. For example, we identify a cluster of regions \revtwo{to the north-east of the galaxy center, along the bar and close to the bar-end, none of which exist in the literature catalogs and may instead be} due to shocks associated with the bar dynamics. 

This comparison demonstrates that overall our technique produces a reliable SNR catalog. Even our SNR candidates are often correctly identifying SNRs. By applying an \revtwo{automated peak-finding algorithm,}  we save time by not having to carry out a visual inspection and have a reproducible, homogeneous selection. However, the choices we have made in how we identify objects and construct our SNR catalog clearly do not produce a complete catalog and are also \revtwo{biased towards selecting only }`typical' SNRs. We do not recover the very young SNR in M83 identified by \cite{blair2014expanded}, although it could in principle be identified by its broadened [\ion{S}{II}] lines. However, it is so compact (1~pc in diameter), that when blended at 100~pc scales with neighboring \ion{H}{ii} regions it is no longer selected by its kinematic signatures, and we expect such young objects are systematically missed in our catalog.    

\section{Results}
\label{sec:results}
\subsection{Comparison with \ion{H}{ii} regions}\label{compare}
\begin{figure}[h!]
    \begin{subfigure}[b]{0.5\textwidth}
    \centering
    \includegraphics[width=\textwidth]{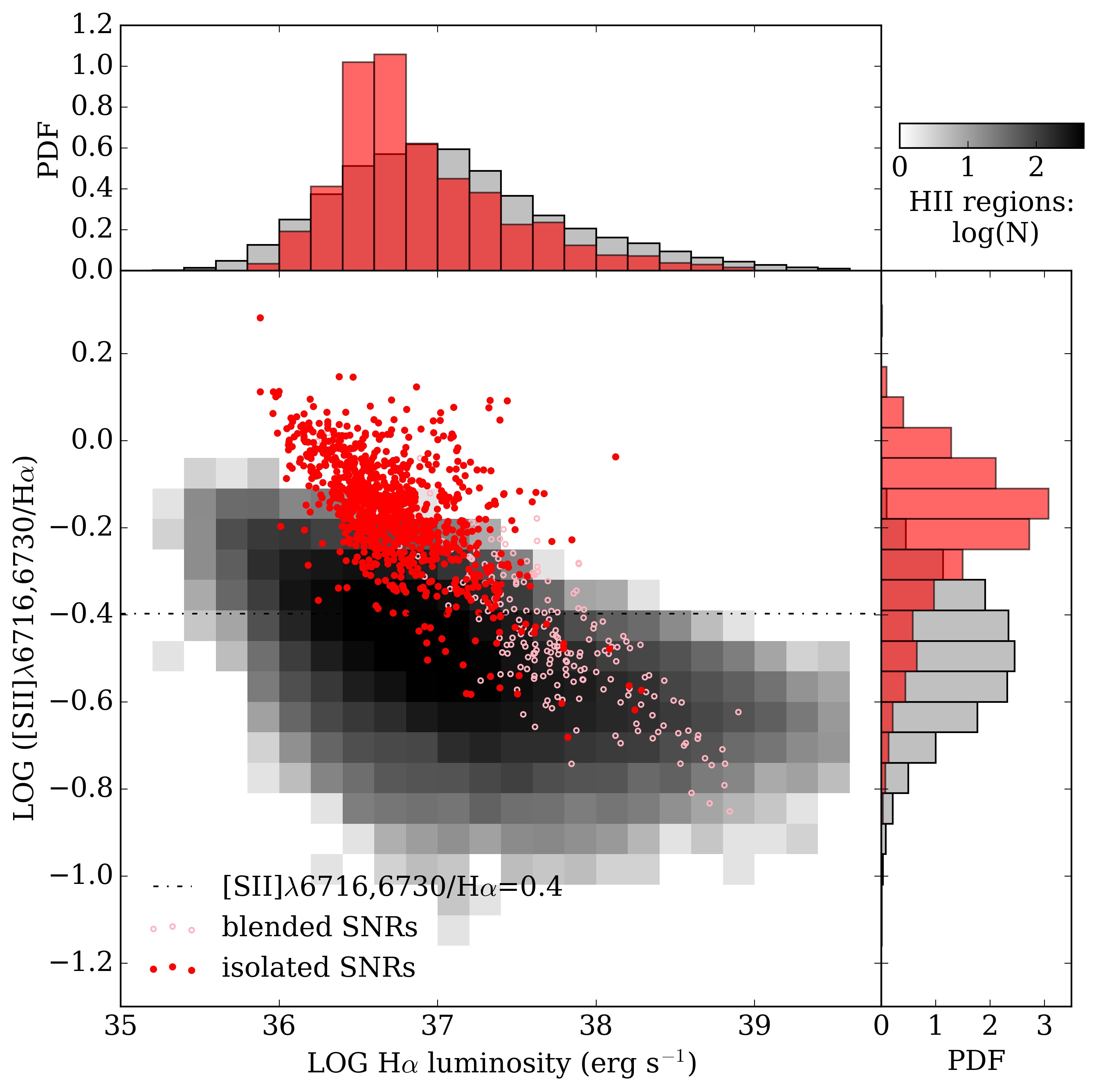}
    \label{fig:SNR_HII_1}
    \end{subfigure}
    \hfill
    \begin{subfigure}[b]{0.5\textwidth}
    \includegraphics[width=\textwidth]{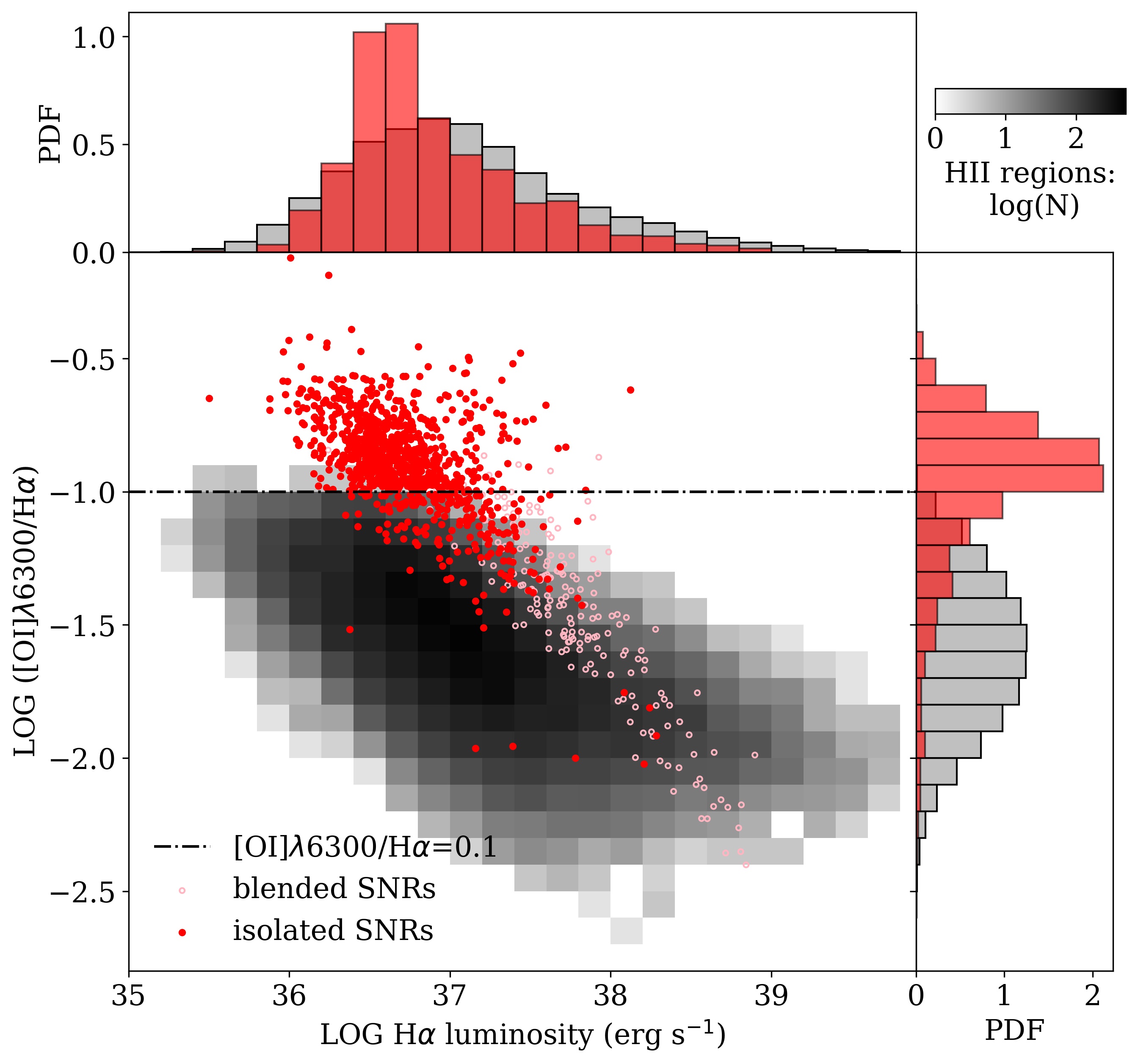}   
    \label{fig:SNR_HII_2}
    \end{subfigure}
    \caption{[\ion{S}{ii}]/H$\alpha$ (top) and [\ion{O}{i}]/H$\alpha$ (bottom) vs. H$\alpha$ luminosity for SNRs as red dots compared with the \ion{H}{ii} regions density distribution in grayscale. For SNRs that overlap with a \ion{H}{ii} region, the symbol is in an empty circle in light red.  Thresholds of [\ion{O}{i}]/H$\alpha$=0.1 and [\ion{S}{ii}]/H$\alpha$=0.4 (horizontal dash-dotted lines) are shown. SNRs overlapping with \ion{H}{ii} regions are selected by other criteria so they can still be identified as SNRs. Emission lines are not corrected for the photoionization.}
    \label{fig: SNR_HII}
\end{figure}

We compare the integrated emission line properties of our sample of SNRs with \ion{H}{ii} regions identified in \citet{groves2023phangs}. As we can see in the comparison in Figure \ref{fig: SNR_HII}, SNRs have higher [\ion{O}{i}]/H$\alpha$ and [\ion{S}{ii}]/H$\alpha$ ratios than \ion{H}{ii} regions at the same  H$\alpha$ luminosity. These ratios decrease with increasing H$\alpha$ luminosity. \revtwo{This reflects our selection biases, as a faint SNR blended with a brighter \ion{H}{ii} region will no longer exhibit the distinctive line ratios and line kinematics that we select for with our methods.} This can be seen directly, as SNRs appearing in projection with \ion{H}{ii} regions (light red circles in the figure) have higher H$\alpha$ luminosities and correspondingly lower [\ion{O}{i}]/H$\alpha$ and [\ion{S}{ii}]/H$\alpha$ ratios.

\begin{figure}[h!]
    \begin{subfigure}[b]{0.5\textwidth}
    \centering
    \includegraphics[width=\textwidth]{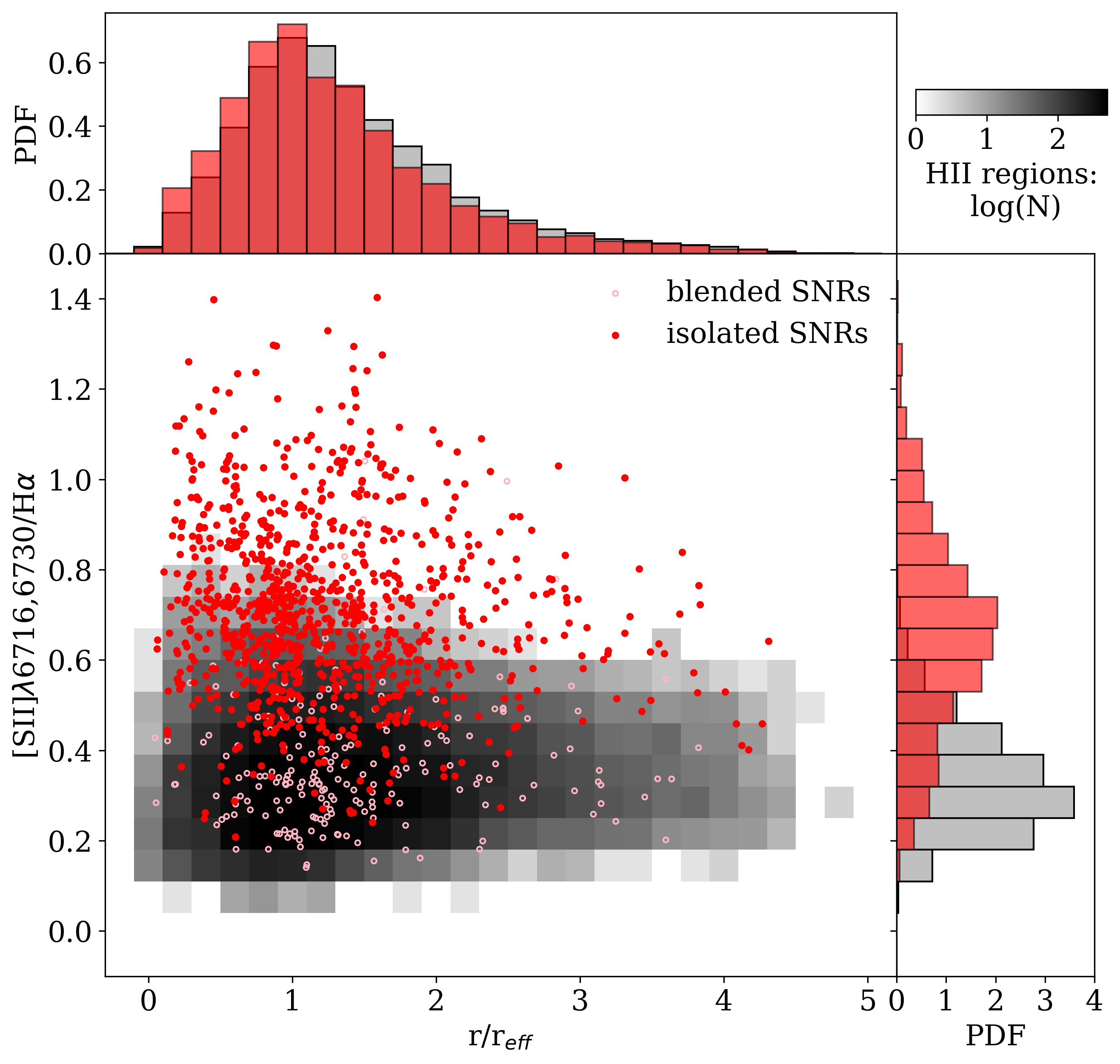}
    \caption{}
    \label{fig:dist_SII}
    \end{subfigure}
    \hfill
    \begin{subfigure}[b]{0.5\textwidth}
    \includegraphics[width=\textwidth]{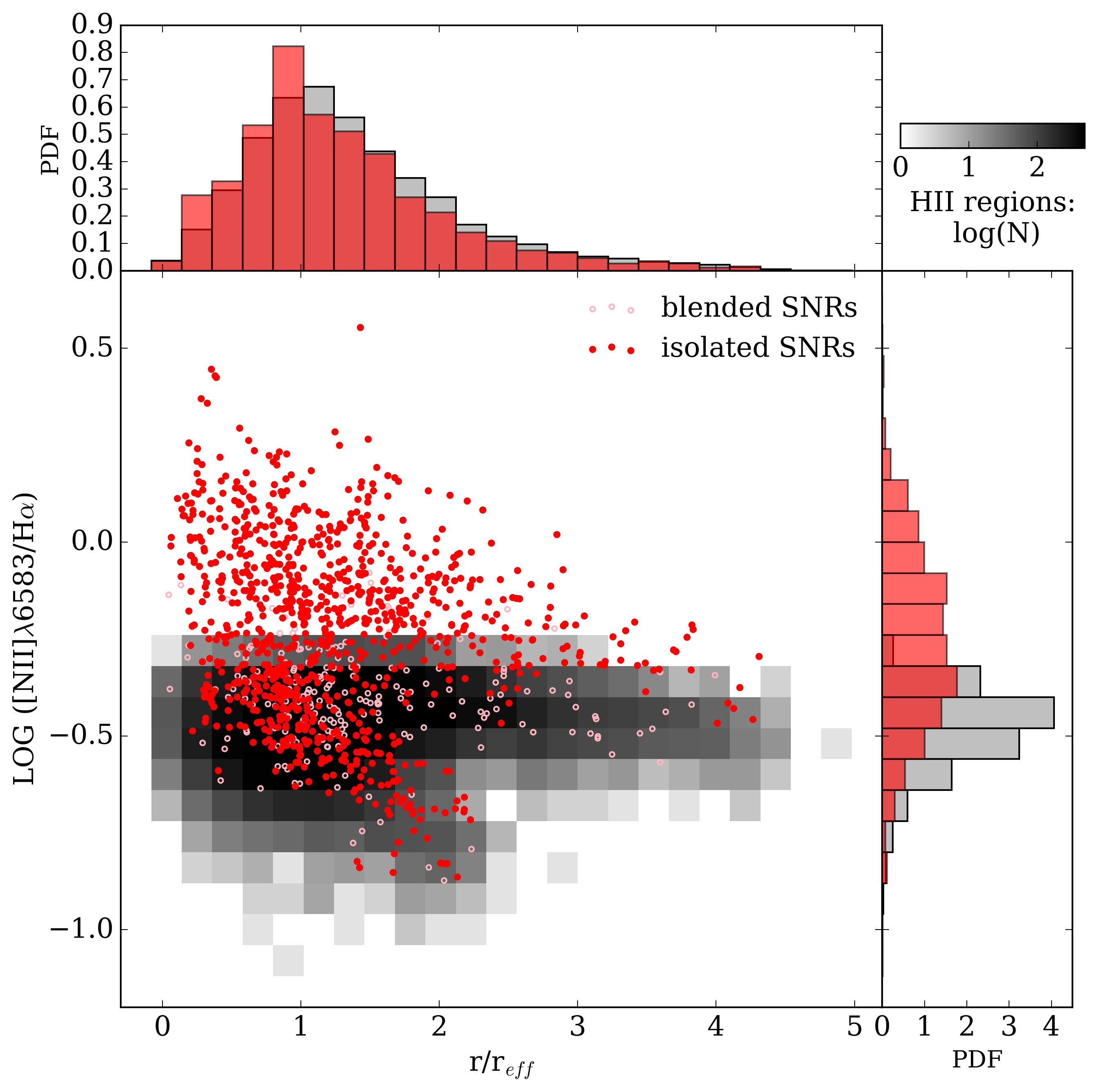}   
    \caption{}
    \label{fig:dist_NII}
    \end{subfigure}
    \caption{Radial trends of SNRs and \ion{H}{ii} regions. r/r$_{\rm eff}$ is the ratio of SNR distance to the effective radius of the corresponding galaxy. \emph{(a)}: [\ion{S}{ii}]/H$\alpha$ changes with the distance to their corresponding galactic centers for SNRs as red dots and \ion{H}{ii} regions in grayscale. \emph{(b)}: [\ion{N}{ii}]$\lambda6583$/H$\alpha$ change with the distance to their corresponding galactic centers for SNRs as red dots and \ion{H}{ii} regions in grayscale. For SNRs that overlap with a \ion{H}{ii} region, the symbol is in an empty circle in light red. The drop in the numbers at low r/r$_{\rm eff}$ is likely influenced by the mask in of the centres of the galaxies. }
    \label{fig:dist}
\end{figure}

\revtwo{In Figure \ref{fig:dist_SII}, we see a concentration of both SNRs and \ion{H}{ii} regions at small r$_{eff}$. This is expected as both closely follow the star formation rate distribution \citep{cronin2021local}, which is more concentrated in the inner regions of galaxies \citep{leroy2021phangs}. We find hints of radial trends in the SNR line ratios, with higher line ratios found in SNRs compared to \ion{H}{ii} regions at all radii, and higher values of [\ion{S}{ii}]/H$\alpha$ in SNRs at smaller radii. }
Generally, the [\ion{S}{ii}] line is brighter for higher shock velocities \citep{allen2008mappings}. \revtwo{Since SNR shocks typically slow down over time, this is indicative of a larger population of younger SNRs. Given the overall larger number of SNRs at small radii, this likely simply reflects the higher probability of detecting these younger SNRs, although } 
higher ambient densities around SNRs at smaller galactic radii could also be playing a role.

We also find similar trends in the metallicity-sensitive [\ion{N}{ii}]$\lambda6583$/H$\alpha$ line ratio with radius (Figure  \ref{fig:dist_NII}). In most previous SNR surveys that have relied on narrow-band imaging, the [\ion{N}{ii}]$\lambda6583$ lines often ended up convolved within the H$\alpha$ filter and not studied separately, however, with MUSE we can cleanly separate the [\ion{N}{ii}]$\lambda6583$ and H$\alpha$$\lambda6562$ lines. 
The higher line ratios at smaller radii could be a consequence of higher ISM metallicities \citep{kreckel2019mapping, pessa2021star}. 

\begin{figure}
    \begin{subfigure}[b]{0.5\textwidth}
    \centering    \includegraphics[width=\textwidth]{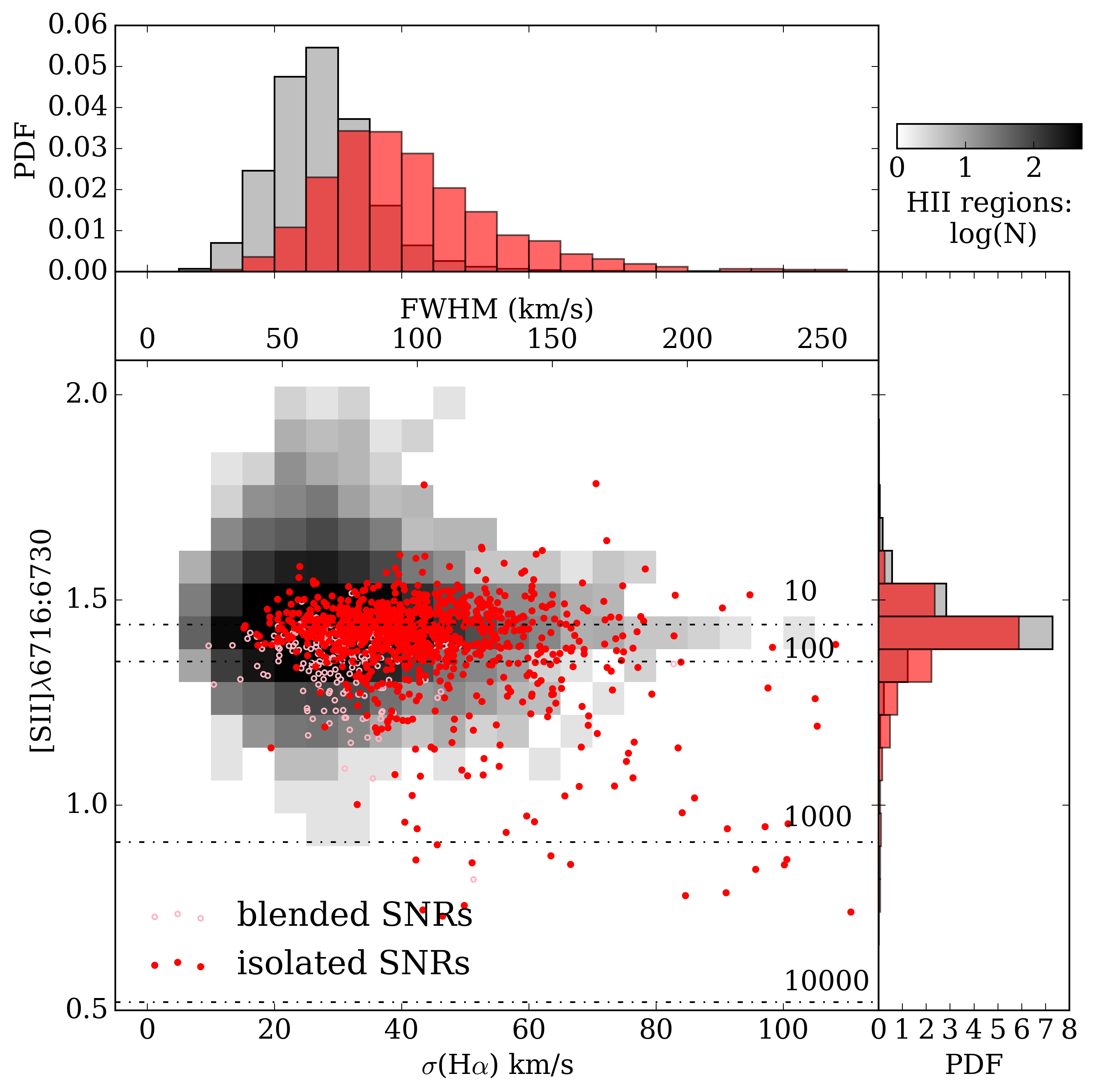}
    \caption{}
    \label{fig:dens_sigma}
    \end{subfigure}   
    \hfill
    \begin{subfigure}[b]{0.5\textwidth}
    \centering    \includegraphics[width=\textwidth]{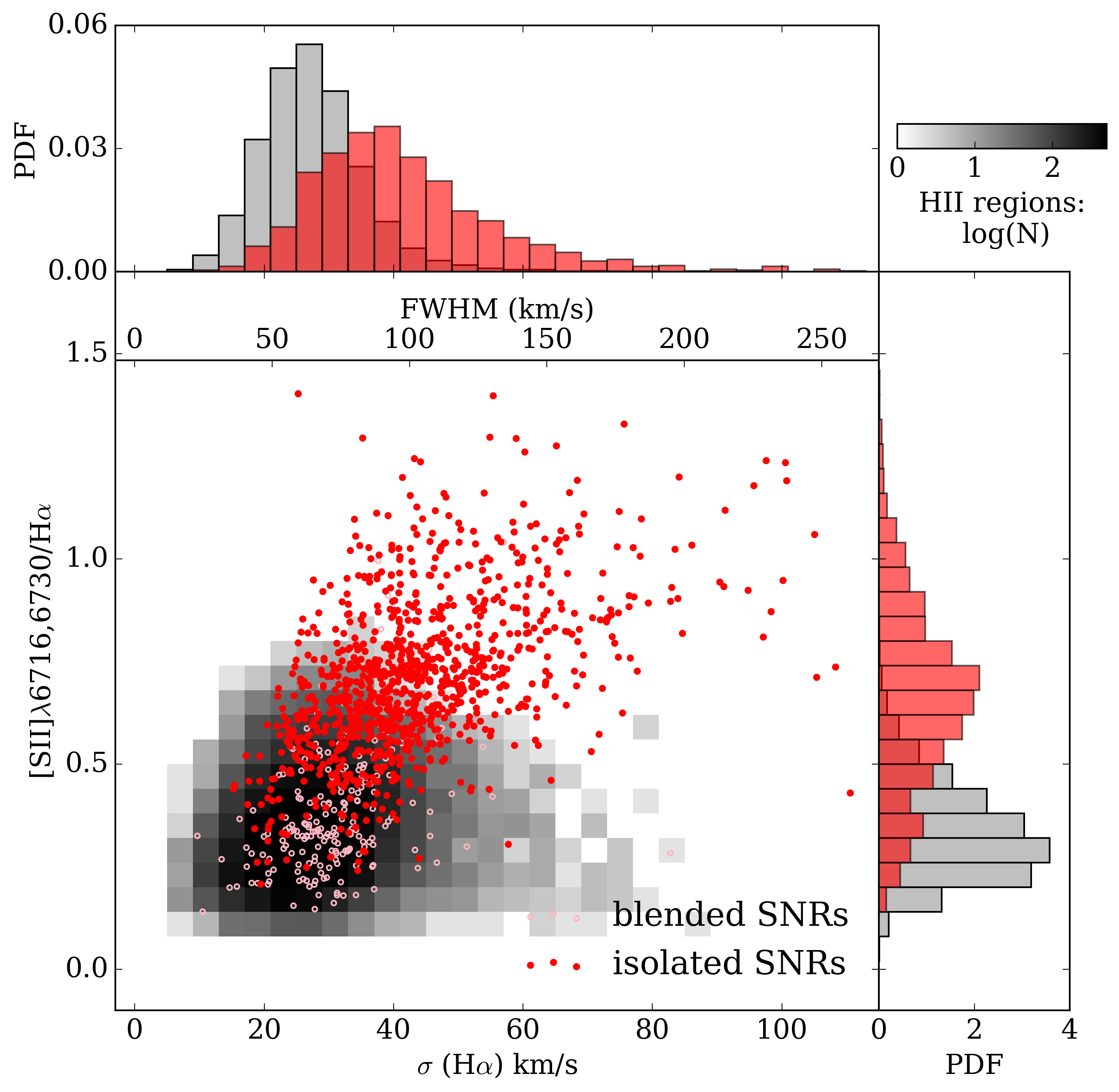}
    \caption{}
    \label{fig:SII_sigma}
    \end{subfigure}
    \caption{ (a): Line ratios of [\ion{S}{ii}] vs. velocity dispersion of SNRs and \ion{H}{ii} regions. The top axis shows the FWHM. The H$\alpha$ velocity dispersion changes with the [\ion{S}{ii}]$\lambda$6716/[\ion{S}{ii}]$\lambda$6730 ratio for SNRs as red dots and \ion{H}{ii} regions in grayscale. The ratios above 1.4 are not physical and due to noise. The horizontal dashed-dotted lines indicate different electron densities with the corresponding value next to it. Bottom: Velocity dispersion of H$\alpha$ changes with [\ion{S}{ii}]$\lambda$6716,6730/H$\alpha$ ratio for SNRs as red dots and \ion{H}{ii} regions in grayscale. For SNRs that overlap with a \ion{H}{ii} region, the symbol is in an empty circle in light red. These two criteria work effectively together in distinguishing between SNRs and \ion{H}{ii} regions.} 
\end{figure}

In Figures \ref{fig:dens_sigma} and \ref{fig:SII_sigma}, we compare the velocity dispersion \revtwo{(corrected for instrumental broadening)} of the H$\alpha$ line and the [\ion{S}{ii}]$\lambda$6716/[\ion{S}{ii}]$\lambda$6730 emission line ratio, which is a density-sensitive diagnostic\footnote{More specifically, the ratio traces the electron density in the post-shock [\ion{S}{ii}] recombination zone, and is proportional to the preshock density \citep{Sutherland2016}.}. We can clearly see that while \ion{H}{ii} regions are concentrated at low-velocity dispersions of  30-40 \kms, SNRs can have velocity dispersions that reach more than 100 \kms, owing to the presence of higher-velocity shocked material in the SNRs. Velocity dispersion is also one criterion that we used in Section \ref{sigma} to identify SNRs. Regarding ambient densities, most of the SNRs \revtwo{have line ratios  consistent at the 3$\sigma$ level with the low-density limit ($\sim$10-100 cm$^{-3}$), with only $\sim$15\% of regions showing ratios corresponding to significantly higher densities ($>$100 cm$^{-3}$). } 
A similar distribution of values has also been seen in the SNRs of nearby galaxies \citep[e.g.][]{long2018mmt, long2022supernova}. No visible trend exists between the [\ion{S}{ii}] doublet ratio and the velocity dispersion of the SNRs. 

Figure \ref{fig:SII_sigma} shows the power of combining velocity dispersion and [\ion{S}{ii}]/H$\alpha$ ratio together to distinguish SNRs from \ion{H}{ii} regions. Our SNRs populate a region of parameter space that has both high-velocity dispersions and high [\ion{S}{ii}]/H$\alpha$ values. The distinction is clearer in this regime, however, it becomes complicated when SNRs overlap with \ion{H}{ii} regions as indicated by light red circles in the figure.

\begin{figure*}
    \centering
    \includegraphics[width=\textwidth]{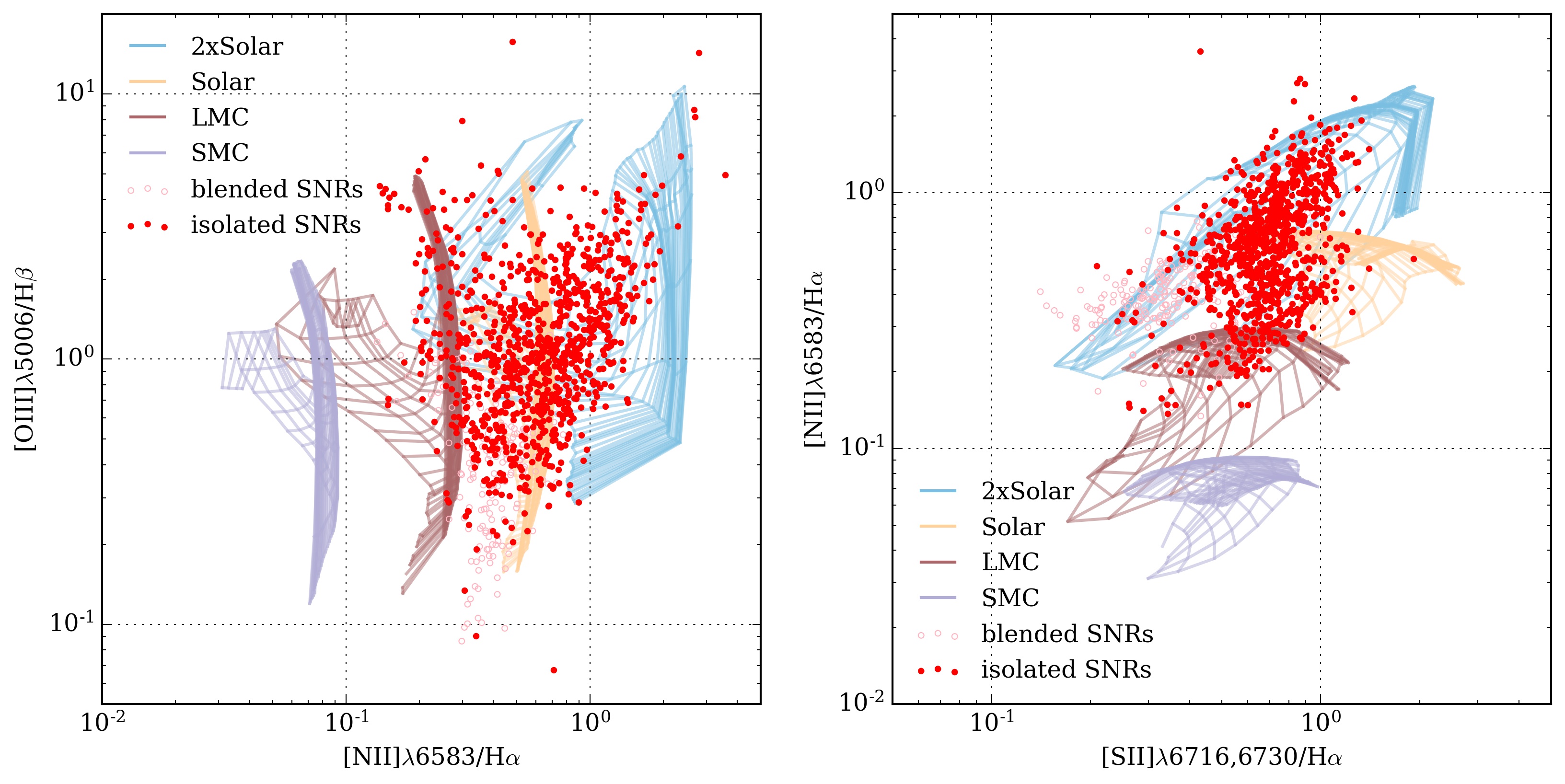}
    \caption{Left: Observed [\ion{O}{iii}]$\lambda$5006/H$\beta$ vs. [\ion{N}{ii}]$\lambda$6583/H$\alpha$ for SNRs as red dots. Right: Observed [\ion{N}{ii}]$\lambda$6583/H$\alpha$ vs. [\ion{S}{ii}]/H$\alpha$ for SNRs as red dots. For SNRs that overlap with a \ion{H}{ii} region, the symbol is in an empty light red circle. Background grids are MAPPINGS models with different metallicities. Our identified SNRs mostly lie between models for LMC and twice solar metallicities.}
    \label{fig:SII_NII}
\end{figure*}

We compare our observed SNRs with the MAPPINGS \citep{allen2008mappings} model in Figure \ref{fig:SII_NII}. Different grids show different ISM metallicities. Here we adapted the \citet{dopita2005modeling} abundance model as opposed to the default solar models. Each grid samples a range of shock velocities and magnetic field strengths (0$\sim$10 $\mu$G).  We see that our SNRs are consistent with grids sampling between LMC (half solar metallicity) and twice solar metallicities (as defined in MAPPINGS model). This is an important validation of our selection process, that we are indeed primarily picking objects that are consistent with the expected parameter space of shocks \citep{long2022supernova}.
This is consistent with the range of metallicities observed in the \ion{H}{ii} regions detected in these galaxies \citep{groves2023phangs}. 

\subsection{Comparison with the literature: SNRs in other galaxies}
In order to further validate our selected SNR population in the 19 MUSE galaxies, we compare their optical line emission properties with the well-studied SNR populations in four nearby galaxies -- M33 \citep{long2018mmt}, M51 \citep{winkler2021optical}, M83 \citep{long2022supernova}, and NGC6946 \citep{long2019new}. The SNR populations in these galaxies have been vetted not only with optical spectroscopy, but also multi-wavelength information from radio, X-ray, and in some cases near-IR lines (e.g. [\ion{Fe}{ii}]1.64 $\mu$m in M83).


In the process of identifying SNRs, a selection bias may exist  
in more distant galaxies, as we are prone to selecting more luminous SNRs. \revtwo{These could either be bright due to being young, or because the explosion happens in a dense environment. 
At larger distances (and lower physical resolution) we are also more likely to miss SNRs, as they can be blended with brighter \ion{H}{ii} regions. } 

We note that within our sample, there is a set of SNRs that exhibit very high H$\alpha$ luminosities ($\geq$ 10$^{38}$ erg s$^{-1}$) and relatively low [\ion{S}{ii}]/H$\alpha$ ratios ($\leq$ 0.4). In many respects, these characteristics more strongly resemble \ion{H}{ii} regions than SNRs, \revtwo{however, as they are selected by multiple criteria we remain confident in their classification as SNRs}. We see that these SNRs are almost all identified as blended with \ion{H}{ii} regions in our sample. If we exclude these, we identify very few SNRs with H$\alpha$ luminosities above 10$^{37.5}$ erg~s$^{-1}$.  

\begin{figure*}
    \centering 
    \includegraphics[width=\textwidth]{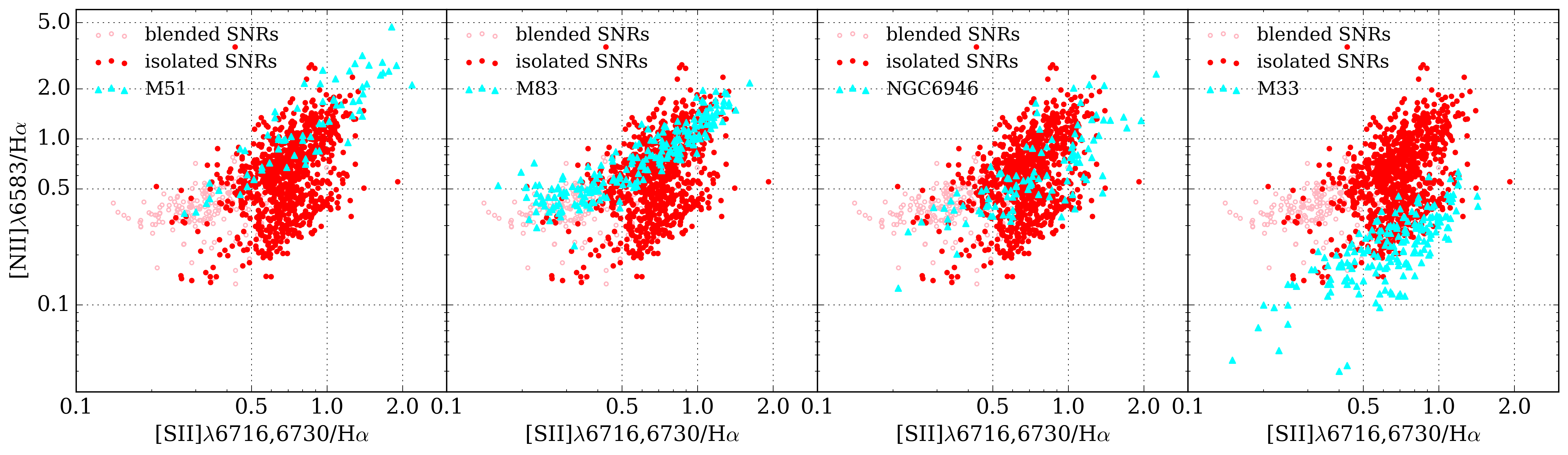}
    \caption{Relation between [\ion{S}{ii}]/H$\alpha$ ratio and [\ion{N}{ii}]/H$\alpha$ ratio for SNRs identified in nearby 4 galaxies from the literature \citep{long2018mmt,winkler2021optical,long2022supernova,long2019new} and 19 MUSE galaxies. SNRs are marked by red dots and for SNRs that overlap with a \ion{H}{ii} region, the symbol is an empty circle in light red. }
    \label{fig: SII_NII_each}
\end{figure*}

In Figure \ref{fig: SII_NII_each}, we compare the line ratios [\ion{S}{ii}]/H$\alpha$ and [\ion{N}{ii}]/H$\alpha$ of the SNRs in our PHANGS MUSE galaxies and the four nearby galaxies. Similar to Figure \ref{fig:SII_NII}, we find that our SNRs show a range of values of different line diagnostics that are also observed in SNR surveys of more nearby well-studied galaxies. All four galaxies exhibit a similar range in [\ion{S}{ii}]/H$\alpha$, which likely implies that we are sampling SNRs within a similar range of shock velocities. 

Interestingly, we find a bimodal distribution for the metallicity-sensitive [\ion{N}{ii}]/H$\alpha$ in our PHANGS-MUSE SNRs. The top branch matches well with SNRs in M83, a high-mass and high-metallicity galaxy, while the bottom branch overlaps with SNRs in M33, a low-mass and low-metallicity galaxy. Some SNRs in M51 occupy the same location in this plot with the top branch of our SNRs but some SNRs have much higher [\ion{N}{ii}]/H$\alpha$ ratios than any other SNRs. For NGC6946, the SNRs lie in between the two branches of our SNRs. 

\begin{figure*}
    \centering 
    \includegraphics[width=0.6\textwidth]{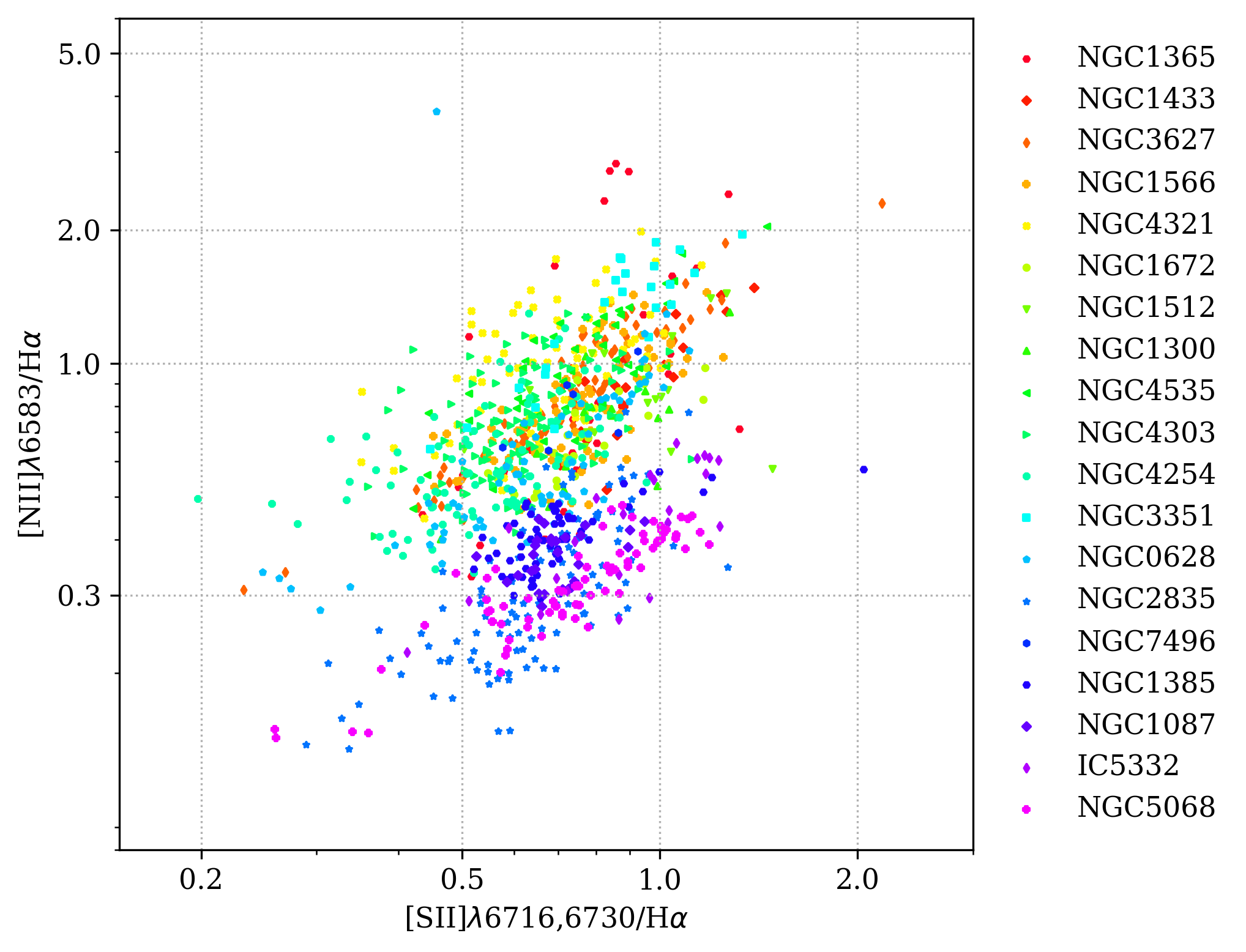}
    \caption{Relation between [\ion{S}{ii}]/H$\alpha$ ratio and [\ion{N}{ii}]/H$\alpha$ ratio for SNRs in 19 galaxies. SNRs in the same galaxy are marked by the same color and symbol. In the legend, from top to bottom, the stellar mass of the galaxy decreases, from 10$^{11}M_\odot$ for NGC~1365 to 10$^{9.41}M_\odot$ for NGC~5068. }
    \label{fig:SII_NII_all}
\end{figure*}

We also attempted to determine whether this trend exists within our 19 galaxies, as shown in Figure \ref{fig:SII_NII_all}. In this figure, SNRs belonging to the same galaxy are represented by the same color, ranging from red to purple as the stellar mass of the galaxy decreases. We observe the same trend as identified from the four literature targets, with higher mass systems exhibiting higher [\ion{N}{ii}]/H$\alpha$ ratios than lower mass galaxies, which likely relates to metallicity offsets between these targets. A more detailed investigation of these trends with metallicity is beyond the scope of this work, but also agrees qualitatively with the trends expected based on shock models (see Figure \ref{fig:SII_NII}). 

In summary, the comparison with MAPPINGS models in Figure \ref{fig:SII_NII} and with the nearby SNR population in Figure \ref{fig: SII_NII_each} provides a crucial validation of our SNR selection process and the purity of our SNR catalog. Despite not being able to resolve their structure, it is clear that we are picking objects that are consistent with the parameter space of shocks, as well as the observed range of properties of well-studied SNRs in nearby galaxies.   

\section{Discussion}
\label{sec:discussion}
\subsection{Expected number of SNRs and inferred SN frequencies}
\label{sec:snrates}
We have identified 2233 objects that we classify as SNRs or SNR candidates in 19 nearby galaxies. To explore the completeness of our sample, we compare it with a theoretical estimation of the number of SNRs we expect in each galaxy. The number of core-collapse (CC) SNRs correlates with the star formation rate (SFR) surface density. To estimate the number of SNRs in each galaxy, we calculate the SN rate per unit area following the formalism presented in \citet{bacchini2020evidence}:
\begin{equation}\label{eqn:snrate}
    R_{\rm cc}=\Sigma_{\rm SFR}f_{\rm cc},
\end{equation}
where $\Sigma_{\rm SFR}$ is the SFR surface density and $f_{\rm cc}\approx1.3\times10^{-2}M_\odot^{-1}$ represents the number of core-collapse SNRs for a unit of stellar mass formed under a \citet{kroupa2002initial} initial mass function. \revtwo{We neglect SN associated with type Ia here, however, as estimates of the ratio of CC to type Ia SNe ranges from about 3:1 to 1:1 \citep{mannucci2005supernova, li2011nearby}, accounting for these} only decreases the value of $f_{\rm cc}$ by a factor of two at most.


We estimate the SFR surface density from the dust-corrected H$\alpha$ luminosity \citep{kennicutt1998star}:

\begin{equation}
    {
    \rm SFR}(M_\odot\, yr^{-1})=7.9\times 10^{-42}L({\rm H}\alpha) 
    ({
    \rm erg\,s}^{-1}).
\end{equation} 

\begin{figure}
    \centering
    \includegraphics[width=0.5\textwidth]{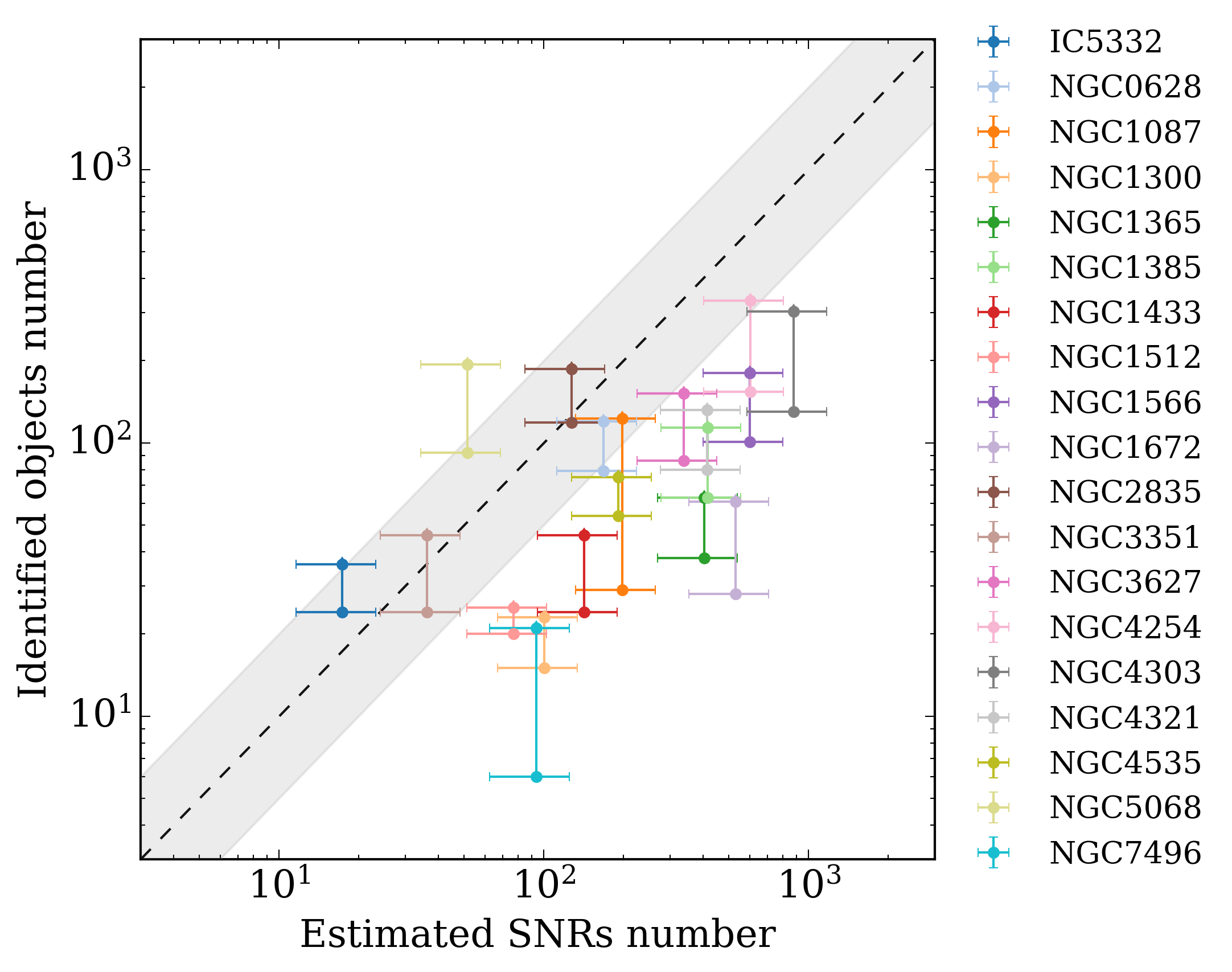}
    \caption{Comparison between the estimated SNR number from the SFR and objects identified in this paper. The lower limit of the identified object is the number of SNRs and the upper limit is the number in the parent sample including SNR candidates. The lower limit of the estimated SNR number is given by the optically visible time of SNRs for 10,000 years and the upper limit corresponds to 20,000 years. The dashed black line is the one-to-one relation. The shaded area provides coverage from two times the recovery rate to 50\% of the recovery rate of SNRs.}
    \label{fig:SNRs-number}
\end{figure}

We account for dust attenuation using the Balmer decrement, assuming a fixed H$\alpha$/H$\beta$ = 2.86 \citep{Storey:95} based on Case B recombination and the Calzetti attenuation curve and R$_V$=3.1 \citep{calzetti2001dust}. Combining SFR, the MUSE FoV area, and the estimated time (10,000 -- 20,000 years; \citealt{badenes2010size,sarbadhicary2017supernova}) for which SNRs remain visible in the optical, we estimate the expected number of SNRs for each galaxy. 
The comparison of the estimated SNR number and identified SNR number is provided in Figure \ref{fig:SNRs-number}.For the majority of galaxies, the number of identified SNRs is around half of the expected number of SNRs. When taking the parent sample as a whole, including both SNRs and SNR candidates, we find a better agreement with our simplistic estimation. This suggests that we are recovering \revtwo{about half of the expected SNRs,  even more if a large number of our SNR candidates are true SNRs.} 

In almost all cases, we find fewer SNRs than would be predicted. This is not surprising, as our approach was not designed to be maximally complete. \revtwo{The few galaxies that show more SNRs than predicted do not show any consistent global properties, although one of these (NGC~5068) is the closest galaxy in our sample and another (NGC~4303) is one of the seven galaxies that hosts an AGN. } 
Within the Milky Way, only $\sim$ 30\% of identified SNRs were chosen based on optical observations \citep{green2014catalogue}, \revtwo{suggesting we can expect to miss a large number of objects.} On the other hand, we cannot distinguish SNRs due to core collapse from Type Ia ones, which might cause a (small) excess of observed SNRs when compared to our theoretical estimation (Equation \ref{eqn:snrate}). Overall, given the tantalizing correlation between our expected and observed SNR numbers, this suggests that our parent sample is surprisingly indicative of the total estimated SNR population in the surveyed areas.  

\begin{figure*}
    \centering
    \includegraphics[width=0.8\textwidth]{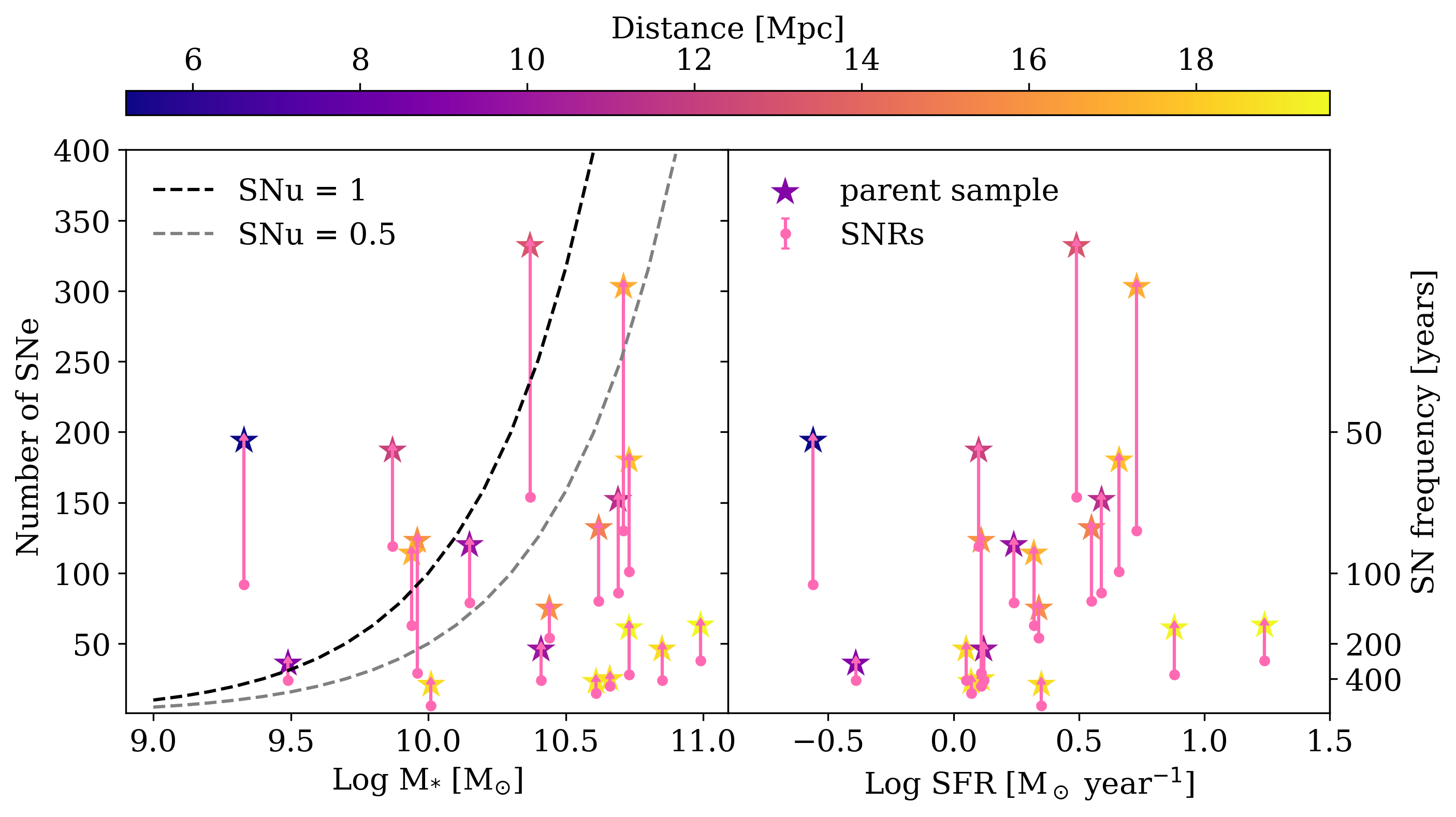}
    \caption{Number of SNe as a function of stellar mass (left) and SFR (right) considering the entire parent sample, colored by galaxy distance and considering galaxy integrated properties measured only within the MUSE field of view. These SN counts can also directly be converted to a SN \revtwo{frequency} by assuming SNRs are visible in the optical for 10,000 years. Lines of constant SN rate per unit mass (SNu) are also shown. The star symbol represents the number for the parent sample while the circle is the number of SNRs. The line between them shows the range of possible true SNRs number.  \label{fig:snrate}}
\end{figure*}

If we believe we are achieving a fairly complete census of SNe, we can also directly compute \revtwo{an upper limit to} the \revtwo{SN frequency} for our sample of galaxies. If we assume SNRs remain visible in the optical for 10,000 years \citep{sarbadhicary2017supernova}, then on average we find a SN frequency of one per 85 years per galaxy, with values of the average time between SN within individual galaxies ranging from 30 to 476 years (Figure \ref{fig:snrate}). The \revtwo{galaxies with less frequent SNe  (lowest number of SN per galaxy)} are found to be our most distant (D $>$ 18 Mpc) targets, which makes sense as we are more likely to have missed SNe in these cases where our physical resolution is worse. The average SN frequency increases to one per 65 years per galaxy if we only consider the 13 galaxies with distances less than 18 Mpc. We find no significant secondary correlation with integrated galaxy properties like global SFR or stellar mass, but our sample is still small. These SN frequencies are only \revtwo{upper limits} due to the uncertainties in the SNR visibility timescales but are  similar to the frequency of one per 40$\pm$10 years estimated for our own Milky Way \citep{tammann1994galactic}, \revtwo{which is reasonably well matched to the median stellar mass and SFR in our sample of galaxies}. 
Scaling this to the stellar mass contained within the MUSE field of view, 
this corresponds to a median SN rate per unit mass (SNu; \citealt{mannucci2005supernova}) of 1 SNe per 10$^{10}$ M$_\sun$ per century. 

\subsection{Lessons learned when identifying SNRs with IFUs}
\label{sec:lessons}

\revtwo{By using multiple criteria to select objects for our parent sample, we find we are able to nearly double the number of objects identified compared to using any one diagnostic alone. The combination of [\ion{O}{i}]/H$\alpha$ residuals and [\ion{S}{ii}]/H$\alpha$ residuals gives the largest number of sources, 1749, corresponding to $\sim$ 78\% of the objects in the parent sample. The combination of [\ion{O}{i}]/H$\alpha$ residuals and BPT (OI-OIII) gives the second largest number of sources, 1703. 
The BPT diagnostics select fewer objects than the line ratio residual approaches (Table \ref{tab:number}), however, they are almost all ($\sim$90\%) identified as robust (isolated) SNRs by our method. Taken together, all five methods have distinct diagnostic powers. }

Our results when using these five different techniques to identify SNRs and SNR candidates reflect the fact that the optical properties of SNRs present a variety of emission line characteristics, particularly when seen in projection against nearby \ion{H}{ii} regions and DIG. There is no singular feature that provides high confidence in our SNR classification, but instead, the emission-line characteristics provide a continuous spectrum, with increasing confidence of a correct classification the more independent features we observe. However, we lack an independent method to confirm our classification, making it challenging to robustly determine `best practices' for SNR determination. 
\revtwo{Shock models like MAPPINGS  provide some guidance (Figure \ref{fig:SII_NII}), however, a richer set of highly resolved emission line maps of SNRs, either via observations \citep[e.g.][]{drory2024} or simulations \citep[e.g.][]{makarenko2023supernova}, would significantly improve our understanding of how to best identify and classify SNRs that are blended with background emission.} 
By introducing an automated approach to source detection, we hope that future work can begin a more systematic exploration of the completeness limits probed by IFU data sets through the injection of simulated objects to test how well SNRs with different characteristics are recovered.

There are other line ratio combinations, not explored in this work, that could also potentially serve as useful diagnostics for the classification of SNRs in IFU datasets. For example, [\ion{Fe}{ii}]$\lambda$8616/Paschen11$\lambda$8862 should be higher in SNRs than in \ion{H}{ii} regions, as iron lines are strong shock indicators \citep{oliva1989infrared,reipurth2000hubble}. Unlike other optical [\ion{Fe}{ii}] lines that are produced by continuum pumping fluorescence (such as $\lambda 4287$ or $\lambda 5158$), $\lambda 8617$ is essentially produced by collisional excitation at low excitation levels \citep{Rodriguez:99,Verner:00,MendezDelgado:21b,Mendoza:23}. This characteristic, combined with the low ionization potential of Fe$^+$, makes it a promising shock indicator analogous to other [\ion{Fe}{ii}] lines at far-infrared wavelengths, such as the bright [\ion{Fe}{ii}] 1.64 $\mu$m line \citep{Antoniucci:14}. Additionally, while Fe is typically depleted into dust grains in HII regions \citep{Rodriguez:02,Izotov:06}, shock waves are capable of efficiently destroying these dust grains and releasing Fe into its gaseous phase \citep{Mesa-Delgado:09,MendezDelgado:22b}. Therefore, in the case of overlapping between SNRs and \ion{H}{ii} regions, this indicator could favor the conditions of the SNRs. We see clear evidence of this line in some of our spectra, however, the strong neighboring skylines make a robust extraction of line fluxes challenging and beyond the scope of this work. 

The [\ion{S}{iii}]$\lambda$9069/[\ion{S}{ii}]$\lambda$6716,6730 ratio might also be used to diagnose the ionization mechanism \citep{long2022supernova}. In \ion{H}{ii} regions, [\ion{S}{iii}]/[\ion{S}{ii}] correlates strongly with ionization parameter \citep{kewley2001theoretical}, but it can span a wide range of values depending on the local ISM density and the radiation field produced by the ionizing source \citep{groves2023phangs}. As shown in L22, high values of [\ion{S}{iii}]/[\ion{S}{ii}] provide strong indications of \ion{H}{ii} regions, but low values are more challenging to interpret as the separation with \ion{H}{ii} regions is not clear and [\ion{S}{iii}] can be too weak.

\subsection{Classification of the candidate sample and blending with HII regions}
\label{sec:overlapHII}
In Section \ref{sec:results}, we only focus on properties of 1166 SNRs, whereas there are still 1067 SNR candidates. Even though we are less certain these are SNRs, they are robust detections using diagnostics that give strong evidence for shock excitation. We also saw in Section \ref{sec:snrates} that including these objects provides better agreement with the expected number of SNRs given the SFR in each galaxy, and in Section \ref{sec:m83} we showed that in M83 about half of our candidates are true SNRs in \cite{long2022supernova}.   

\begin{figure}
    \centering
    \includegraphics[width=0.5\textwidth]{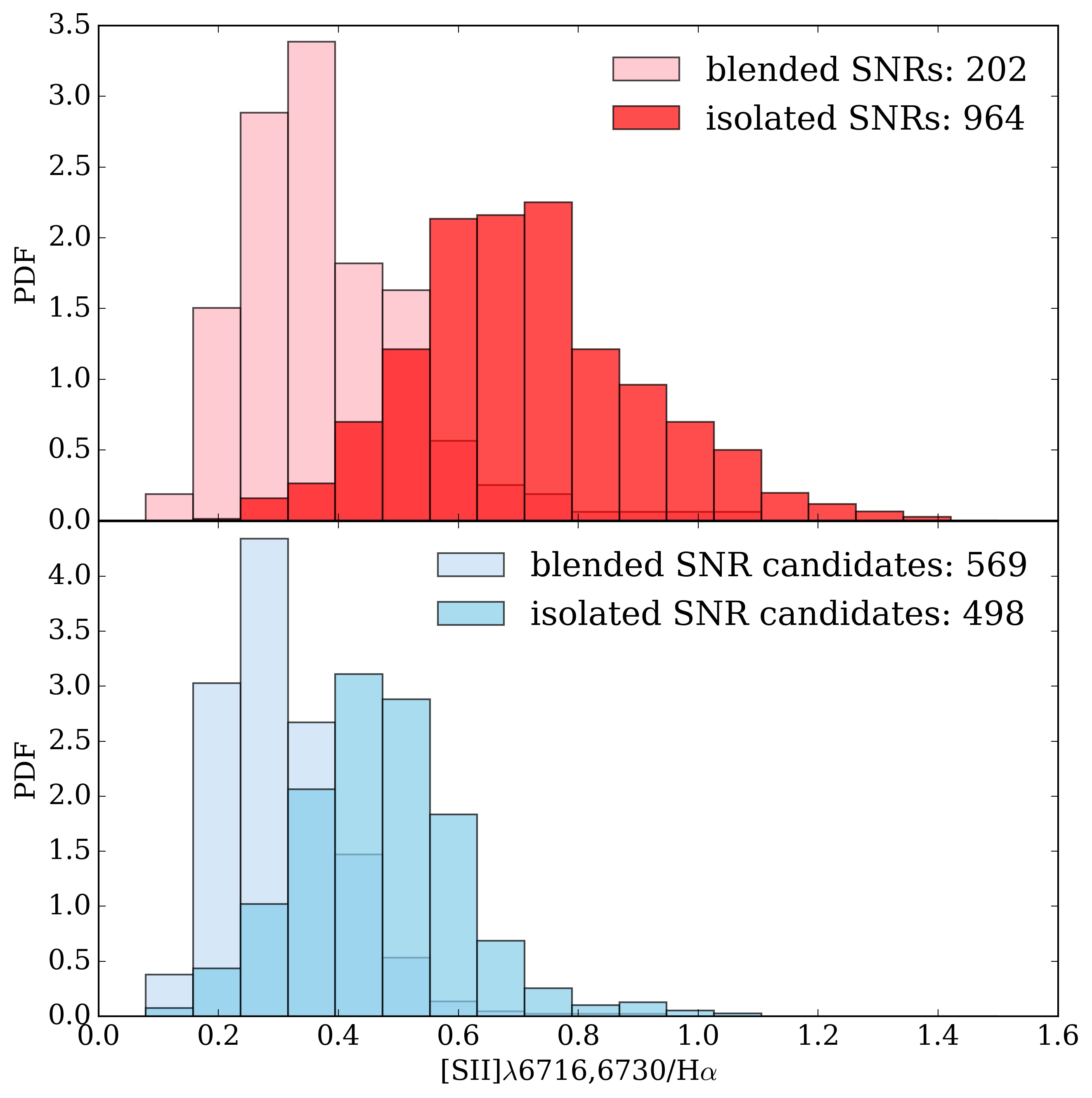}
    \caption{The probability distribution functions of [\ion{S}{ii}]/H$\alpha$ for SNRs and SNR candidates that are isolated or overlap with \ion{H}{ii} region. More SNR candidates (bottom) overlap with \ion{H}{ii} regions (blended) and have lower [\ion{S}{ii}]/H$\alpha$ values than SNRs (top). }
    \label{fig:can_HII}
\end{figure}

One reason it is difficult for us to confirm the nature of individual objects is due to the blending of emission arising from different ionizing sources within our $\sim$ 100 pc apertures. 
As shown in Table \ref{tab:number}, 202 SNRs (17\%) and 569 SNR candidates (53\%) are blended with \ion{H}{ii} regions. In Figure \ref{fig:can_HII} we see that more than half of the SNR candidates overlap with \ion{H}{ii} regions and that these objects exhibit systematically lower [\ion{S}{ii}]/H$\alpha$ line ratios. This is expected if bright photoionized emission, which exhibits systematically lower ratios, is blended with an unresolved SNR. Clearly, we have more confidence in classifying objects as SNRs when they are isolated. At our resolution, we cannot determine if blended objects are physically associated with \ion{H}{ii} regions or they just appear in projection against the \ion{H}{ii} region along our line of sight. In total, 35\% of objects in the parent sample are not isolated sources. 

As is shown in Section \ref{compare}, the majority of SNRs that exhibit large H$\alpha$ luminosities (Figure \ref{fig: SNR_HII}) 
are found to be blended with \ion{H}{ii} regions and have the lowest [\ion{S}{ii}]/H$\alpha$ and [\ion{O}{i }]/H$\alpha$ line ratios. Interestingly, if we remove these blended SNRs from the sample then we find a relatively sharp upper H$\alpha$ luminosity threshold of $\sim$10$^{37.5}$ erg s$^{-1}$.  At these distances, it remains a challenge to properly decompose SNRs from the surrounding photoionized emission, but from our combination of multiple line diagnostics and novel use of the [\ion{S}{ii}]/H$\alpha$ residual to account for blended photoionized emission, we have succeeded in identifying a large sample of objects that would have been missed by previous searches.  

This can be seen more directly in the four SNR examples shown in Figure \ref{fig:SNRs_zoom}. In the third row, it is clear from the H$\alpha$ intensity that this object sits in a complex star-forming environment, as reflected by the extended photoionization-dominated nebulae outlined in red (left panel). In the [\ion{S}{ii}]/H$\alpha$ map this object is not pronounced, but in the residual map it becomes more distinct from the background. Detailed deblending of such objects is beyond the scope of this work but it could be improved in the future by the use of higher-resolution imaging of the ionized hydrogen line emission from HST \citep{barnes2022linking} and JWST \citep{barnes2023phangs}. 


\subsection{Characterizing the shocked regions identified in Congiu et al. (2003) and Egorov et al. (2023)}
\label{sec:shocked_regions}
\revtwo{\citet{congiu2023phangs} and \citet{egorov2023phangs}, have both presented catalogs of objects classified as `shocked' using the same MUSE data for the same sample of 19 galaxies, and here we compare our catalog with their results.}

C23 used a contour-based algorithm to identify peaks in a weighted average of [\ion{O}{iii}]$\lambda5007$, H$\alpha$ and [\ion{S}{ii}]$\lambda6716,6731$ flux maps. Regions are defined by the surface brightness contours around the closest peak. In this way, the same set of 19 MUSE galaxies are segmented into 40,920 regions and \revtwo{by comparing with different model grids the} regions are classified into five types: \ion{H}{ii} regions, shocked regions (6336), ambiguous, planetary nebula, and unknown. We cross-matched these regions with our parent sample and compare their classification.

\revtwo{Our sample is 35\% smaller in total than the `shocked' sample in C23.  1676 objects (75\%) of our parent sample can be associated with the C23 segmentation mask. The 25\% of objects missed is comparable to the $\sim$35\% of objects that had not been identified in \citet{groves2023phangs}, suggesting that many SNRs are missed when selecting by emission line flux and morphology alone. 
Of the objects that can be cross-matched to the C23 catalog, 676 (38\%) of our regions are classified as shocked. This is only about 10\% of the total number of shocked regions in C23. The SNRs sample shows a better overall agreement, with 512 (56\%) of our SNRs identified as shocked in C23. 
The differences in classification likely reflect a difference between our empirical diagnostics and the model grids used in that work. 
Moreover, line ratios from our integrated spectra can appear consistent with photoionization (Figure \ref{fig:OI_SII_can}), given the frequent blending with \ion{H}{ii} regions and the contamination from the DIG background, which would result in a C23 classification of `\ion{H}{ii} region'. }


To further demonstrate that the usage of a combination of emission line ratios and gas kinematics helps to improve the identification of SNRs, we compare our objects with about 1500 regions of locally elevated H$\alpha$ velocity dispersion identified by E23 in the same sample of galaxies. Only 23\% of objects in our catalog reside in the regions identified by E23. This indicates that the kinematic information alone is usually not sufficient for SNR identification. Meanwhile, about 320 unresolved regions with high-velocity dispersion from E23 are considered to be probable SNR candidates because of their high [\ion{S}{ii}]/H$\alpha$ line ratio. 65\% of these regions coincide with the objects in our catalog. Visual inspection of the remaining objects demonstrates that the majority of them reside in very 
low surface brightness regions making their classification uncertain, and leaving us with only 25 (about 7\%) SNR candidates from E23 that are not present in our catalog.

\section{Conclusions}
\label{sec:conculsion}

Using optical IFU data from the PHANGS-MUSE survey \citep{emsellem2022phangs}, we have constructed a new catalog identifying 1,166 SNRs plus 1,067 SNR candidates in 19 nearby galaxies, doubling the number of known SNRs in the local universe. 

We summarize the key results below:
\begin{itemize}
    \item We use a total of five criteria to identify objects with \texttt{astrodendro} in our $\sim$50~pc resolution observations. Four criteria are based on line-ratio diagnostics ([\ion{S}{ii}]/H$\alpha$, [\ion{O}{i}]/H$\alpha$, and two BPT diagnostics), and the fifth criterion is based on the [\ion{S}{ii}] emission line velocity dispersion. H$\alpha$ morphology is not used as a criterion in our selection. 
    \item Instead of identifying sources directly in the line ratio maps, we construct [\ion{S}{ii}]/H$\alpha$ and [\ion{O}{i}]/H$\alpha$ residual maps, where we can identify objects that emit brightly at these line ratios given their H$\alpha$ surface brightness. This allows us to naturally account for background and foreground photoionized emission, consisting of contributions from both \ion{H}{ii} regions and DIG.
    \item Using multiple diagnostics increases our sample size and provides a natural mechanism to robustly construct a SNR catalog. Our SNR catalog consists of all objects identified by at least two criteria, or having integrated line ratios meeting both of the criteria [\ion{S}{ii}]/H$\alpha$ $>$ 0.4 and [\ion{O}{I}]/H$\alpha$ $>$ 0.1. The remaining objects are classified as SNR candidates. 
    \item We validate our method by applying it to MUSE observations of M83 \citep{dellabruna2022}. All of the SNRs we identify using our method are found in literature catalogs (L22), along with 77\% of our SNR candidates, giving us confidence that our method is robustly identifying SNRs. 
    \item We demonstrate that the SNRs in our sample have line ratios and line kinematics distinct from \ion{H}{ii} regions, are consistent with MAPPINGS shock models and are similar to literature SNR samples. Variations between galaxies in our sample are consistent with the expected variations with metallicity, given the difference in galaxy stellar mass found in our sample. 
    \item We estimate the total number of expected SNRs based on theoretical models, given the current SFR in each galaxy, and we find this agrees within a factor of two with the total number of objects identified. This suggests that while our sample is incomplete, it is illustrative of the SNR population.
    \item By using line ratios and line kinematics to identify our objects, we are able to locate SNRs that are seen in projection against bright star-forming regions. 17\% of our SNRs are blended with \ion{H}{ii} regions, while 52\% of the SNR candidates are blended. This blending makes confirmation of SNR candidates as true SNRs challenging.
\end{itemize}

Building statistical samples of SNRs provides a new avenue to quantify the impact of stellar feedback, however, it remains difficult to estimate how complete these SNR catalogs are. By studying a large sample of galaxies with IFU data, we have begun exploring how  automatic identification of objects allows us to systematically produce large, uniform catalogs. Based on the multiple different criteria explored in this work, [\ion{O}{i}]/H$\alpha$ is the most successful at selecting objects for our parent sample (62\%). BPT diagnostics select fewer objects, but they are almost all ($\sim$90\%) identified as robust (isolated) SNRs. 

With these techniques, it is easy to imagine that the next steps can include the injection of simulated objects with different characteristics to quantitatively determine how many can be recovered. High physical resolution simulations have also already begun to provide insights into the interpretation of these variable diagnostics, and the importance of accounting for contaminating background emission \citep{makarenko2023supernova}. Given the growing number of galaxies observed with multiple emission lines mapped at $\sim$100~pc scales, these automated approaches to identifying SNRs have great potential to significantly increase our SNR samples in the coming years. 

\begin{acknowledgements}
We thank Dr. Angela Adamo for generously sharing M83 data.
We extend our sincere gratitude to Dr. Eva Grebel and Dr. Andreas Sander for their valuable discussions and contributions that inspired the concept of the supernova rate (SN rate) in our research.\\

JL, KK, J.E.M-D and OE gratefully acknowledge funding from the Deutsche Forschungsgemeinschaft (DFG, German Research Foundation) in the form of an Emmy Noether Research Group (grant number KR4598/2-1, PI Kreckel) and the European Research Council’s starting grant ERC StG-101077573 (“ISM-METALS").

G.A.B. acknowledges the support from the ANID Basal project FB210003. 

F.B. acknowledges support from the INAF Fundamental Astrophysics program 2022.

SCOG and RSK acknowledge financial support from the European Research Council via the ERC Synergy Grant ``ECOGAL'' (project ID 855130),  from the German Excellence Strategy via the Heidelberg Cluster of Excellence (EXC 2181 - 390900948) ``STRUCTURES'', and from the German Ministry for Economic Affairs and Climate Action in project ``MAINN'' (funding ID 50OO2206). 

Based on observations collected at the European Southern Observatory under ESO programmes 094.C-0623 (PI: Kreckel), 095.C-0473,  098.C-0484 (PI: Blanc), 1100.B-0651 (PHANGS-MUSE; PI: Schinnerer), as well as 094.B-0321 (MAGNUM; PI: Marconi), 099.B-0242, 0100.B-0116, 098.B-0551 (MAD; PI: Carollo) and 097.B-0640 (TIMER; PI: Gadotti). \\

This research made use of \texttt{astrodendro}, a Python package to compute dendrograms of Astronomical data (http://www.dendrograms.org/). Other main Python Packages that have been used are \texttt{ASTROPY} \citep{robitaille2013astropy,price2018astropy,price2022astropy}, \texttt{NUMPY} \citep{harris2020array}, \texttt{MATPLOTLIB} \citep{hunter2007matplotlib} and \texttt{PYNEB} \citep{luridiana2015pyneb}.\\

Table 1 lists the distances that were compiled by \citet{anand2021distances} from \citet{freedman2001final,nugent2006toward,jacobs2009extragalactic,kourkchi2017galaxy,shaya2017action,kourkchi2020cosmicflows} and \citet{scheuermann2022planetary}.
\end{acknowledgements}

%
\bibliographystyle{aa} 
\bibliography{aanda} 
%
\begin{appendix}

\section{Masked regions within each galaxy}
\label{sec:env_masks}
Environmental masks that applied to each criterion for 19 galaxies are described in Table \ref{tab:mask}. Environment masks are defined in \citet{querejeta2021stellar}, 1 is a 'center' mask, 2 is a 'bar' mask, 3 is a 'bar ends' mask, and 9 is the 'interbar' mask. They are applied to avoid over-selecting shocked regions around the center or along the stellar bar. For NGC 1566 and NGC 3351, their center is not sufficient to cover the above shocked regions, so the convolved mask of center and bar masks were applied, with corresponding Gaussian kernels as listed. 
\setcounter{table}{0}
\renewcommand{\thetable}{A.\arabic{table}}
\begin{sidewaystable*}
\caption{Environmental masks that applied to each criterion for 19 galaxies. Environment masks are defined in \citet{querejeta2021stellar}, 1 is a 'center' mask, 2 is a 'bar' mask, 3 is a 'bar ends' mask, and 9 is 'interbar' mask. They are applied to avoid over-selected shocked regions around the center or along the bar. For NGC 1566 and NGC 3351, their center is not sufficient to cover the above shocked regions, so the convolved mask of center and bar masks were applied, with corresponding Gaussian kernels as listed. \label{tab:mask}}
\centering
\begin{tabular}{ccccccc}
\hline\hline\noalign{\vskip 0.05in}
		gname&	OI residual&	SII residual&	 $\sigma$(SII)&	BPT\_OI&	BPT\_SII	&comments\\
	\hline\noalign{\vskip 0.05in}
IC5332&	1	&1	&1	&1	&1	\\
NGC628&	1	&1&	1&	1&	1\\	
NGC1087	&2\&9	&2\&9	&2\&9	&2\&9	&2\&9	\\
NGC1300	&1\&2&	1\&2&	1&	1&	1\\	
NGC1365&	1\&2	&1\&2	&1\&2	&1\&2	&1\&2\\	
NGC1385&	-	&-&	-&	-&	-\\	
NGC1433&	1	&1\&2	&1	&1	&1\\	
NGC1512&	1&	1\&2&	1&	1&	1\\	
NGC1566&	conv(1\&2)&	conv(1\&2)&	conv(1\&2)&	conv(1\&2)&	conv(1\&2)&	kernel = Gaussian2DKernel(x\_stddev=5), conv = convolve(env\_masks.data, kernel)\\
NGC1672&	1\&2&	1\&2&	1\&2&	1\&2&	1\&2\\	
NGC2835&	1&	1&	1&	1&	1\\	
NGC3351&	conv(1\&2)&	conv(1\&2)&	conv(1\&2)&	conv(1\&2)&	conv(1\&2)&	kernel = Gaussian2DKernel(x\_stddev=10), conv = convolve(env\_masks.data, kernel)\\
NGC3627&	1\&2\&3&	1\&2\&3&	1\&2\&3&	1\&2\&3&	1\&2\&3\\	
NGC4254&	1	&1&	1&	1&	1\\	
NGC4303&	1	&1	&1	&1	&1	\\
NGC4321&	1&	1&	1&	1&	1\\	
NGC4535&	1&	1&	1&	1&	1	\\
NGC5068&	1&	1&  1	&1	&1	\\
NGC7496&	1\&2&	1\&2&	1\&2	&1\&2	&1\&2	\\
\hline
\end{tabular}
\end{sidewaystable*}

\section{Atlas of SNRs locations  PHANGS-MUSE galaxies}\label{rest}
\begin{figure*}
    \centering
    \includegraphics[width=\textwidth]{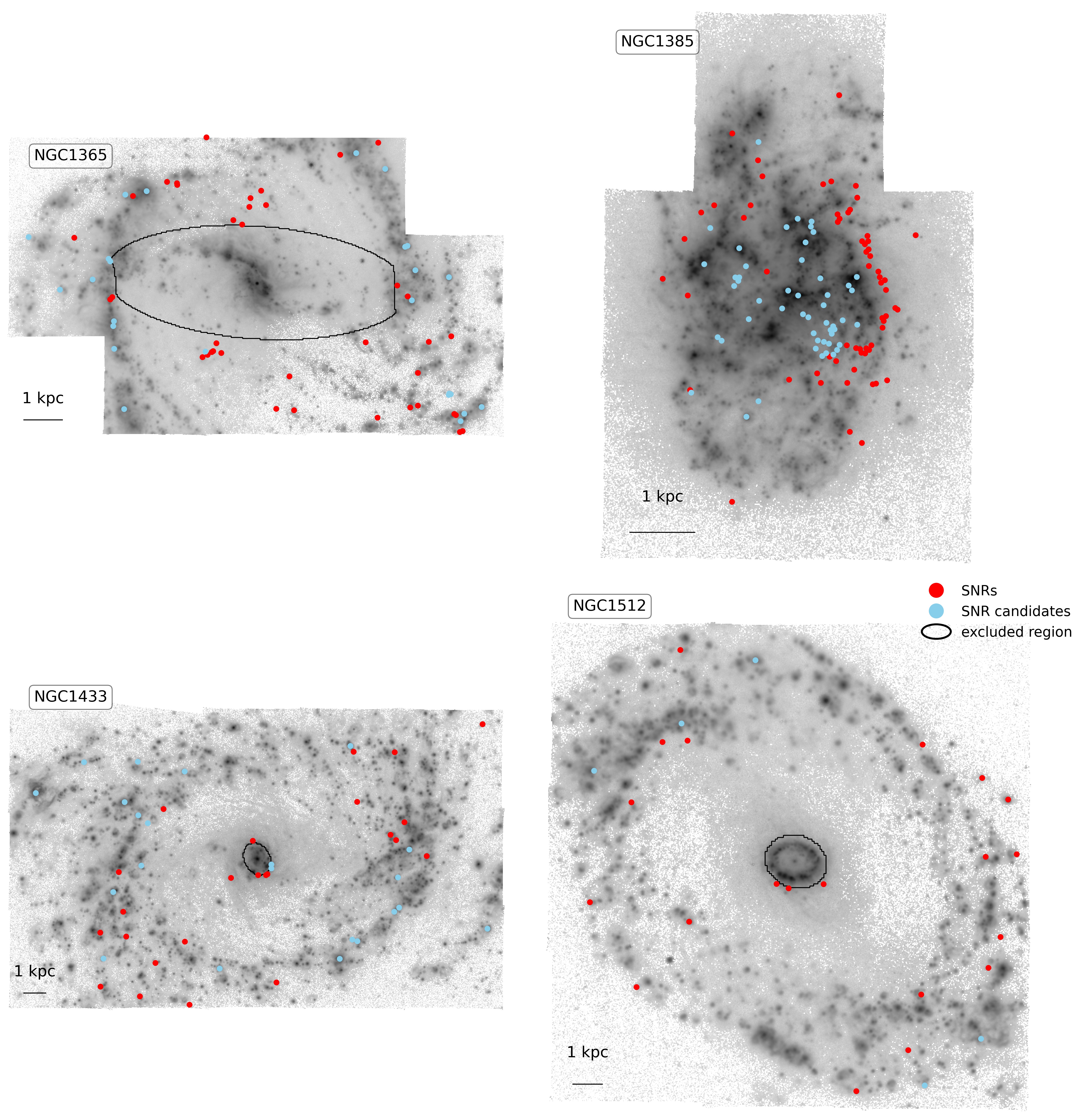} 
     \caption{The same as in Figure \ref{fig:19gal}.\label{fig:19gala}}
\end{figure*}

\begin{figure*}
    \centering
    \includegraphics[width=\textwidth]{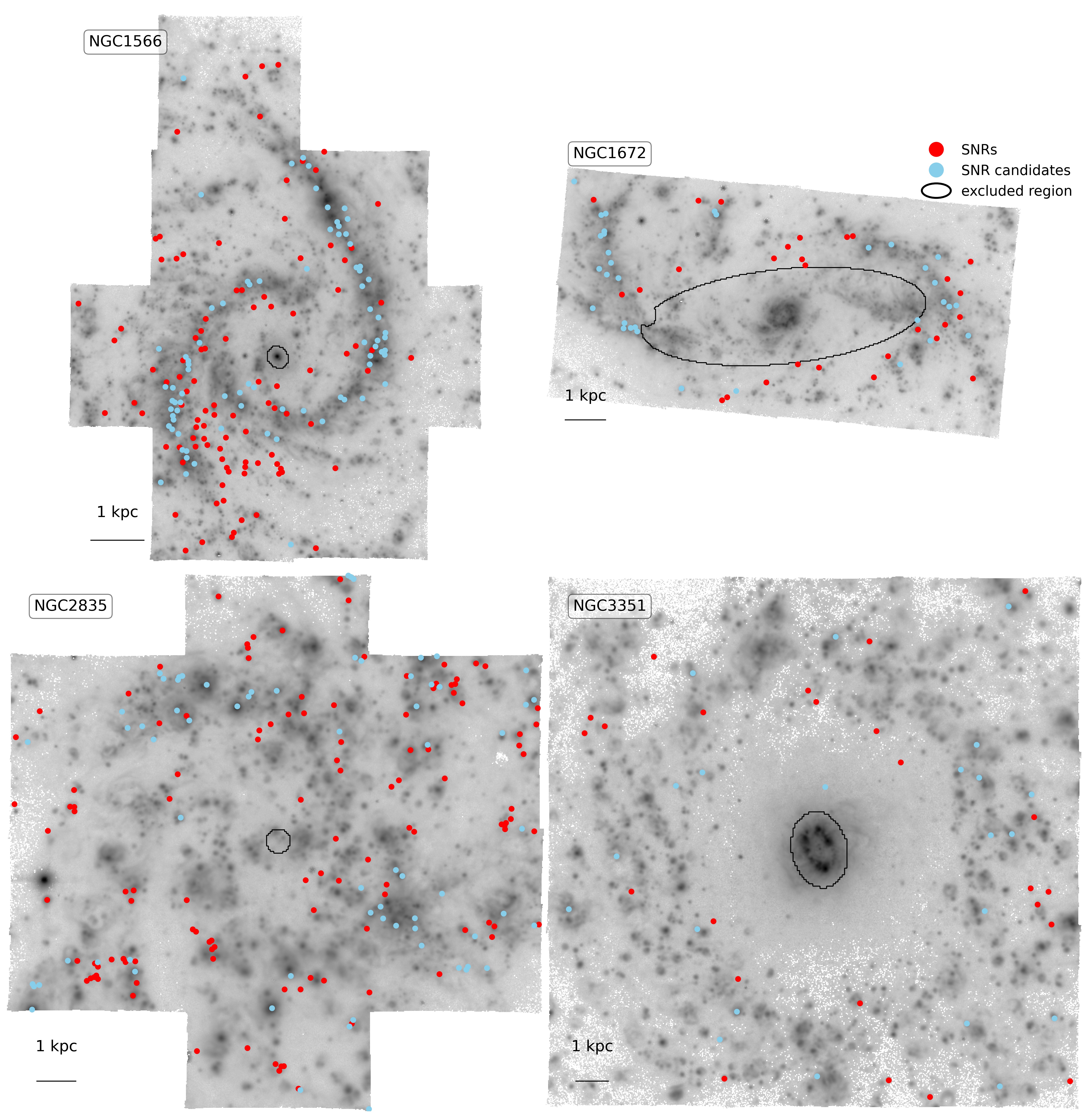}
    \caption{The same as in Figure \ref{fig:19gal}.\label{fig:19galb}}
\end{figure*}

\begin{figure*}
    \centering
    \includegraphics[width=\textwidth]{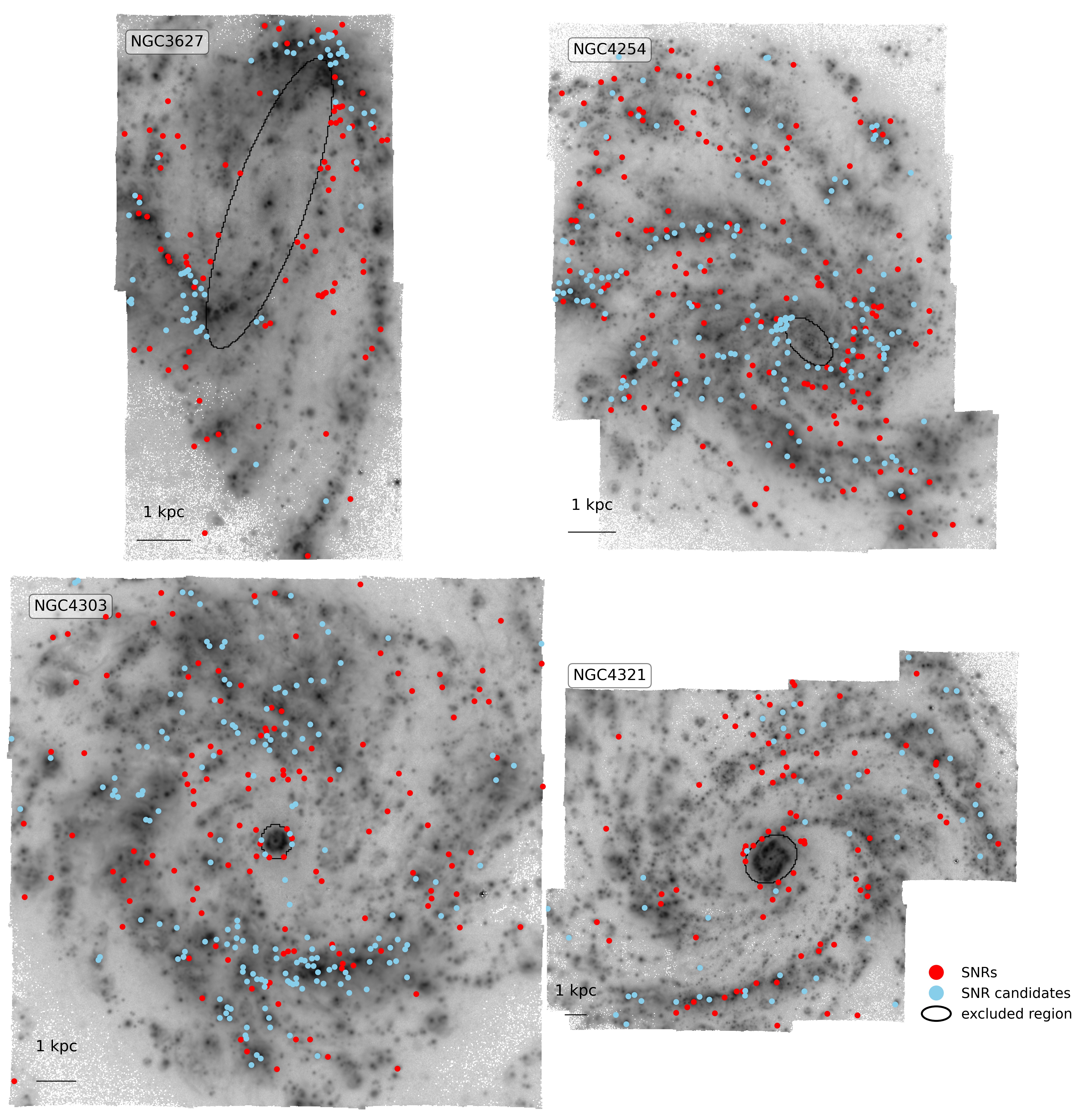}
    \caption{The same as in Figure \ref{fig:19gal}.\label{fig:19galc}}
\end{figure*}

\begin{figure*}
    \centering
    \includegraphics[width=\textwidth]{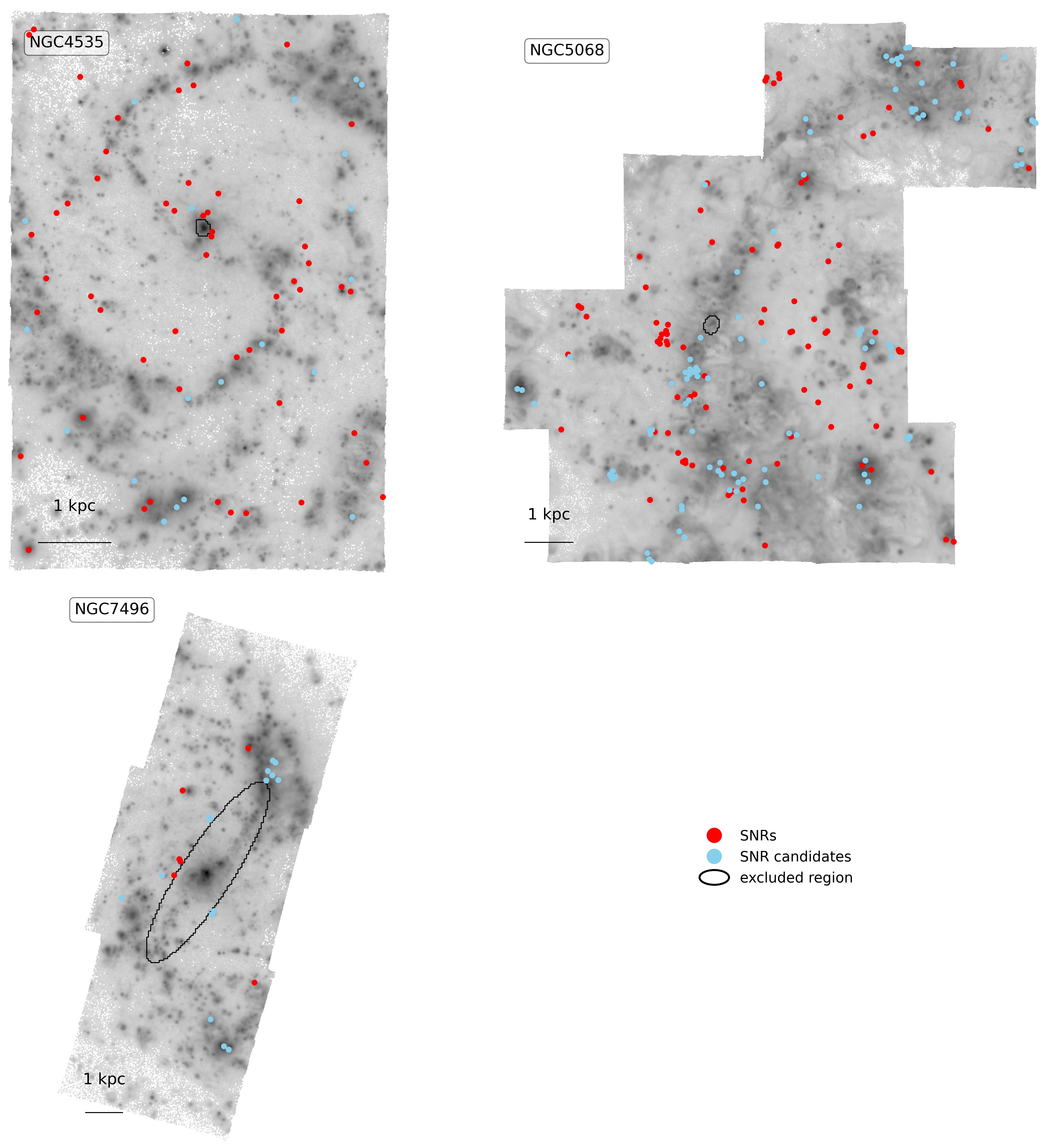}
    \caption{The same as in Figure \ref{fig:19gal}.\label{fig:19gald}}
\end{figure*}

2233 objects are identified in 19 PHANGS-MUSE galaxies, only three of them are shown in Figure \ref{fig:19gal}. The remaining 16 galaxies are shown in Figures \ref{fig:19gala}-\ref{fig:19gald}. They display the distribution of identified objects across different galaxies. 
\section{Fitted emission lines and masked skylines while running DAP.}\label{lines}
In Table \ref{tab:lines}, we listed the fitted emission lines, skylines that have been masked, ISM absorption regions that need to be masked while running the Data Analysis Pipeline in the first column. Not all lines are fitted independently, their kinematic dependency is shown in the second column.
\setcounter{table}{0}
\renewcommand{\thetable}{C.\arabic{table}}
\begin{table*}
	\centering
	\caption{Fitted emission lines and masked skylines while running DAP.}
	\begin{tabular}{c  c} 
	\hline\hline\noalign{\vskip 0.05in}
		Line name& Dependency\\
  \hline\noalign{\vskip 0.05in}
  \multicolumn{2}{c}{Forbidden and recombination emission lines}\\
  \hline\noalign{\vskip 0.05in}
  \ion{He}{ii}$\lambda$4685.70&kinematics depend on [\ion{O}{iii}]$\lambda$5006.84\\
  H$\beta$$\lambda$4861.35&kinematics depend on H$\alpha$$\lambda$6562.79\\
  \ [\ion{O}{iii}]$\lambda$4958.91&kinematics and amplitude depend on [\ion{O}{iii}]$\lambda$5006.84\\
  \ [\ion{O}{iii}]$\lambda$5006.84&None\\
  \ [\ion{N}{i}]$\lambda$5197.90&kinematics depend on [\ion{N}{ii}]$\lambda$6583.45\\
  \ [\ion{N}{i}]$\lambda$5200.26&kinematics depend on [\ion{N}{ii}]$\lambda$6583.45\\
  \ [\ion{N}{ii}]$\lambda$5754.59&kinematics depend on [\ion{N}{ii}]$\lambda$6583.45\\
  \ion{He}{i}$\lambda$5875.61&kinematics depend on H$\alpha$$\lambda$6562.79\\
  \ [\ion{O}{i}]$\lambda$6300.30&kinematics depend on [\ion{N}{ii}]$\lambda$6583.45\\
  \ [\ion{S}{iii}]$\lambda$6312.06&kinematics depend on [\ion{O}{iii}]$\lambda$5006.84\\
  \ [\ion{O}{i}]$\lambda$6363.78&kinematics and amplitude depend on [\ion{O}{i}]$\lambda$6300.30\\
  \ [\ion{N}{ii}]$\lambda$6548.05&kinematics and amplitude depend on [\ion{N}{ii}]$\lambda$6583.45\\
  H$\alpha$$\lambda$6562.79&None\\
  \ [\ion{N}{ii}]$\lambda$6583.45&None\\
  \ [\ion{S}{ii}]$\lambda$6716.44&kinematics depend on [\ion{N}{ii}]$\lambda$6583.45\\
  \ [\ion{S}{ii}]$\lambda$6730.82&kinematics depend on [\ion{N}{ii}]$\lambda$6583.45\\
  \ [\ion{Fe}{ii}]$\lambda$8616.95?&kinematics depend on [\ion{N}{ii}]$\lambda$6583.45\\
  \ [\ion{Fe}{iii}]$\lambda$4881.00?&kinematics depend on [\ion{N}{ii}]$\lambda$6583.45\\
  HPaschen 11$\lambda$8862.89&kinematics depend on H$\alpha$$\lambda$6562.79\\
  \ [\ion{S}{iii}]$\lambda$9068.6 &kinematics depend on [\ion{O}{iii}]$\lambda$5006.84\\
  \hline\noalign{\vskip 0.05in}
  \multicolumn{2}{c}{Masked lines}\\
  \hline\noalign{\vskip 0.05in}
  \ion{He}{i}$\lambda$6678.15&kinematics depend on H$\alpha$$\lambda$6562.79\\
  \hline\noalign{\vskip 0.05in}
  \multicolumn{2}{c}{Sky lines are fitted and used to remove sky emission}\\
  \hline\noalign{\vskip 0.05in}
  $\lambda$5577.34\\
  $\lambda$6300.30 \\
  $\lambda$6363.78\\
  \hline\noalign{\vskip 0.05in}
  \multicolumn{2}{c}{Masked regions due to the ISM absorption}\\
  \hline\noalign{\vskip 0.05in}
  \ion{Na}{i}$\lambda$5889.95\\
  \ion{Na}{i}$\lambda$5895.92\\
	\hline\noalign{\vskip 0.05in}
		\hline
	\end{tabular}
\label{tab:lines}
\end{table*}
\section{Catalog of Objects identified in M83}
\label{app:m83}
In Section \ref{sec:m83} we validate our method by applying it to archival MUSE observations of M83. The galaxy is two to four times closer (D = 4.61~Mpc; \citealt{saha2006cepheid}) than galaxies in our PHANGS-MUSE sample, and the line maps and line kinematic maps used are taken from \cite{dellabruna2022}, which used a different approach to line fitting and spatial binning. For these reasons, we do not consider this catalog completely equivalent to what was produced for the PHANGS-MUSE galaxies, but we provide the object locations and classifications (SNR or SNR candidate) for all 125 objects in the following table. 
\clearpage 


\section*{Supplementary material}

Table D.1: Identified objects in M83.

| RA | DEC | OI residual | SII residual | $\sigma$(SII) | BPT_OI | BPT_SII | match |
|---|---|---|---|---|---|---|---|
| 204.2644509 | -29.90044891 | TRUE | FALSE | FALSE | FALSE | FALSE | 1 |
| 204.2684764 | -29.89644006 | TRUE | TRUE | FALSE | FALSE | FALSE | 2 |
| 204.2217992 | -29.89035411 | TRUE | FALSE | FALSE | TRUE | FALSE | 2 |
| 204.2794457 | -29.88916123 | TRUE | TRUE | FALSE | TRUE | TRUE | 4 |
| 204.2300562 | -29.88472005 | TRUE | TRUE | FALSE | TRUE | FALSE | 3 |
| 204.2294845 | -29.8846112 | TRUE | TRUE | TRUE | TRUE | FALSE | 4 |
| 204.2317025 | -29.884343 | TRUE | TRUE | TRUE | FALSE | FALSE | 3 |
| 204.2311406 | -29.88429423 | TRUE | FALSE | FALSE | FALSE | TRUE | 2 |
| 204.2411654 | -29.8840859 | TRUE | TRUE | TRUE | TRUE | FALSE | 4 |
| 204.2458424 | -29.88369318 | TRUE | TRUE | FALSE | FALSE | FALSE | 2 |
| 204.2146209 | -29.88367883 | TRUE | FALSE | FALSE | FALSE | FALSE | 1 |
| 204.2121447 | -29.88300623 | TRUE | TRUE | FALSE | FALSE | FALSE | 2 |
| 204.2342854 | -29.88199423 | TRUE | TRUE | FALSE | FALSE | FALSE | 2 |
| 204.2219733 | -29.8800032 | TRUE | TRUE | FALSE | TRUE | TRUE | 4 |
| 204.2095214 | -29.8797877 | TRUE | TRUE | FALSE | FALSE | FALSE | 2 |
| 204.2234408 | -29.87949562 | TRUE | TRUE | TRUE | TRUE | TRUE | 5 |
| 204.2311919 | -29.87875809 | TRUE | TRUE | FALSE | TRUE | TRUE | 4 |
| 204.2221012 | -29.8784649 | TRUE | TRUE | FALSE | FALSE | FALSE | 2 |
| 204.2135546 | -29.87803204 | TRUE | FALSE | FALSE | FALSE | FALSE | 1 |
| 204.2293475 | -29.87765827 | TRUE | TRUE | TRUE | FALSE | TRUE | 4 |
| 204.2119211 | -29.87767374 | TRUE | TRUE | FALSE | TRUE | FALSE | 3 |
| 204.2994947 | -29.87094228 | TRUE | TRUE | TRUE | TRUE | TRUE | 5 |
| 204.2256829 | -29.86925442 | TRUE | TRUE | FALSE | FALSE | TRUE | 3 |
| 204.250294 | -29.86908168 | TRUE | TRUE | FALSE | TRUE | TRUE | 4 |
| 204.2509311 | -29.86839083 | TRUE | TRUE | TRUE | TRUE | TRUE | 5 |
| 204.2479135 | -29.86776394 | TRUE | TRUE | TRUE | TRUE | TRUE | 5 |
| 204.2854061 | -29.86719197 | TRUE | TRUE | FALSE | FALSE | FALSE | 2 |
| 204.2501895 | -29.86712891 | TRUE | TRUE | FALSE | FALSE | FALSE | 2 |
| 204.2586728 | -29.86621778 | TRUE | TRUE | TRUE | TRUE | TRUE | 5 |
| 204.2859831 | -29.86485631 | TRUE | TRUE | FALSE | FALSE | FALSE | 2 |
| 204.2977868 | -29.86147724 | TRUE | TRUE | TRUE | TRUE | TRUE | 5 |
| 204.3044806 | -29.86067448 | TRUE | TRUE | FALSE | FALSE | FALSE | 2 |
| 204.285712 | -29.85975022 | TRUE | TRUE | TRUE | TRUE | TRUE | 5 |
| 204.2811515 | -29.85925031 | TRUE | TRUE | TRUE | TRUE | TRUE | 5 |
| 204.2919771 | -29.85782428 | TRUE | TRUE | FALSE | FALSE | FALSE | 2 |
| 204.2600849 | -29.85723198 | TRUE | TRUE | FALSE | TRUE | TRUE | 4 |
| 204.2833441 | -29.85453299 | TRUE | FALSE | FALSE | FALSE | FALSE | 1 |
| 204.2571478 | -29.85368573 | TRUE | TRUE | TRUE | TRUE | TRUE | 5 |
| 204.2820262 | -29.85278978 | TRUE | FALSE | FALSE | FALSE | FALSE | 1 |
| 204.2299161 | -29.84480477 | TRUE | TRUE | FALSE | TRUE | FALSE | 3 |
| 204.229708 | -29.84459285 | TRUE | FALSE | FALSE | FALSE | TRUE | 2 |
| 204.226026 | -29.8411766 | TRUE | TRUE | TRUE | TRUE | FALSE | 4 |
| 204.2271069 | -29.84061925 | TRUE | TRUE | FALSE | FALSE | FALSE | 2 |
| 204.2768417 | -29.84028213 | TRUE | TRUE | FALSE | TRUE | TRUE | 4 |
| 204.2286799 | -29.83850856 | TRUE | TRUE | FALSE | TRUE | TRUE | 4 |
| 204.2570585 | -29.90280692 | FALSE | TRUE | FALSE | TRUE | TRUE | 3 |
| 204.2669936 | -29.90066473 | FALSE | TRUE | FALSE | FALSE | FALSE | 1 |
| 204.2420095 | -29.89587097 | FALSE | TRUE | FALSE | FALSE | FALSE | 1 |
| 204.2604374 | -29.89571024 | FALSE | TRUE | FALSE | FALSE | FALSE | 1 |
| 204.2310394 | -29.89468833 | FALSE | TRUE | FALSE | FALSE | FALSE | 1 |
| 204.27771 | -29.89289502 | FALSE | TRUE | FALSE | FALSE | FALSE | 1 |
| 204.2381533 | -29.89269151 | FALSE | TRUE | FALSE | TRUE | TRUE | 3 |
| 204.2499832 | -29.88996355 | FALSE | TRUE | FALSE | FALSE | FALSE | 1 |

| | | | | | | | |
|---|---|---|---|---|---|---|---|
| 204.2516662 | -29.88972639 | FALSE | TRUE | FALSE | FALSE | FALSE | 1 |
| 204.2671822 | -29.88784274 | FALSE | TRUE | FALSE | FALSE | FALSE | 1 |
| 204.2820848 | -29.88374798 | FALSE | TRUE | FALSE | FALSE | FALSE | 1 |
| 204.230436 | -29.88153219 | FALSE | TRUE | FALSE | FALSE | TRUE | 2 |
| 204.2585384 | -29.8803596 | FALSE | TRUE | FALSE | FALSE | FALSE | 1 |
| 204.2293476 | -29.88012847 | FALSE | TRUE | FALSE | FALSE | FALSE | 1 |
| 204.2740267 | -29.87947508 | FALSE | TRUE | FALSE | FALSE | FALSE | 1 |
| 204.2362217 | -29.87874652 | FALSE | TRUE | FALSE | FALSE | FALSE | 1 |
| 204.2112029 | -29.87823226 | FALSE | TRUE | FALSE | FALSE | FALSE | 1 |
| 204.2365291 | -29.87801058 | FALSE | TRUE | FALSE | FALSE | FALSE | 1 |
| 204.2409307 | -29.8777841 | FALSE | TRUE | FALSE | FALSE | FALSE | 1 |
| 204.2258752 | -29.87764806 | FALSE | TRUE | FALSE | FALSE | FALSE | 1 |
| 204.2501167 | -29.87762316 | FALSE | TRUE | FALSE | FALSE | FALSE | 1 |
| 204.223054 | -29.87734384 | FALSE | TRUE | FALSE | FALSE | FALSE | 1 |
| 204.2391076 | -29.87725384 | FALSE | TRUE | FALSE | FALSE | FALSE | 1 |
| 204.2443313 | -29.87690358 | FALSE | TRUE | FALSE | FALSE | FALSE | 1 |
| 204.2442496 | -29.87431932 | FALSE | TRUE | FALSE | FALSE | FALSE | 1 |
| 204.2454162 | -29.87396653 | FALSE | TRUE | FALSE | FALSE | FALSE | 1 |
| 204.2529456 | -29.87279832 | FALSE | TRUE | FALSE | FALSE | FALSE | 1 |
| 204.281562 | -29.87202737 | FALSE | TRUE | FALSE | FALSE | TRUE | 2 |
| 204.2207596 | -29.87107315 | FALSE | TRUE | FALSE | FALSE | TRUE | 2 |
| 204.2534307 | -29.87015945 | FALSE | TRUE | TRUE | TRUE | TRUE | 4 |
| 204.2465962 | -29.86962522 | FALSE | TRUE | FALSE | FALSE | FALSE | 1 |
| 204.2525727 | -29.86930212 | FALSE | TRUE | TRUE | FALSE | FALSE | 2 |
| 204.2484159 | -29.86907761 | FALSE | TRUE | FALSE | FALSE | FALSE | 1 |
| 204.2568639 | -29.86822971 | FALSE | TRUE | FALSE | FALSE | FALSE | 1 |
| 204.2578842 | -29.86700242 | FALSE | TRUE | FALSE | FALSE | FALSE | 1 |
| 204.2584275 | -29.86428098 | FALSE | TRUE | FALSE | FALSE | FALSE | 1 |
| 204.2639809 | -29.86392586 | FALSE | TRUE | FALSE | FALSE | FALSE | 1 |
| 204.2465503 | -29.86329952 | FALSE | TRUE | FALSE | FALSE | TRUE | 2 |
| 204.2949993 | -29.86241028 | FALSE | TRUE | FALSE | FALSE | FALSE | 1 |
| 204.264702 | -29.86149558 | FALSE | TRUE | FALSE | FALSE | FALSE | 1 |
| 204.2544683 | -29.86157964 | FALSE | TRUE | FALSE | FALSE | FALSE | 1 |
| 204.2592479 | -29.86127812 | FALSE | TRUE | FALSE | FALSE | FALSE | 1 |
| 204.286481 | -29.86047591 | FALSE | TRUE | TRUE | TRUE | TRUE | 4 |
| 204.2412203 | -29.85932485 | FALSE | TRUE | FALSE | FALSE | TRUE | 2 |
| 204.2877302 | -29.85929447 | FALSE | TRUE | FALSE | FALSE | FALSE | 1 |
| 204.2930491 | -29.85811013 | FALSE | TRUE | TRUE | FALSE | FALSE | 2 |
| 204.2615187 | -29.85741316 | FALSE | TRUE | FALSE | FALSE | FALSE | 1 |
| 204.2658561 | -29.85746575 | FALSE | TRUE | FALSE | FALSE | FALSE | 1 |
| 204.2404214 | -29.85724456 | FALSE | TRUE | FALSE | FALSE | TRUE | 2 |
| 204.2660981 | -29.85723297 | FALSE | TRUE | FALSE | FALSE | FALSE | 1 |
| 204.2292471 | -29.85689938 | FALSE | TRUE | FALSE | FALSE | FALSE | 1 |
| 204.2398281 | -29.85651329 | FALSE | TRUE | FALSE | FALSE | FALSE | 1 |
| 204.2420434 | -29.85632066 | FALSE | TRUE | FALSE | FALSE | FALSE | 1 |
| 204.2670838 | -29.85640905 | FALSE | TRUE | FALSE | FALSE | FALSE | 1 |
| 204.2674814 | -29.85633153 | FALSE | TRUE | FALSE | FALSE | FALSE | 1 |
| 204.2426143 | -29.85610931 | FALSE | TRUE | FALSE | FALSE | FALSE | 1 |
| 204.2514087 | -29.85576722 | FALSE | TRUE | FALSE | TRUE | TRUE | 3 |
| 204.2325097 | -29.85544753 | FALSE | TRUE | FALSE | FALSE | FALSE | 1 |
| 204.2662443 | -29.85551952 | FALSE | TRUE | FALSE | FALSE | FALSE | 1 |
| 204.2713089 | -29.8552245 | FALSE | TRUE | FALSE | FALSE | FALSE | 1 |
| 204.2662435 | -29.85499119 | FALSE | TRUE | FALSE | FALSE | FALSE | 1 |
| 204.2658529 | -29.85400158 | FALSE | TRUE | FALSE | FALSE | FALSE | 1 |
| 204.2446542 | -29.85012319 | FALSE | TRUE | FALSE | TRUE | TRUE | 3 |
| 204.2316606 | -29.84998917 | FALSE | TRUE | FALSE | FALSE | FALSE | 1 |

| | | | | | | | |
|---|---|---|---|---|---|---|---|
| 204.2542134 | -29.84900798 | FALSE | TRUE  | FALSE | FALSE | TRUE  | 2 |
| 204.2306176 | -29.84827798 | FALSE | TRUE  | FALSE | TRUE  | TRUE  | 3 |
| 204.2183088 | -29.84554202 | FALSE | TRUE  | FALSE | FALSE | FALSE | 1 |
| 204.2305171 | -29.84366947 | FALSE | TRUE  | FALSE | TRUE  | FALSE | 2 |
| 204.2181362 | -29.84254466 | FALSE | TRUE  | FALSE | FALSE | FALSE | 1 |
| 204.2487567 | -29.84236869 | FALSE | TRUE  | FALSE | FALSE | FALSE | 1 |
| 204.2265473 | -29.83821292 | FALSE | TRUE  | FALSE | FALSE | FALSE | 1 |
| 204.2563517 | -29.83739678 | FALSE | TRUE  | FALSE | FALSE | FALSE | 1 |
| 204.2769334 | -29.83505468 | FALSE | TRUE  | FALSE | FALSE | FALSE | 1 |
| 204.2367137 | -29.83047398 | FALSE | TRUE  | FALSE | TRUE  | TRUE  | 3 |
| 204.2124283 | -29.8737564  | FALSE | FALSE | TRUE  | FALSE | FALSE | 1 |
| 204.2457516 | -29.86550793 | FALSE | FALSE | TRUE  | FALSE | FALSE | 1 |
| 204.2474556 | -29.86504903 | FALSE | FALSE | TRUE  | FALSE | FALSE | 1 |
| 204.2491144 | -29.86303519 | FALSE | FALSE | TRUE  | FALSE | FALSE | 1 |
| 204.2454392 | -29.86759737 | FALSE | FALSE | FALSE | FALSE | TRUE  | 1 |
| 204.2467429 | -29.82869852 | FALSE | TRUE  | FALSE | FALSE | FALSE | 1 |

\label{tab:m83}

\end{appendix}
\end{document}